\documentclass[12pt,authoryear]{elsarticle}
\usepackage{subfigure}
\usepackage{hyperref}

\usepackage{natbib}
\usepackage{amssymb}
\usepackage{amsfonts}
\usepackage{dsfont}
\usepackage{amsmath}
\usepackage{multirow}

\usepackage{upgreek}
\input{mystyle.sty}

\newcommand{\be}{\begin{equation}}
\newcommand{\ee}{\end{equation}}
\newcommand{\bq}{\begin{eqnarray}}
\newcommand{\eq}{\end{eqnarray}}

\newcommand{\bc}{\begin{center}}
\newcommand{\ec}{\end{center}}
\newcommand{\beq}{\begin{equation}}
\newcommand{\eeq}{\end{equation}}
\newcommand{\bea}{\begin{eqnarray}}
\newcommand{\eea}{\end{eqnarray}}

\newcommand{\ket}[1]{\ensuremath{| #1 \rangle}} 
 
\newcommand{\braket}[2]{\ensuremath{\langle #1 | #2 \rangle}}

\journal{Advances in Atomic, Molecular and Optical Physics}

\begin{document}

\begin{frontmatter}

\title{Direct Laser Cooling of Polyatomic Molecules}

\author{Benjamin L. Augenbraun}
\ead{augenbraun@g.harvard.edu}
\author{Lo{\"i}c Anderegg}
\author{Christian Hallas}
\author{Zack D. Lasner}
\author{Nathaniel B. Vilas}
\author{John M. Doyle}
\ead{jdoyle@g.harvard.edu}
\address{Department of Physics, Harvard University, Cambridge, MA 02138, USA}
\address{Harvard-MIT Center for Ultracold Atoms, Cambridge, MA 02138, USA}

\begin{abstract}
Over the past decade, tremendous progress has been made to extend the tools of laser cooling and trapping to molecules. Those same tools have recently been applied to polyatomic molecules (molecules containing three or more atoms). In this review, we discuss the scientific drive to bring larger molecules to ultralow temperatures, the features of molecular structure that provide the most promising molecules for this pursuit, and some technical aspects of how lasers can be used to control the motion and quantum states of polyatomic molecules. We also present opportunities for and challenges to the use of polyatomic molecules for science and technology.
\end{abstract}


\end{frontmatter}

DRAFT: February 18, 2023

\maketitle

\tableofcontents

\section{Introduction}
\label{sec:intro}

Detailed understanding of polyatomic molecules (those containing three or more atoms) is central to such diverse fields as chemistry, biology, and interstellar science. Beyond the inherent interest in their structures and interactions, physicists also hope to fully control polyatomic molecules at the single quantum state level for next-generation explorations. The diversity of electronic structures, geometries, and atomic constituents present in polyatomic molecules may provide powerful building blocks for quantum information processing and precision tests of fundamental physics, e.g. searching for dark matter or for new particles that help explain the matter-antimatter asymmetry of the universe. However, using complex molecules for these applications requires us to confront one of their most defining---and daunting---characteristics: their immensely rich and varied internal structures. Attempts to tame polyatomic molecules, for example by controlling their internal (e.g. vibrational/rotational) and external (motional) states have a long history both in atomic, molecular, and optical (AMO) physics and in physical chemistry. It is the purpose of this review to describe recent advances that have introduced direct laser cooling as a new element in the toolkit of polyatomic molecular control.

The already achieved exquisite control over certain quantum systems has been realized in no small part by using optical photons---this is a hallmark of modern quantum science and physical chemistry. The most recent wave of advances with atoms and molecules relies on the ability to cool, control, and detect molecules efficiently (and, ideally, nondestructively). Optical cycling, a process in which molecules are made to rapidly and repeatedly scatter many hundreds or thousands of photons, can be an effective way to carry out these tasks. Using photon cycling, scientists have exploited the mechanical effects of light to prepare and interrogate individual atoms and diatomic molecules in pristine and dynamically controllable traps (\cite{barredo2016atom,endres2016atom}). Optical photons also allow scientists to probe the fragile quantum effects that form the heart of modern quantum technologies (\cite{haroche2013nobel,wineland2013nobel}). The creation of quantum gases of atoms and the production of useful architectures for quantum computing also rely on these experimental feats. These wide ranging impacts span many frontiers of quantum science, as well as cold chemistry, and precision searches for new fundamental (particle) physics.

\subsection{Polyatomic molecules for quantum science}
Ultracold molecules are a promising platform for quantum simulation and quantum information processing due to their large electric dipole moments and the intrinsically long coherence times achievable in low-lying rotational states. Heteronuclear molecules have molecule-frame electric dipole moments, typically on the order of several debye (D), that can be accessed in the electronic ground state, eliminating the need for the short-lived excited electronic levels often employed in atomic systems. The molecules may interact via the electric dipole-dipole interaction, whose long-range and anisotropic behavior enables access to a diverse variety of Hamiltonians for quantum simulation, as well as enabling entangling gates for quantum information applications. While not always required to achieve interactions, it can be advantageous to realize a molecular dipole moment in the laboratory frame by aligning the atoms or molecules with an external electric field (either DC or microwave).

\begin{figure}[tb]
    \centering 
    \includegraphics[width=1\columnwidth]{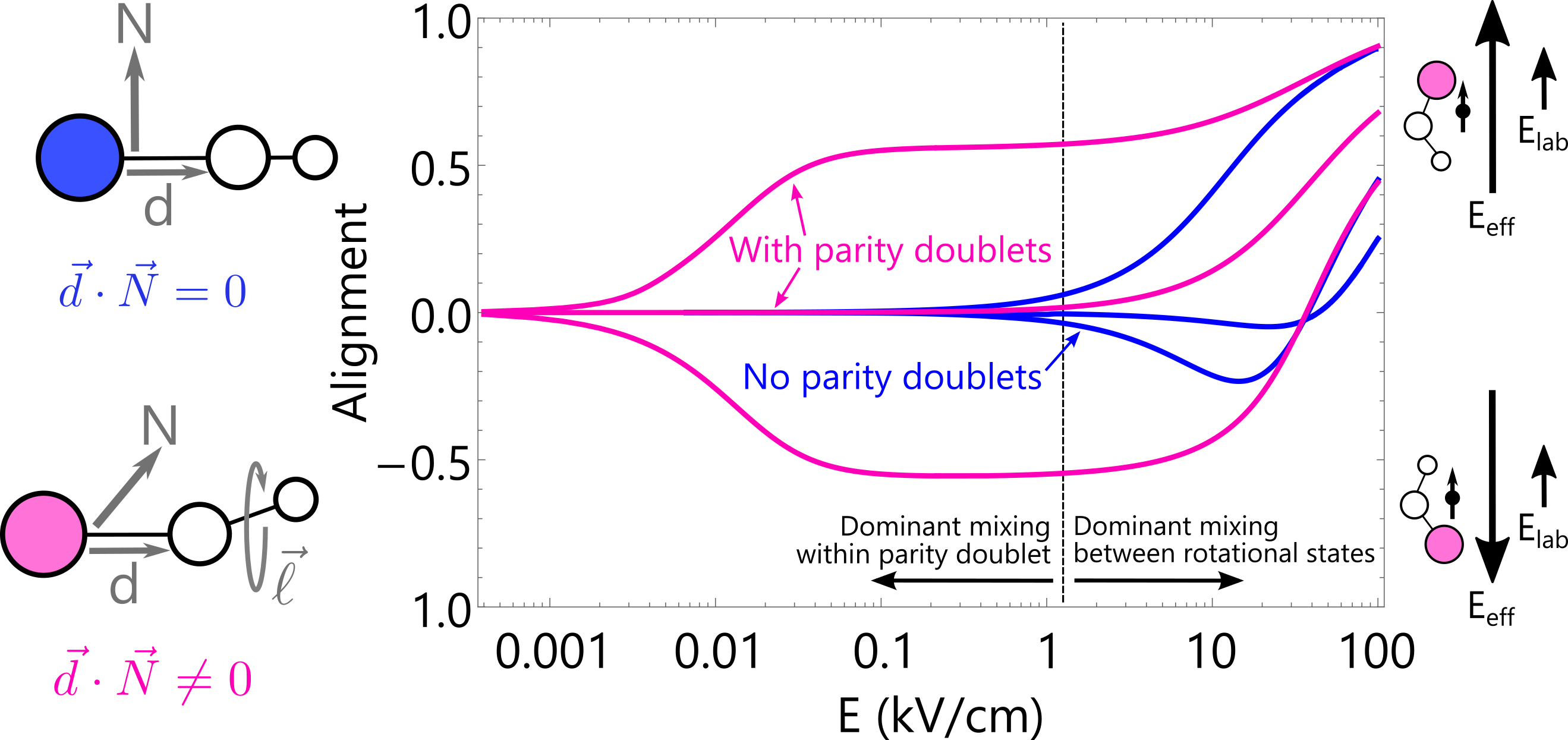}
    \vspace*{-5mm} 
    \caption{ 
    Molecular polarization as a function of applied laboratory electric field. Parameters are typical of YbOH molecules. Blue lines show alignment in a state without parity doubling ($N \leq 2$, $\abs{M_N} \leq 1$ plotted). Pink lines show alignment of $N = 1$ sublevels in a vibrational level that has parity doubling, e.g. the fundamental bending vibrational state. We indicate two electric‐field regimes: one where mixing occurs within a single rotational state and another where mixing occurs among many rotational states.
}
    \label{fig:PolarizationTriatomic}
\end{figure}

While polyatomic molecules, compared to diatomic molecules, possess greater complexity and additional degrees of freedom that need to be controlled, many of the structures present in polyatomic molecules have no analogue in atoms or diatomic molecules.  
One especially appealing feature of polyatomic molecules is the existence of parity-doubled states in the ground electronic manifold. These ``parity doublets'' comprise nearly-degenerate pairs of quantum states with opposite parity that can be mixed by a small DC electric field (often $\lesssim 100$ V/cm depending on the molecule), enabling the molecule to be easily polarized in the laboratory frame. Moreover, because this method of polarizing the molecule does not mix end-over-end rotational levels (which requires much larger electric fields, typically $>$1 kV/cm), the polarized molecules contain states with positive, negative, and near-zero lab-frame dipole moment (see Fig.~\ref{fig:PolarizationTriatomic}).

The existence of parity-doubled states is a general feature of molecules that have a nonzero projection of the total angular momentum along the molecule-frame dipole moment. In particular, for a molecule with total angular momentum $J$, whose projection onto the laboratory $Z$ axis is $M$ and whose projection onto the molecular symmetry axis is $K$, the states $2^{1/2}\lvert \pm \rangle = \lvert J, K, M \rangle \pm (-1)^{J-K} \lvert J, -K, M \rangle$ have opposite parity. States with $K\neq 0$ are universally present in polyatomic molecules: linear polyatomic molecules have angular momentum about the internuclear axis in vibrational bending modes, while nonlinear polyatomic molecules have nonzero moments of inertia about the symmetry axis even in the electronic and vibrational ground state. The degeneracy of these parity doublets is lifted by a range of mechanisms, including Coriolis interactions, nuclear spin-rotation interactions, hyperfine interactions, or molecular asymmetry (\cite{Klemperer1993Can}). The resulting energy splitting can be as large as tens of MHz (notably in linear polyatomic molecules) or as small as tens of Hz or kHz (e.g., in singlet symmetric top molecules).

For the parity doublet states $\lvert \pm \rangle$ described above, the Hamiltonian under the influence of an external DC electric field $\mathcal{E}$ is
\begin{equation}
    H =
    \begin{pmatrix}
    -\hbar \Delta / 2 & -d\mathcal{E} \\
    -d\mathcal{E} & \hbar \Delta /2
    \end{pmatrix}
\end{equation}
where $\Delta$ is the parity doublet splitting and
\begin{equation}
    d\mathcal{E} \equiv \langle + | \vec{d} \cdot \vec{\mathcal{E}} | - \rangle = d_\text{0}\mathcal{E}\frac{KM}{J(J+1)}
\end{equation}
where $d_0$ is the permanent dipole moment of the molecule. The energy eigenvalues are therefore
\begin{equation}
    E_\pm = \pm \frac{1}{2}\hbar \Delta \left[1 + 4 \left(\frac{d\mathcal{E}}{\hbar \Delta}\right)^2 \right]^{1/2}
\end{equation}
At low electric fields $d\mathcal{E}/\hbar \Delta \ll 1$ the states split apart quadratically, while at high electric fields $d\mathcal{E}/\hbar \Delta \gg 1$ the molecule becomes aligned in the laboratory frame and follows linear Stark shifts:
\begin{equation}
    E \approx -d_\text{0}\mathcal{E}\frac{KM}{J(J+1)},
\end{equation}
where $K$ and $M$ are signed quantities. 
The structure of the molecule in this ``polarized'' regime is amenable to a number of interesting quantum simulation and quantum information applications.

\subsubsection{Quantum information processing}

Since the seminal proposal by \cite{DeMille2002}, ultracold polar molecules have generated significant interest as a platform for quantum information applications, see e.g., \cite{Yelin2006, Ni2018, Sawant2019}. The key element of these proposals is the fact that polar molecules have many long-lived rotational and vibrational degrees of freedom which enable storage of quantum information, while dipole-dipole interactions enable transfer of information and entanglement between individual molecules. Recent progress toward this goal has been realized with diatomic molecules by \cite{Holland2022OnDemand} and \cite{Bao2022Dipolar}, who demonstrated dipolar coupling and entanglement between CaF molecules trapped in a tweezer array.

The linear Stark shifts discussed above for polyatomic molecules make them especially appealing for certain quantum computing and entanglement schemes, including those proposed by \cite{Andre2006Coherent, wei2011entanglement, Yu2019, Zhang2017}. For example, \cite{wei2011entanglement} studied theoretically entanglement generation in polar symmetric top molecules and identified two sets of polarized states that could be used as qubits in a quantum computer. They additionally proposed a method for realizing a CNOT entangling gate between two such molecules. Interestingly, it was also pointed out that polarized symmetric top molecules share certain similarities with the NMR quantum computing platform (see, e.g., \cite{Cory2000}). However, molecuels in optical tweezer arrays offer the possibility of isolating individual qubits and controlling entanglement on demand more cleanly than has been done with liquid-phase NMR and with intrinsic scalability.

\cite{Yu2019} proposed a quantum computing scheme harnessing the parity-doublet structure of symmetric top molecules, wherein the qubit states are contained within the $M=0$ manifold, which eliminates first-order electric field sensitivity, thereby improving robustness to external perturbations. Interactions are switched on using a third state in either the $M=+1$ or $M=-1$ manifold, and entanglement is generated using an interaction blockade mechanism. Using the linear Stark shift structure to switch dipole-dipole interactions on and off in this manner is a key advantage of polyatomic molecules.

The large number of rotational and vibrational degrees of freedom in polyatomic molecules also has potential advantages for quantum information applications. For instance, error-correcting logical qubits could be constructed from coherent superpositions of rotational states, as proposed by \cite{albert2019robust}. The large number of internal states in polyatomic molecules could also make them uniquely amenable to ``qudit'' schemes where multiple bits of information are stored in the same molecule, as described by \cite{Tesch2002, Sawant2019}. Such schemes could significantly increase the speed and fidelity of molecular quantum computers by performing the majority of gates using single-molecule operations, and only coupling molecules via dipole-dipole interactions when necessary. Full cooling and control of larger polyatomic molecules with many nuclear spins could also enable platforms where NMR-based quantum computing is performed within a single molecule, while scalability is achieved by coupling individual molecules with dipole-dipole interactions. More theoretical work is required to explore the feasibility of such schemes.

\subsubsection{Quantum simulation}

The long range and anisotropic dipolar interactions between polar molecules lend themselves to quantum simulation applications, in particular to simulation of quantum magnetism models where molecules pinned in a lattice act as effective spins (\cite{Carr2009, bohn2017cold}). \cite{wall2015quantum} provide a review of this topic. To date, much of this work has been focused on diatomic molecules, which are naturally suited to simulation of Heisenberg XY models, as first demonstrated experimentally by \cite{Yan2013} using KRb molecules. In theoretical work by \cite{micheli2006toolbox}, it was shown that microwave-dressed diatomic molecules with unpaired electron spins can be used to simulate even more general spin models, though the technical requirements of this proposal have not yet been realized in present day experiments.

Compared to diatomic molecules, polyatomic molecules are naturally suited to simulation of a greater diversity of quantum magnetism models, such as those described by \cite{Wall2013, Wall2015, wall2015quantum}. For instance, the unique rotational structure of molecules with parity doublet structure enables simulation of XYZ spin models via microwave dressing, as proposed by \cite{Wall2015}. Critically, the technical requirements of this proposal are significantly reduced compared to the method of \cite{micheli2006toolbox} for simulating an XYZ Hamiltonian with diatomic molecules, which requires more microwave frequencies and significantly smaller lattice spacings to compensate for the molecules' sparser internal structure. Another benefit of large polyatomic molecules may be their small fine and hyperfine splittings, which can be made comparable to the dipole-dipole interaction energy at larger molecular spacings. For instance, operating in the regime where the dipolar interaction energy is comparable to the spin-rotation interaction energy in molecules with an unpaired spin would enable simulation of a diverse array of lattice spin models, as described in \cite{micheli2006toolbox}. These small energy-level splittings could, however, present a challenge for achieving full control of individual quantum states in the molecule, e.g., due to off-resonant excitations during control pulses.

\cite{Wall2013} have also proposed to use the (nearly) linear Stark shifts present in many polyatomic molecules as a way to simulate lattice spin models involving polarized dipoles. Because this proposal involves using electric dipole moments to simulate magnetic dipoles, the experimental interaction strengths can be several orders of magnitude larger. Importantly, this analogy can be extended to molecular states with arbitrarily large $J$, enabling quantum simulation of magnetic dipoles with integer spin $S=J$. These features could have interesting applications in simulation of lattice spin models as well as in the study of bulk dipolar gases (see \cite{Lahaye2009} for a review on the subject), which have previously been studied using magnetic atoms with much smaller interaction energies.

\subsection{Precision measurements}

Polyatomic molecules hold further promise for a variety of precision
measurements probing the Standard Model (SM) and beyond-Standard-Model
(BSM) physics (see \cite{hutzler2020} for a focused review). At least three distinctive features of polyatomic molecules
can be leveraged for improved precision measurements. The first such
feature is the closely-spaced pairs of opposite-parity states described
above, which can be directly mixed by electric fields to orient molecules
in the lab frame, or brought to degeneracy via magnetic fields. 

Mixing opposite-parity states to orient a molecule in the lab frame
is useful for electron electric dipole moment (eEDM) searches, as
well as Schiff moment and magnetic quadrupole moment searches operating
in a similar manner. The proposals by \cite{kozyryev2017PolyEDM,Maison2019,Oleynichenko2022laser,Yu2021probing,Zakharova2021ptodd} describe experiments using polyatomic molecules in which this capability is useful. The basic principle of an
eEDM measurement using oriented molecules is as follows. If the electron
possesses an electric dipole moment, $\vec{d_{e}}$, then it must
be oriented along or against the electron spin, so that $|\vec{d_{e}}\cdot\vec{S}|\neq0$.
In a given molecular state, the energy shift associated with the eEDM
is $\langle H_{d_{e}}\rangle=d_{e}\langle\vec{S}\cdot\mathcal{\vec{E}}_{{\rm eff}}\rangle$,
where the ``effective electric field'' vector $\mathcal{\vec{E}}_{{\rm eff}}$
is a state-dependent constant oriented in the molecular frame, for
example along an internuclear axis. In a parity eigenstate, the electron
spin has no average orientation in the molecular frame, and $\langle\vec{S}\cdot\mathcal{\vec{E}}_{{\rm eff}}\rangle$
vanishes. The simplest way to obtain a non-vanishing eEDM interaction
is to orient the molecule in the lab frame so that $\vec{\mathcal{E}}_{{\rm eff}}||\hat{Z}$,
where $\hat{Z}$ is the lab $z$-axis, and to simultaneously orient
the electron spin along or against the same axis (e.g., via angular
momentum selection rules on electronic transitions) so that $M_{S}\neq0$.
In this ideal case, $\langle\vec{S}\cdot\mathcal{\vec{E}}_{{\rm eff}}\rangle$
will have maximal magnitude and $\langle H_{d_{e}}\rangle$ can be
probed. More generally, as long as the parity of a molecular state
is at least partially mixed, then $\langle\vec{S}\cdot\mathcal{\vec{E}}_{{\rm eff}}\rangle\neq0$
and the eEDM can be measured via $\langle H_{d_{e}}\rangle$. Heteronuclear diatomic molecules in $^2\Sigma$ electronic states have rotational states of both positive and negative parity, but they
are generally spaced by tens of GHz and require electric fields on
the order of tens of kV/cm to saturate the energy shifts associated
with the eEDM. By contrast, the parity doublets generically found
in polyatomic molecules can be fully mixed at fields at or below 1
kV/cm. Furthermore, in contrast to rotational states, the structure
of parity doublets found in polyatomic molecules, for example $K$-doublets
in symmetric top molecules or $\ell$-doublets in vibrational bending
modes of polyatomic molecules, enables the orientation of molecules
to be spectroscopically reversed in a fixed external electric field.
This feature can also be found in $\Lambda$- or $\Omega$-doublets of diatomic molecules,
and it has already been a valuable tool for systematic error
rejection in ThO and HfF$^{+}$, such as the experiments described in \cite{ACME2018,cairncross2017precision}. The special feature of polyatomic molecules is that such parity doublets can be obtained
irrespective of the electronic structure of the selected polyatomic species. 

Another case where near-degenerate opposite-parity states are useful
is in probing intrinsic parity-violating Standard Model (SM) effects
such as the vector electron-axial nucleon electroweak current coupling
and the nuclear anapole moment. The total parity-violating interaction
in a given electronic state is characterized by the constant $W_{p}$.
By determining the value of $W_{p}$ in multiple nuclei of the same
molecular species, the contribution of each SM effect could be independently
determined. A sensitive method to probe $W_{p}$ is Stark interference (see \cite{DeMille2008Using}),
where an electric dipole transition drives a molecule between opposite-parity
states separated by energy $\Delta$. The population transfer contains
an interference term between the driving electric field $E_{0}$ and
the parity-violating interaction $W$ (which is proportional to $W_{p}$
but contains additional factors from the molecular state). By comparing
measurements with opposite phases of the driving electric field, a
measured quantity proportional to $W/\Delta$ can be obtained. Thus
the experimental signal is enhanced when the states under consideration
are brought to near-degeneracy. Whereas rotational states of diatomic
molecules can be brought to near-degeneracy with Tesla-scale magnetic
fields, \cite{Norrgard2019nuclear} showed that in a large class of linear polyatomic molecules, degeneracy
can be achieved with fields of 10 mT or less, dramatically reducing
the experimental complexity of operating an experiment within the
bore of a superconducting magnet. By measuring nuclear-spin-independent
parity-violating effects in light molecules, where calculations of
Standard Model effects are not prohibitively challenging, parity-violating
interactions in the SM may be probed and, perhaps, distinguished from BSM effects.

A second feature of polyatomic molecules that can be exploited in
precision measurements of BSM physics is the multiplicity of rotational
and vibrational modes. Whereas diatomic molecules have one rotational
mode and one vibrational mode, every polyatomic molecule contains
at least three vibrational modes and up to three rotational modes.
Thus accidental near-degeneracies between rovibronic states can be
commonly found at low energies (e.g., below 1000 cm$^{-1}$), and
are nearly guaranteed at higher rovibronic energies where the density
of states increases. As described by \cite{jansen2014perspective}, these near-degeneracies are useful for probing
potential variation of the proton-to-electron mass ratio, $\mu\equiv m_{p}/m_{e}$,
over time: as $\mu$ changes, each rovibrational level shifts since
rotational and vibrational energies depend directly on the masses
of atomic nuclei according to $\delta E=\partial E/\partial\mu\times\delta\mu$. (Here, changes in physical quantities associated with changes in $\mu$ are indicated by the prepended symbol $\delta$.)
Pure rotational energies and anharmonic vibrational energies obey
$\delta E=-E\times\delta\mu/\mu$, while pure harmonic vibrational
energies obey $\delta E=-\frac{1}{2}E\times\delta\mu/\mu$. In each
case, the \emph{absolute} energy sensitivity $\delta E$ to $\mu$
variation scales with the overall energy $E$. In cases of accidental
near-degeneracies, it is possible to obtain $\omega\approx0$ even
when the absolute frequency sensitivity of a transition $\delta\omega\equiv\delta E_{2}-\delta E_{1}$
does not vanish because $E_1$ and $E_2$ depend differently on $\mu$. For example, rovibronic transitions between near-degenerate
states can exploit the relatively large absolute frequency shifts
of vibrational energy levels ($E\sim10$ THz) while being amenable
to the technical convenience of lower-frequency microwave sources
($\omega\sim10$ GHz) where stable frequency references are readily
available and certain systematic errors like Doppler shifts are suppressed.
Transitions where $\delta\omega/\omega\gg1$ have been used to set
limits on $\mu$ variation in molecules including methanol, ammonia,
and KRb (where a degeneracy between an excited vibrational state and
a metastable electronic state was used by~\cite{Kobayashi2019}). Laser-cooled
polyatomic molecules possess convenient rovibronic
near-degeneracies to sensitively probe $\mu$ variation in a platform
offering ultracold temperatures, long trap lifetimes, and full quantum
control. \cite{kozyryev2021enhanced} identified a promising near-degeneracy in the energy levels of SrOH that could be used for such an experiment. 

The third, and final, feature of polyatomic molecules that we note for precision measurements is chirality. Because chirality requires
three distinct molecular axes, it can only be found in molecules with
at least four atoms. Parity-violating effects in the Standard Model
are predicted to result in energy splittings between chiral molecules
at the level of $\sim$mHz to Hz in various species of experimental
interest (\cite{Cournol2019}). In principle,
even in the absence of parity violation, any chiral molecule could
convert to its enantiomer via the tunneling of some set of nuclei
through a vibrational energy barrier, resulting in a double-well-type
energy structure and associated splitting $\Delta E_{\pm}$ between
molecular eigenstates. 
In many molecules such as hydrogen peroxide (HOOH) or the chiral isotopologue of ammonia (NHDT), the tunneling splitting dominates parity-violating effects, $\Delta E_{\pm} \gg \Delta E_{\rm{pv}}$, and the symmetry breaking between left- and right-handed molecules occurs only ``\emph{de facto}," i.e. due to initial conditions described by \cite{Quack2022perspectives}. Nevertheless, in this case it is possible to measure the effect of parity-violating interactions, for example by observing an initial parity eigenstate acquire a non-zero amplitude of an opposite-parity eigenstate upon free evolution (\cite{Quack1986onthemeasurement}). 
The other limiting case is where the parity-violating energy shifts are large compared to the tunneling splitting, $\Delta E_{\rm{pv}} \gg\Delta E_{\pm}$, so that the symmetry breaking between enantiomers is ``\emph{de lege},'' i.e. due to intrinsic dynamics of the molecular energies. In this case, the energies of two enantiomers can be directly spectroscopically measured (e.g., via high-sensitivity infrared absorption experiments) and compared. Most molecules with heavier constituents, including species of interest such as CHFClBr and S$_{2}$Cl$_{2}$, are expected to exhibit ``\emph{de lege}'' parity violation due to the large reduced tunneling mass (\cite{Quack2008}). Chiral molecules are also sensitive to parity-violating cosmic fields associated with certain dark matter candidates, as pointed out by \cite{Gaul2020a}. Thus precise measurements of enantiomeric energy splittings can be a definitive probe of parity-violating weak interactions and beyond-Standard Model interactions in suitably chosen molecules. Laser-cooled polyatomic chiral molecules (for example CaOCHDT, a chiral
analogue of CaOCH$_{3}$) could potentially enable measurements of
these parity-violating effects for the first time due to the possibility of longer interaction times and full quantum control (\cite{Augenbraun2020ATM}).

\subsection{Collisions and chemistry}
Collisions of polyatomic molecules---both with atoms and other molecules---are of great scientific interest to a number of disciplines across physics and chemistry. A distinction is typically made between two different temperature regimes, namely ``cold'' collisions at temperatures of $\lesssim$$1~\text{K}$, and the ``ultracold'' collision regime, defined by temperatures sufficiently low that only a single partial wave participates in the collision. The ultracold collision regime is sometimes defined informally as $\lesssim$$1~\text{mK}$, though the actual temperature for collisions to only include a single partial wave can be much lower for heavier molecules. In addition, there are subtleties involved in defining the single-partial-wave regime for dipolar collisions where all partial waves may contribute even at low collisional energies, see \cite{Chomaz2023}. Numerous techniques have been developed for studying collisions at cold temperatures, and cold collision dynamics for a number of molecules, including polyatomic species, have already been investigated. These experiments have typically relied on beam-based approaches, though a variety of different experimental techniques have been employed, including slowing techniques to reduce collisional temperatures, e.g., Stark (\cite{vandeMeerakker2009}), Zeeman (\cite{plomp2020high}), and ``cryofuge'' (\cite{Wu2017cryofuge}) deceleration. Trapping of the molecular species---from slowed or buffer-gas-cooled (\cite{hummon2011cold}) molecular beams---has allowed for increased molecular interaction times, and has lead to novel studies of cold dipolar collisions. Stark deceleration followed by magnetic trapping has, for example, been used to study dipolar collisions between $\text{OH}$ and $\text{ND}_3$ molecules in a magnetic trap (\cite{Sawyer2011}), and cryofuge deceleration has recently been combined with electric trapping to study dipolar collisions between $\text{CH}_3\text{F}$ molecules (\cite{koller2022electric}).
Collision studies at cold temperatures play a particularly important role in astrophysics, specifically for studying the astrochemical reactions that result in the molecular compositions observed in interstellar environments, see \cite{herbst2009complex}. For more extensive reviews of cold collision studies, we refer interested readers to the articles by \cite{toscano2020cold, heazlewood2021towards}. The ultracold regime remains largely unexplored due to the difficulty in producing ultracold molecules, especially for polyatomic molecules. The extension of direct laser cooling techniques to polyatomic molecules could potentially provide a path towards realizing collision studies at ultracold temperatures. Below we provide a few key highlights of some of the novel collision dynamics and chemistry that are expected to arise at ultracold temperatures and which could potentially be explored with ultracold polyatomic molecules.

In the ultracold regime, the de Broglie wavelength of the colliding molecules exceeds their classical size, requiring a quantum-mechanical description of the collision process. Molecular interactions in this regime are therefore characterized by quantum statistics and quantum threshold behavior, and a host of different quantum collision and chemistry phenomena are expected to arise as a result (\cite{Carr2009}). For example, due to the delocalized wavelike nature of molecules in this regime, long-range dipolar interactions are expected to be essential for determining collision rates between polar molecules at ultracold temperatures. This bears important implications for chemical reaction rates. While reaction rates are ``classically'' expected to increase with temperature, many chemical reactions instead proceed at much accelerated rates in the limit of absolute zero (\cite{richter2015ultracold}). Some of these quantum effects are already starting to be explored with ultracold diatomic molecules. Seminal experiments with ultracold $\text{KRb}$ molecules created by photoassociation of laser-cooled atoms have, for example, probed long-range interactions in ultracold $\text{KRb}{-}\text{KRb}$ (\cite{ospelkaus2010quantum, ni2010dipolar, hu2021nuclear}) and $\text{KRb}{-}\text{Rb}$ (\cite{nichols2022detection}) collisions. Ultracold collision experiments have since also been performed with a range of other bialkali species (\cite{ye2018collisions, gregory2021molecule, bause2021collisions}). Similar experiments could be imagined with polyatomic molecules, which would grant still broader access to new quantum collision and chemistry phenomena. As an example, it has been proposed that collective many-body effects in a Bose degenerate gas of triatomic molecules may lead to a ``Bose-stimulated'' photodissociative process in which branching to either of two decay channels, $ABC \rightarrow AB + C$ or $ABC \rightarrow A + BC$, can be significantly amplified (\cite{moore2002bose}). Ultracold collision experiments with polyatomic molecules would also, along with experiments with diatomic molecules, provide important experimental benchmarks for new scattering theories and quantum chemistry at ultracold temperatures.

An essential implication of the quantum nature of the ultracold regime is that both elastic and inelastic collision rates are expected to have a strong dependence on the exact molecular quantum states (\cite{Carr2009}). External fields, too, can have strong and potentially very different effects on elastic and inelastic rates, for example by changing the relative orientation of two colliding molecules (\cite{tscherbul2006controlling, tscherbul2009magnetic, brouard2014taming, tscherbul2015tuning}). This provides the capability for separately ``tuning'' elastic as well as inelastic collision rates, which, as we discuss below, may be important for the realization of degenerate gases of polyatomic molecules. In the context of chemical reactions, this provides a path to controlled chemistry at ultracold temperatures in which chemical reactions can be very precisely studied and, possibly, engineered (\cite{krems2008cold, balakrishnan2016perspective, bohn2017cold}). These points are especially pertinent to polyatomic molecules, whose many internal degrees of freedom may provide additional tools for external field control, and, in particular, allow for easy orientation of the molecules. Recent experiments with CaF molecules have demonstrated that direct laser cooling in combination with optical trapping is a feasible approach for realizing the level of control required to characterize the quantum state and field dependencies of collision rates (\cite{Cheuk2020, anderegg2021observation}). Optical trapping of ultracold polyatomic molecules additionally opens the door to probing collisional dynamics in confined geometries in which collision dynamics take on qualitatively distinct behavior compared to that of an unconfined gas (\cite{Carr2009}).

Finally, understanding the collisional processes of polyatomic molecules at ultracold temperatures is likely to have a fundamental impact on the potential realization of Bose or Fermi degenerate polyatomic gases. In particular, the experiments with bialkali molecules mentioned earlier have demonstrated that, in the ultracold regime, even molecules that are not chemically reactive can exhibit large inelastic collision losses due to forming long-lived complexes in so-called ``sticky collisions'' (\cite{bause2022ultracold}). This prevents efficient evaporative cooling, which is the typical approach used for creating degenerate atomic gases (\cite{Pethick2002}). The tunability of the reaction rates mentioned above is in this regard important and can enhance the elastic-to-inelastic collision ratio by several orders of magnitude using external fields. Several ``shielding'' techniques have already been demonstrated for diatomic molecules  (\cite{matsuda2020resonant, valtolina2020dipolar, anderegg2021observation, li2021tuning}). Recently, \cite{Schindewolf2022} showed that quantum degeneracy could be reached for diatomic species using this shielding technique. Similar methods are likely required for the realization of degenerate gases of polyatomic molecules. A scheme for field-induced shielding in collisions between CaOH molecules has been proposed by \cite{augustovicova2019ultracold}. Alternative prospects for collisional cooling using sympathetic molecule-atom collisions have also recently been explored by \cite{wojcik2019interactions} for complex polyatomic molecules such as benzene and azulene.

\subsection{Experimental approaches besides direct laser cooling}
Achieving any of the diverse goals summarized above requires exquisite control over the molecules to be probed. In order to reap the benefits of polyatomic molecules, we must cool them to cold ($\mylesssim 4$~K) or ultracold ($\mylesssim 1$~mK) temperatures. This cooling is required to ``compress'' the molecular population into a small number of quantum states and to slow their thermal velocities to tens of meters per second; the former increases an experiment's signal-to-noise ratio and the latter enables long interrogation times to improve achievable precision and control. Because polyatomic molecules contain many internal degrees of freedom, cooling them can be a very difficult task. Many research groups have tackled this problem, and we review some relevant methods here. A variety of approaches have been explored to realize this control. 

A very successful method of producing ultracold diatomic molecules involves laser cooling atoms and then binding these pre-cooled atoms together (using photoassociation, magnetoassociation, etc.); see \cite{ni2008high, Molony2014, Park2015, Guo2016, Rvachov2017Long, Cairncross2021Assembly}. While this has produced a number of ultracold diatomic molecules in single quantum states, including degenerate gases, it is not clear it can be generalized to produce \textit{polyatomic} molecules. However, very recent work by \cite{Yang2022} has shown some evidence for production of triatomic molecules in a mixture of $^{23}$Na$^{40}$K and $^{40}$K. It is also unclear whether this exciting result can be extended to other species (or larger ones), especially those containing difficult-to-cool atomic species such as O, C, and/or H.

Optoelectrical Sisyphus cooling, described by \cite{Zeppenfeld2012Sisyphus, prehn2016optoelectrical}, uses state-dependent energy shifts and repeated microwave transitions/optical pumping steps to cool molecules as they move through an electric trap. Energy is removed by ensuring molecules move away from the trap's center along a ``steep'' potential and return to the center along a ``shallow'' potential. This technique relies on the linear Stark shifts that can be achieved in symmetric or asymmetric top molecules. It has been used to produce trapped samples of CH$_3$F and H$_2$CO at $\sim$mK temperatures and could potentially be used to observe molecule-molecule collisions in the trap. To date, the molecules that have been cooled using this method do not offer convenient optical transitions, so vibrational transitions with relatively long lifetimes have been used instead. The method could likely be adapted to make use of the optical transitions offered by some of the molecules discussed in this review, speeding up the cooling rates considerably.

\subsection{Direct laser cooling}
\begin{figure}[tb]
    \centering 
    \includegraphics[width=1\columnwidth]{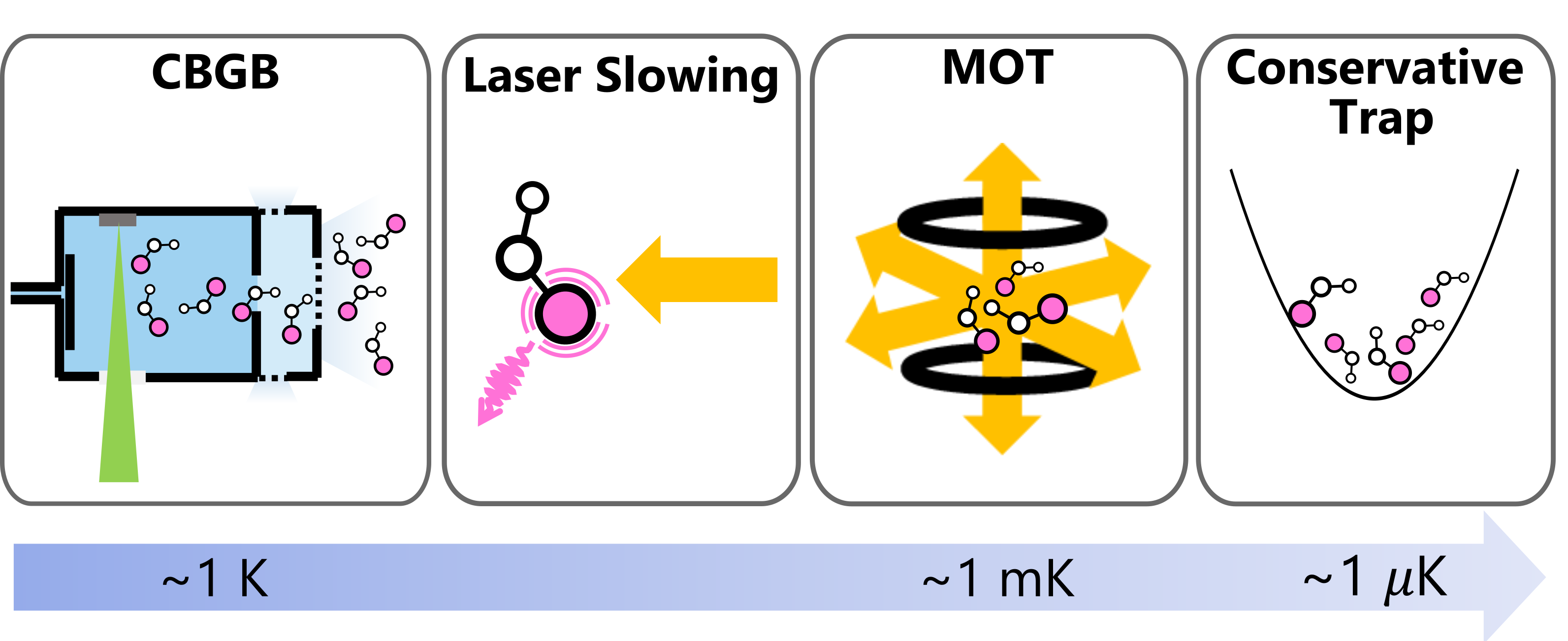}
    \caption{ 
    Overview of an idealized molecular laser cooling sequence. Molecules are produced in a cryogenic buffer-gas beam (CBGB) source, decelerated, trapped in a magneto-optical trap, and then transferred into a conservative trap for a particular science goal. Reproduced from \cite{AugenbraunThesis}.
}
    \label{fig:ExperimentalSequence}
\end{figure}

Direct laser cooling is a promising approach because it may be widely applicable to diverse structures of molecules and brings with it the possibility of high-efficiency quantum state preparation and readout using optical photons. The direct laser cooling approach is the focus of this review paper. Figure~\ref{fig:ExperimentalSequence} presents a schematic overview of an idealized molecular laser cooling experiment. Because many of these techniques were honed in the context of diatomic molecules, the interested reader should refer to the reviews by \cite{hutzler2012buffer,Tarbutt2018, McCarron2018,Fitch2021LaserCooled}.

\subsubsection{Forming closed cycling transitions}
The photon scattering process is one in which molecules go through a series of photon absorption and spontaneous emission cycles that can be described as a Bernoulli sequence. Suppose we have applied laser repumpers such that a molecule has a probability $p$ to decay to a state that is \textit{not} addressed by laser light. The probability $P_n$ for a molecule to experience $n$ absorption-emission cycles is given by $P_n = (1-p)^n$. The average number of photons scattered by molecules is $\bar{n} = \frac{1}{1-p}$. We often refer to this as the ``photon budget,'' and it sets the (exponential) scale for how many photons can be scattered before significant fractions of the population are lost. For example, if laser slowing requires 10,000 photon scattering events and we would like 90\% of the initial population to remain after slowing, we require $p \approx 10^{-5}$. Clearly, understanding branching ratios as small as $1$ part in $10^5$ can be crucial to achieve efficient laser cooling.

A typical laser cooling experiment requires scattering many thousands of optical photons.  To repeatedly scatter this number of photons without the molecules accumulating in states that do not couple to applied laser light (``dark states''), it is necessary to form a ``closed cycling transition.'' In a closed cycle, a molecule that is driven to an electronically excited state is guaranteed to decay back to the same state that it started in. In reality, no optical transition is fully closed and the molecule has a finite probability to decay to a state that is different from the initial state. Such a molecule needs to be “repumped” into the cycling transition. The higher the probability of decaying to other states, the more repumping transitions must be addressed. This quickly becomes an experimental limitation, as each decay typically requires a separate laser to be added to the experiment. This points to the importance of selecting atoms and molecules whose branching ratios are favorable. The general guidelines for selecting molecules and transitions for laser cooling were first pointed out by \cite{dirosa2004laser}, who highlighted the need for (1) strong transitions, (2) highly diagonal vibrational branching, and (3) no intermediate (i.e., metastable electronic) states between the two states used for optical cycling.

The electronic transitions driven in laser cooling are typically electric dipole (E1) transitions. Since the dipole operator is odd under parity transformations, the parity of the excited and ground states must be opposite. As a result, selection rules for changes to the total angular momentum $J$ are $\Delta J =0, \pm 1 $ (with $J'=0$ to $J=0$ forbidden). We can use these selection rules to our advantage by cycling from $J=1$ to $J'=0$, as originally recognized by \cite{Stuhl2008}, which must decay back to $J=1$, leading to rotational closure.

One consequence of driving from $J=1$ to $J'=0$ to attain rotational closure is the presence of dark states. This is a generic problem in the laser cooling of molecules, where the excited state often has the same number of sublevels (or fewer) than the ground state. This means that for any fixed laser polarization, dark states exist. Molecules that collect in these dark states can be returned to bright states in a number of different ways, most commonly using DC magnetic fields, polarization modulation, or microwave pulses (\cite{Berkeland2002}).

\subsubsection{Effects of multilevel systems on laser cooling} \label{sec:EffectiveScatteringRate}
For a system that can decay to multiple states, the prototypical two-level scattering rate equation is modified. For resonant light where all the transitions are driven with equal intensity, \cite{Williams2017} show that the maximum achievable scattering rate is modified to
\begin{equation}
R_\text{scat}^\text{max} =\frac{n_e}{n_g+n_e}\Gamma
\end{equation}
where $n_e$ is the number of excited states and $n_g$ is the number of ground states. 
As an example, in CaF there are 12 Zeeman sublevels in the ground state, 4 in the excited state, and 12 in the $v$=1 ground state. Hence $n_e=4$ and $n_g=24$. If lasers are tuned to drive both $\tilde{X}(v=0)$ and $\tilde{X}(v=1)$ to the $\tilde{A}$ state, the achievable scattering rate will reduce by a factor of about 4. This decrease in the scattering rate would significantly hinder the laser cooling of a molecule. A commonly employed technique to circumvent this limitation is to repump the molecules through a different excited state than the one used for the ``main'' transition. This approach is also crucial for polyatomic molecules as more states must be repumped in a laser cooling scheme.

With closed cycling transitions established, diatomic molecules have been laser slowed and cooled (\cite{Shuman2009,Shuman2010,Barry2012,Zhelyazkova2014,Hemmerling2016,truppe2017CaF,Lim2018}) and trapped in red-detuned magneto-optical traps (\cite{hummon20132d,Barry2014,McCarron2015,norrgard2015sub,Steinecker2016,chae2017,Truppe2017b,anderegg2017CaFMOT,Williams2017,Collopy2018}) and blue-detuned magneto-optical traps (\cite{burau2022blue}), cooled to sub-Doppler temperatures (\cite{Truppe2017b,Cheuk2018lambda,Caldwell2019,Ding2020Sub}), and loaded into magnetic (\cite{Williams2018magtrap,McCarron2018magtrap}) and optical traps (\cite{anderegg2018laser,langin2021,wu2021,lu2022}) and tweezers (\cite{anderegg2019,lu2022,Holland2022OnDemand,Bao2022Dipolar}).

\section{Molecular structure for laser cooling experiments} \label{sec:PolyatomicStructure}
Compared to diatomic molecules, polyatomic molecules can have significantly more complex internal structures. Here, we review the structure of polyatomic molecules as is relevant to laser cooling and trapping experiments. We generally work within the Born-Oppenheimer approximation, separating the electronic, vibrational and rotational motions, but molecular physics beyond the Born-Oppenheimer approximation is also introduced as necessary to understand its impact on laser cooling and trapping experiments.

\subsection{Electronic structure} \label{sec:ElectronicStructure}

For the molecules of interest to laser cooling experiments, the largest relevant energy scale corresponds to excitation of a valence electron that is used for optical cycling (changing the principle quantum number). Typical electronic excitation energies for several classes of molecules are in the visible region of the electromagnetic spectrum (400--800 nm). Understanding the origin and nature of these electronic states is needed to predict which molecules are favorable for laser cooling experiments.

We will focus, for the moment, on molecules in which an alkaline-earth atom, $M$, bonds to some electronegative ligand, $L$ (e.g., F, OH, OCH$_3$, SH, etc.), because these molecules are expected to have the structure desired for laser cooling (\cite{kozyryev2016MOR, Augenbraun2020ATM, Isaev2017, Ivanov2019Rational, isaev2015polyatomic}). As explained by \cite{Ellis2001}, the alkaline-earth atoms tend to form ionically-bonded molecules due to their low ionization energies. Whether $M$ transfers one or both of its valence electrons to the bonding partner depends on whether $L$ can form singly or doubly charged anions. For the examples of $L$ listed above, singly-charged anions form, so $M$ retains one electron even after forming a bond.

\begin{figure}
	\centering
	\includegraphics[width=0.9\linewidth]{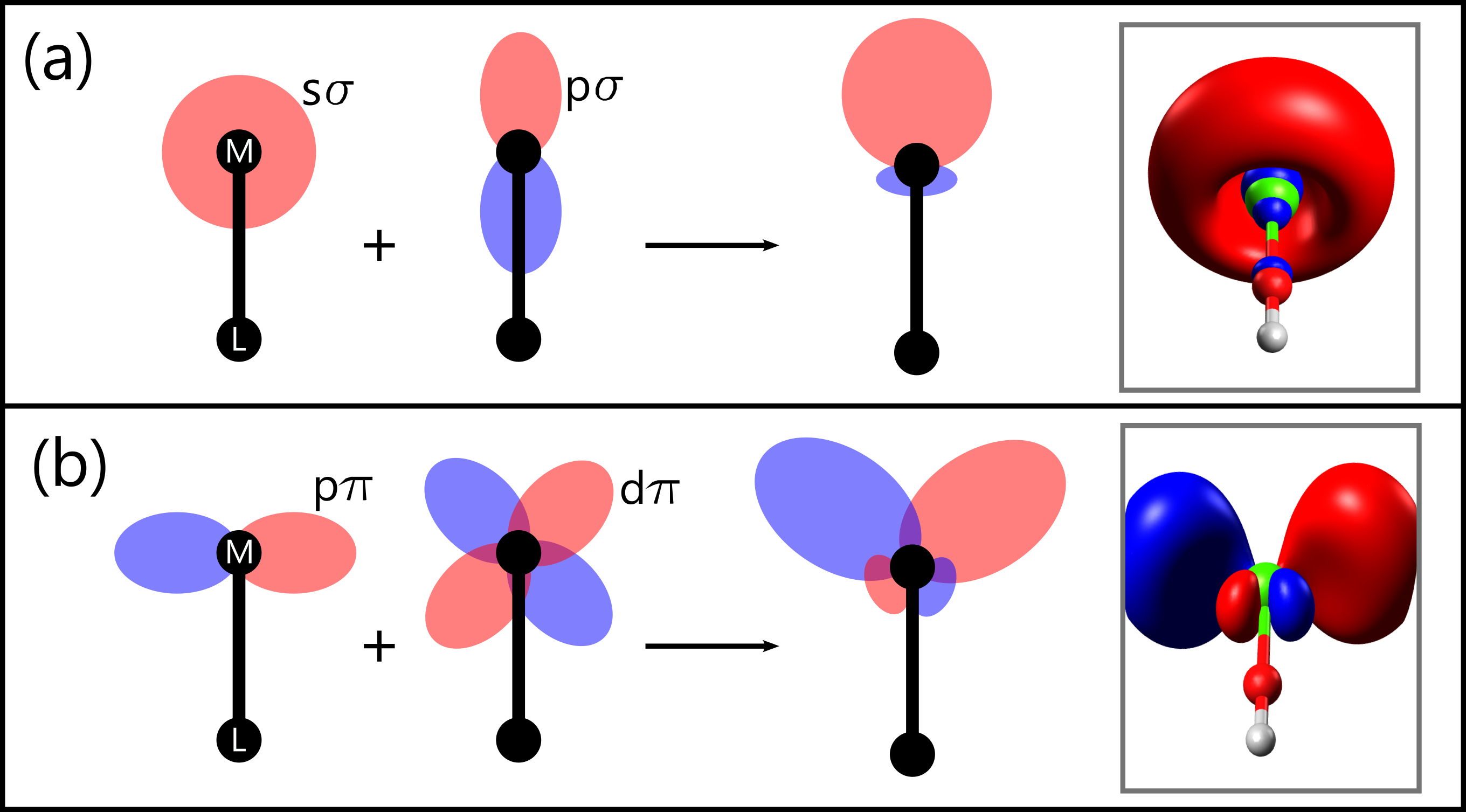}
	\caption[Orbital mixing in alkaline-earth pseudohalides]{Illustration of orbital mixing reducing the interaction between valence electron and negatively charged ligand. (a) Mixing of $s\sigma$ and $p\sigma$ orbitals to generate the $\tilde{X}\,^2\Sigma^+$ state. (b) Mixing of $p\pi$ and $d\pi$ orbitals to generate the $\tilde{A}\,^2\Pi$ state. In both cases, the rightmost image shows quantum chemical calculations of the electronic distribution confirming this simple orbital mixing picture, using CaOH as an example. Reproduced from \cite{AugenbraunThesis}.}
	\label{fig:OrbitalMixing}
\end{figure}

The relevant structural details can largely be extracted from a simple picture involving just three ingredients: a positively charged metal ion, $M^{2+}$, an optically active ``valence'' electron near the metal, and a negatively charged ligand, $L^-$. The presence of an unpaired electron leads to low-lying, metal-centered electronic excitations that can be used for optical cycling and laser cooling. One would naively expect the unpaired electron to have dominant $s$ orbital character in the electronic ground state. More detailed ligand-field theory calculations (e.g., those by \cite{Rice1985,Allouche1993}) show that this is largely true, but also that the interaction with the negatively charged ligand deforms the valence electron to minimize electron-electron repulsion (\cite{Ellis2001}). For example, in the ground state this deformation is realized via mixing of $s\sigma$ and $p\sigma$ orbitals on $M$, a process shown schematically in Fig.~\ref{fig:OrbitalMixing}. The orbital notation will be described more in the next paragraph. It has been found that in CaF, the ground electronic state (\XSigma) arises from a mixture of approximately $80\%$ of the $4s\sigma$ orbital and about $20\%$ of the $4p\sigma$ orbital, while the lowest electronic state $\tilde{A}\,^2\Pi$ is made up of about $70\%$ $4p\pi$ and $25\%$ $3d\pi$ character (\cite{Rice1985}). Similar values are found for Ba-containing monohalides (\cite{Allouche1993}) and larger monomethoxide species (\cite{Augenbraun2021Observation}).

An energy level diagram of the low-lying electronic states can be constructed from these ideas (\cite{Dick2007Spectroscopy, Ellis2001}). The basic idea is to consider the Hamiltonian $\mathbf{H} = \mathbf{H}_{M^{+}} + \mathbf{H}_{L^{-}} + \mathbf{H}'$, where $\mathbf{H}_{M^{+}}$ and $\mathbf{H}_{L^{-}}$ describe the energy levels of the free ions and $\mathbf{H}'$ describes the interaction between the optically active valence electron and the ligand (\cite{Rice1985, Allouche1993}). When the ligand is treated as a point charge perturbation, it has three qualitative effects on the spectrum (\cite{Dick2007Spectroscopy}):
\begin{enumerate}
	\item It can shift the atomic ion's energy levels.
	\item It can split the $m_l$ components of each atomic $nl$ state. We can think of this as arising from strong electrostatic forces along the bond, producing a Stark effect that resolves $m_l$ components along the bond axis. As is typical for the Stark effect, we do not resolve $+m_l$ and $-m_l$ components, and therefore label states as $\lambda = \abs{m_l} = 0, 1, 2, \ldots$. States with $\lambda = 0, 1, 2, \ldots$ are denoted by $\sigma, \pi, \delta, \ldots$, respectively.\footnote{We are using lower-case letters here because we are describing the single valence electron. Below, we will use capital letters to describe the total electronic state.} 
	\item It can mix orbitals obeying the selection rule $\Delta m_l = 0$. This may seem innocuous at first, but is actually quite important because this effect means each molecular orbital is a linear combination of atomic ion orbitals. Not only does this greatly affect the ordering of molecular states, it also means the molecular parameters of a given state will appear as an ``average'' of the atomic states that it comprises.
\end{enumerate}

\begin{figure}
	\centering
	\includegraphics[width=0.97\linewidth]{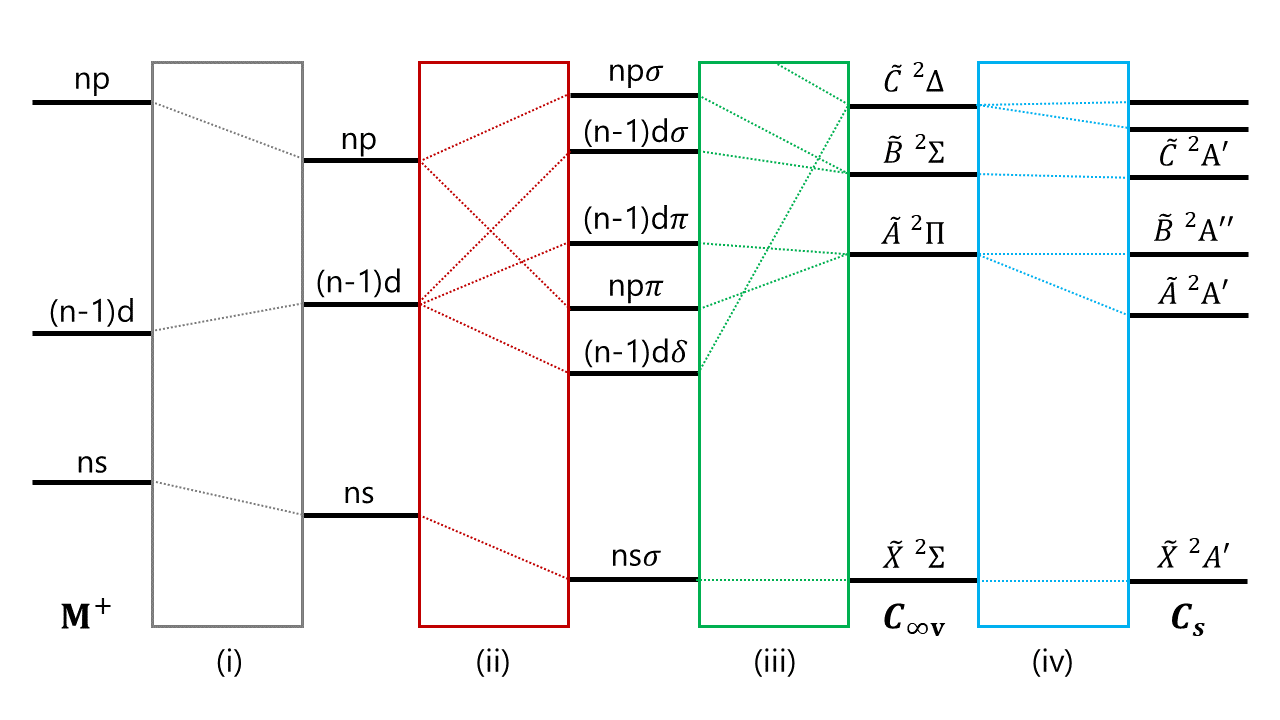}
	\caption[Molecular energy levels under various degrees of symmetry breaking]{Correlation of low-lying electronic states from atomic ion ($M^+$) to linear ($C_{\infty v}$) and nonlinear ($C_s$) molecules. The labeled regions correspond to the following qualitative processes: (i) Shifting of atomic ion levels, (ii) Splitting of atomic $m_l$ components, (iii) Mixing of energy levels with the same $m_l$, (iv) Cylindrical symmetry breaking due to a nonaxial ligand. Modeled after diagrams in \cite{Ellis2001, Dick2007Spectroscopy}.}
	\label{fig:LFLevelDiagram}
\end{figure}

Figure~\ref{fig:LFLevelDiagram}(i-iii) shows the development of the energy levels as these effects are sequentially added. This diagram also shows how we name electronic states. For a linear molecule ($C_{\infty v}$ point group symmetry), we label electronic states by letters: $\tilde{X}$ for the ground state, $\tilde{A}, \tilde{B}, \ldots$ for the first, second, ..., electronically excited states, respectively. Note that (for historical reasons) the alphabetical ordering usually, but does not always, match the energetic ordering of energy levels.\footnote{Electronic states with multiplicity different from than the ground state are labeled by lower case letters $\tilde{a}, \tilde{b}, \ldots$, usually in order of increasing energy.} In addition, because in most cases the electronic states can be identified by their electronic orbital angular momentum ($\Lambda$) and spin multiplicity ($2S+1$), we use the labeling scheme $^{2S+1}\Lambda$. For states with definite projection of spin onto the molecular axis, we can also add a subscript ($\Omega$) to specify the \textit{total} angular momentum projection on the internuclear axis (see also discussion of angular momentum coupling below). 

\begin{figure}
	\centering
	\includegraphics[width=0.7\linewidth]{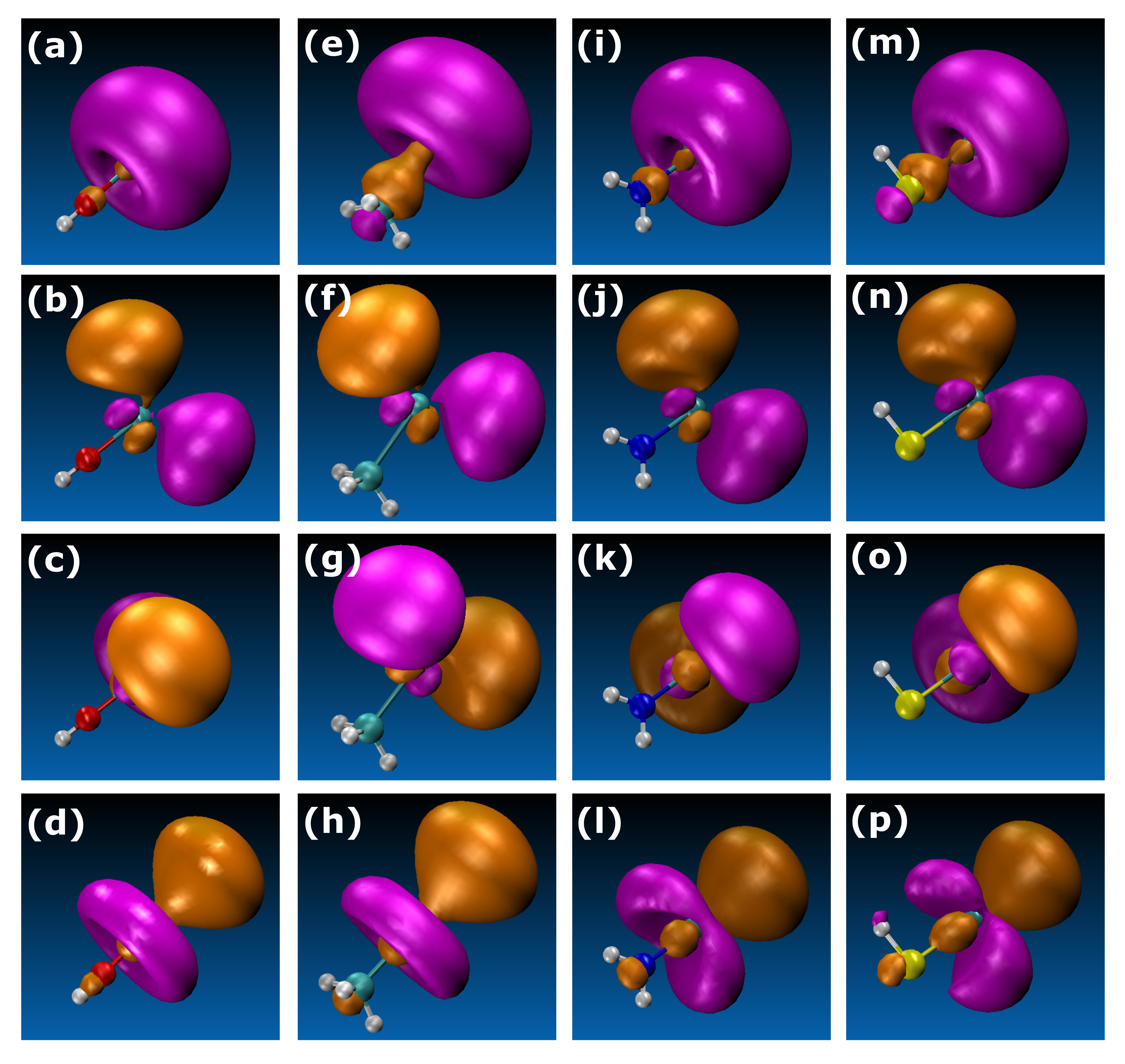}
	\caption{Molecular orbitals of the lowest several electronic states for Ca-containing molecules as symmetry is systematically lowered from $C_{\infty v}$ (CaOH) to $C_{3v}$ (CaCH$_3$) to $C_{2v}$ (CaNH$_2$) to $C_{s}$ (CaSH). The distortion of the valence electron as a function of ligand asymmetry is clearly visible. Panels (a-d) show, respectively, the HOMO, LUMO, LUMO+1, and LUMO+2 for CaOH; (e-h) show these for CaCH$_3$; (i-l) for CaNH$_2$; and (m-p) for CaSH. Figure reproduced from \cite{Augenbraun2020ATM}.}
	\label{fig:ATMOrbitals}
\end{figure}

This description has so far assumed the ligand can be treated as a point charge, meaning the $M^+L^-$ system has cylindrical symmetry about the $M-L$ axis. Such symmetry is exact only for linear molecules (diatomic or polyatomic). If $ML$ is a nonlinear molecule but still retains some axial symmetry (e.g., MOCH$_3$) then the picture is largely unchanged except that there will be no formal distinction between the symmetries of degenerate electronic states (e.g., for $C_{3v}$ symmetry, $\Pi, \Delta, \ldots$ states are all classified as having $E$ symmetry). If the ligand breaks the axial symmetry (e.g., MSH), then the orbital degeneracy described in steps (2) and (3) just above is no longer guaranteed, and degenerate electronic states can split. For example, a $\Pi$ electronic state of CaOH will correlate to states of $A'$ and $A''$ symmetry for CaSH, as shown in Fig.~\ref{fig:LFLevelDiagram}(iv). Figure~\ref{fig:ATMOrbitals} shows the clear role that asymmetry of the ligand has on slightly deforming the valence electron that remains localized on the metal optical cycling center. Despite the asymmetry, in all cases plotted the optical cycling properties are preserved. 

\subsection{Vibrational structure} \label{sec:VibrationalStructure}

During laser cooling of a polyatomic molecule, a number of vibrational states may be excited due to the lack of a perfectly closed cycling transition. These states are generally close to the bottom of the potential energy surface, which can be Taylor expanded as
\begin{align}
    V = V_0 &+ \frac{1}{2}\sum_{i=1}^{3N}\sum_{j=1}^{3N} \left(\frac{\partial^2 V}{\partial q_i \partial q_j} \right)_{q_i = q_j = 0} q_i q_j \nonumber \\
    &+ \frac{1}{3!}\sum_i\sum_j\sum_k \left(\frac{\partial^3 V}{\partial q_i \partial q_j \partial q_k} \right)_0 q_i q_j q_k + \ldots
    \label{eqn:pes}
\end{align}
Here we express the potential energy as a function of the $3N$ nuclear coordinates $q_i$, which describe the 3-dimensional motion of all $N$ nuclei. To a good approximation, the low-lying vibrational states in this potential may be described within the harmonic approximation, in which case the vibrational Hamiltonian is (\cite{Demtroder2003,BernathTextbook})
\begin{equation}
    H_v = \sum_{i=1}^{3N-6(5)} \left( -\frac{\hbar^2}{2}\frac{\partial^2}{\partial Q_i^2} + \frac{1}{2}\lambda_i Q_i^2\right)
\end{equation}
where $Q_i$ is a mass-weighted normal coordinate describing the $i$th normal mode of vibration and $\sqrt{\lambda_i}$ is the vibrational frequency of the mode. Notice that for a polyatomic molecule there are $3N-6$ (or $3N-5$ for a linear molecule) normal vibrational coordinates; the other 6(5) normal coordinates are taken up by the 3 translational and 3(2) rotational degrees of freedom of the molecule.

Vibrations along each of the normal coordinates are fully separable, so the vibrational wavefunction can be expressed as
\begin{equation}
    \psi(Q_1, Q_2, \ldots, Q_{3N-6}) = \prod_{i=1}^{3N-6} \psi_{v_i}(Q_i)
    \label{eqn:shostates}
\end{equation}
where $\psi_{v_i}(Q_i)$ is the simple harmonic oscillator wavefunction (\cite{BernathTextbook}) with vibrational quantum number $v_i$. The energy of the state in the harmonic approximation is
\begin{equation}
    E_v = \sum_{i=1}^{3N-6} \hbar \omega_i\left(v_i + \frac{1}{2} \right)
\end{equation}

Reintroducing the anharmonic terms in the potential energy surface has two effects. The first is to couple the harmonic oscillator eigenstates from Eq.~\ref{eqn:shostates}, as discussed below. The second effect is to alter the state energies, which can be expressed as (\cite{Demtroder2003, BernathTextbook, HerzbergVol3})
\begin{align}
    G(v_1,v_2,\ldots,v_p) = \sum_i \omega_i \left(v_i + \frac{d_i}{2}\right) &+ \sum_{j\leq i} x_{ij} \left(v_i + \frac{d_i}{2}\right) \left(v_j + \frac{d_j}{2}\right) \nonumber \\
    &+ \sum_{j\leq i}g_{ij}\ell_i \ell_j + \ldots
\end{align}
where $\omega_i$ is the frequency of the $i$th mode, $d_i$ is its degeneracy, $x_{ij}$ and $g_{ij}$ are anharmonicity constants, and $\ell$ is a quantum number describing the vibrational angular momentum along the molecular axis.

\subsubsection{Vibrational state notation}

\textbf{Linear triatomic molecules.} Linear triatomic molecules, including the alkaline earth monohydroxides CaOH, SrOH, and YbOH, have 4 normal vibrational modes, one of which (the bending mode) is doubly degenerate. The vibrational state is labeled by $(v_1 v_2^\ell v_3)$, where $v_1$ is the quantum number for the symmetric stretching mode, $v_2$ describes the bending mode, $v_3$ describes the antisymmetric stretching mode, and $\ell$ gives the vibrational angular momentum in the bending mode. Note that this angular momentum arises because linear combinations of the degenerate bending vibrations along two perpendicular (linear) axes can be formed which have elliptical trajectories. This can be thought of as producing a nuclear orbital angular momentum about the molecular axis. Doubly degenerate bending modes such as this one are formally described by the 2D harmonic oscillator, and $\ell$ can take the values $v_2, v_2-2, \ldots, -v_2+2, -v_2$ (see \cite{BernathTextbook}).

\textbf{Larger polyatomic molecules.} For larger molecules, labeling vibrational states as $(v_1, v_2,\ldots,v_N)$ becomes unwieldy. Instead, we use the notation $i_{v_i}$, where $i$ labels the vibrational mode and $v_i$ is the vibrational quantum number for that mode. Only modes with $v_i \neq 0$ are labeled, and $i$ is enumerated from 1 to $n$, where $n$ is the total number of modes. Typically, the modes are ordered first by the symmetry of the vibrational mode and then in order of decreasing energy, as described in \cite{HerzbergVol3}. For example, in the symmetric top molecule CaOCH$_3$, the vibrational state $4_1$ corresponds to one excitation of the 4th vibrational mode (Ca--O stretch), while $3_14_2$ corresponds to 2 excitations of the Ca--O stretch mode and one excitation of the O--C stretch mode (3rd vibrational mode). In this notation, unspecified vibrational modes are assumed to have $v_i = 0$.

\subsubsection{Anharmonic coupling}
As mentioned above, one effect of the anharmonic terms in the potential energy surface (Eq. \ref{eqn:pes}) is to mix the harmonic oscillator wavefunctions, such that the eigenstates of the vibrational Hamiltonian are admixtures of different harmonic oscillator basis states. This effect is most prominent for vibrational states of the same symmetry that are nearby in energy, and is often referred to as \emph{Fermi resonance}. The Fermi resonance interaction is described in detail in \cite{hougen1962}.

One example of Fermi resonance occurs in CaOH, whose bending mode frequency is approximately half the stretching mode frequency. This means that, for example, the $(100)$ and $(02^00)$ states are nearly degenerate and strongly mixed by a cubic term in the potential energy surface, $V_{122} = k_{122}q_1q_2^2$, which mixes states with $\abs{\Delta v_1} = 1$ and $\abs{\Delta v_2} = 2$. Here $k_{122}$ is an anharmonic force constant. One practical implication of this interaction is that excited electronic states that would decay to $\widetilde{X}(100)$ can also decay to $\widetilde{X}(02^00)$ at an enhanced rate; this is important for understanding vibrational branching ratios in polyatomic molecules (see Sec.~\ref{sec:VibrationalTransitions}). In this example, $(100)$ mixes with $(02^00)$ but not $(02^20)$ because states with different $\ell$ have different symmetry. For example, in a $\Sigma$ electronic state, $\ell = 0$ vibrational levels have $\Sigma$ symmetry, $\ell = 1$ vibrational levels have $\Pi$ symmetry, and so on. For a more detailed discussion of symmetry in polyatomic molecules, see \cite{BunkerJensen}.

\subsection{Rotational structure} \label{sec:RotationalStructure}

The rotational structure of molecules can be described, to leading order, by a rigid rotor model. The Hamiltonian is $H_{{\rm rot}}=\frac{R_{a}^{2}}{2I_{a}}+\frac{R_{b}^{2}}{2I_{b}}+\frac{R_{c}^{2}}{2I_{c}}$,
where $a,b,c$ denote the three principal axes in the molecular body-fixed
frame, $I_{a}\leq I_{b}\leq I_{c}$ are their corresponding moments
of inertia, and $R_{a},R_{b},R_{c}$ are the corresponding projections
of the molecule's rotational angular momentum. It is useful to introduce
the molecular constants $A=(2I_{a})^{-1}$, $B=(2I_{b})^{-1}$, and
$C=(2I_{c})^{-1}$. The energy level structure resulting from this
Hamiltonian depends on the relative values of $I_{a},\,I_{b},$ and
$I_{c}$. We describe each case in turn, initially neglecting the
effects of electronic, nuclear, and vibrational angular momentum.
In this scenario, the quantum number $N=J-S$ is equivalent to the
rigid-body rotational angular momentum $R$. We will denote the rotational
angular momentum by $N$ to make the notation compatible with the
more general case treated later.

\subsubsection{Linear molecules} 
A linear molecule has $I_{a}=0$ and $I_{b}=I_{c}$
so that $B=C$. All diatomic molecules, and a small but important
class of polyatomic molecules including alkaline-earth hydroxides
like CaOH and SrOH, are linear. In the limit that $I_{a}\rightarrow0$,
any angular momentum about the $a$-axis requires infinite energy,
so we take $N_{a}=0$. Then $H_{{\rm rot}}=B(N_{b}^{2}+N_{c}^{2})=BN^{2}$.
The energy levels of this Hamiltonian form a quadratic ladder with
eigenvalues $E_{N}=BN(N+1)$. Each $N$ manifold contains $2N+1$
degenerate states, distinguished by the quantum number $M\equiv N_{Z}$,
the projection of $N$ on the lab-fixed $Z$-axis.

\subsubsection{Spherical top molecules} 
In the special case that $I_{a}=I_{b}=I_{c}$,
so that $A=B=C$, the Hamiltonian also reduces to the form $H_{{\rm rot}}=BN^{2}$,
again with eigenvalues $E_{N}=BN(N+1)$. In this case, however, the
additional structure leads to a set of $(2N+1)^{2}$ degenerate states,
distinguished by independent quantum numbers $M$ and $K\equiv N_{a}$.
Spherical top molecules include species such as CH$_{4}$ (methane),
SF$_{6}$ (sulfur hexafluoride), and C$_{60}$ (buckminsterfullerene),
but to our knowledge no laser-coolable spherical top molecules have
been proposed to date and we do not consider them further here.

\subsubsection{Symmetric top molecules}
A molecule with exactly two equal
moments of inertia is classified as a symmetric top. This case is
further subdivided into the prolate (``cigar-shaped'') symmetric
top, where $I_{a}<I_{b}=I_{c}$, and the oblate (``pancake-shaped'')
symmetric top, where $I_{a}=I_{b}<I_{c}$. We explicitly consider
the prolate case first; in the oblate case, one must only substitute
$a\rightarrow c$ and $A\rightarrow C$ in all formulas. Then the
Hamiltonian is $H_{{\rm rot}}=BN^{2}+(A-B)K^{2}$, where as before
$K\equiv N_{a}$. The energy levels are $E_{N,K}=BN(N+1)+(A-B)K^{2}$,
with the restriction that $K\leq N$. There are $2N+1$ degenerate
states with energy $E_{N,0}$, distinguished by $M$. When $K\neq0$,
there are $2(2N+1)$ degenerate states distinguished by both $M$
and the sign of $K=\pm|K|$. Examples of prolate symmetric top molecules
include CH$_{3}$F (methyl fluoride) and the laser-coolable species
CaOCH$_{3}$ (calcium monomethoxide). Examples of oblate molecules
include C$_{6}$H$_{6}$ (benzene) and NH$_{3}$ (ammonia); to date,
no oblate symmetric top molecules have been proposed for laser cooling.
For molecules where $A\gg B$, typical of known laser-coolable molecules,
there is a quadratic ladder of widely separated ``$K$-stacks,''
each of which contains a quadratic ladder of more finely separated
$N$ states.

\subsubsection{Asymmetric top molecules}
In an asymmetric top molecule, all
moments of inertia are unequal. The resulting energy level structure
is quite complex, and no projection of $N$ on the molecule-frame
axes is a good quantum number. For each value of $N$, there are $2N+1$
non-degenerate states. The degree of asymmetry can be characterized
by the Ray's asymmetry parameter, $\kappa=\frac{2B-A-C}{A-C}$. In
the prolate (oblate) symmetric top limit, $B=C(A)$, we obtain $\kappa=1(-1)$.
In the maximally asymmetric case, $B=(A+C)/2$, we obtain $\kappa=0$.
Eigenstates can be designated by labels $N_{K_{a}K_{c}}$, where $K_{a}$
and $K_{c}$ are the projections of $N$ on the $a$-axis and $c$-axis
in the limit that a molecule is deformed to the prolate and oblate
symmetric top limits, respectively. In the general case, $-1<\kappa<1$,
neither $K_{a}$ nor $K_{c}$ are rigorously good quantum numbers.

Laser-coolable asymmetric top molecules proposed in the literature
to date (for example, CaSH, CaOCHDT, and SrOC$_{6}$H$_{5}$) are
prolate ($\kappa<0$), though there is no apparent reason that oblate
molecules ($\kappa>0$) are necessarily inconsistent with laser cooling.
An example of an oblate asymmetric top molecule is C$_{4}$H$_{4}$N$_{2}$
(pyrimidine). In the case of an asymmetric top molecule near the prolate
symmetric top limit, $\kappa\approx-1$, the energy level structure
closely resembles that of a prolate symmetric top. A similar situation holds for asymmetric top molecules near the oblate limit.

\subsubsection{Dependence on spin and electronic
angular momenta}
The ground states of proposed laser-coolable polyatomic
molecules generally have vanishing electronic angular momentum, and can be described
in a Hund's case (b) basis where $N$ is a good quantum number. Corrections
to the rotational structure described above occur due to electron
spin, nuclear spin, and vibrational angular momentum (for example
in the bending vibrational mode of a linear triatomic molecule). The
last of these shifts the overall energy of the ladder of eigenstates
but does not otherwise change the rotational energy progression.

Coupling of the electron spin and molecular rotation is generally
more complicated. The spin-rotation Hamiltonian takes the general
form 
\begin{equation}
H_{{\rm SR}}=\frac{1}{2}\sum_{\alpha,\beta}\epsilon_{\alpha\beta}(N_{\alpha}S_{\beta}+S_{\beta}N_{\alpha}),
\end{equation}
where $\alpha,\beta$ takes values $a,b,c$ and $\epsilon$ is a symmetric
tensor described in, e.g., \cite{Hirota1985}. In a linear molecule, only the components $\epsilon_{bb}=\epsilon_{cc}\equiv\gamma$
take nonzero values and the simpler form 
\begin{equation}
H_{{\rm SR}}=\gamma N\cdot S
\end{equation}
is
obtained. For linear molecules with $S=1/2$, this term mixes
adjacent $N$ levels, and both $N$ and $S$ cease to be good quantum
numbers, though $\vec{J}=\vec{N}+\vec{S}$ is preserved. When $\gamma\ll B$, as is typical of the alkaline-earth pseudohalides,
the effect is to split $N$ into a pair of states with $J=N\pm1/2$.

The situation is more complicated for nonlinear molecules. In a symmetric top molecule, the spin-rotation Hamiltonian (neglecting
contributions that can mix different values of $\Lambda$ in a $^{2}E$
state) reduces to 
\begin{equation}
H_\text{SR} = \epsilon_{aa}N_{a}S_{a}+\frac{1}{4}(\epsilon_{aa}+\epsilon_{bb})(N_{+}S_{-}+N_{-}S_{+}),
\end{equation}
assuming the prolate case (see \cite{Hougen1980double}).
The asymmetric top case is generically much more complicated, and spin-rotation
interactions can mix states that differ in $N$ by up to 1, and differ
in $K$ by up to 2 (\cite{Hirota1985}). In alkaline-earth
pseudohalides, the ground-state spin-rotation splittings are typically
no larger than $\sim$100 MHz at low $N$, and increase with larger
$N$. In most laser-coolable molecules used so far, the size of the spin-rotation
constant is dominated by a second-order perturbation via spin-orbit
coupling rather than the direct coupling of electron spin and molecular
rotation. Excited states with spin-rotation structure, for example
the $\tilde{B}\,^{2}\Sigma^+$ state in MOH molecules, can have spin-rotation
constants comparable to the rotational constant due to the closer
proximity to electronic states with $\Delta \Lambda = \pm 1$. Because
excited-state spin-rotation structure is large compared to the natural linewidth, and because the ground-state spin-rotation splittings are $\sim$100
MHz or less, the spin-rotation structure
is easily addressed with frequency sidebands added to the laser beam, e.g. using acousto-optic or electro-optic
modulation.

In states with orbital angular momentum, for example a $^{2}\Pi_{1/2}$
electronic state of a linear molecule, electrostatic interactions
couple the orbital angular momentum $L$ to the internuclear axis
and spin-orbit interactions couple $L$ with $S$. In this case, $N$
is not a good quantum number. The rotational energies follow a quadratic
ladder in $J$ as $H_{{\rm rot}}=BJ(J+1)$, with each $J$ value split
into a pair of opposite-parity states by a $\Lambda$-doubling Hamiltonian,
which arises from spin-orbit mixing with states of different $\Lambda$.
In symmetric top and asymmetric top molecules, similar interactions
split pairs of opposite-parity states in electronic manifolds with
non-zero orbital angular momentum. 

\subsubsection{Hyperfine effects and nuclear spin statistics}
In many of the polyatomic laser cooling experiments pursued to date, the valence
electron responsible for optical cycling is localized on the spin-0
nucleus of an alkaline-earth (or alkaline-earth-like) metal atom, and a spin-0 oxygen nucleus
serves as a ``linker'' to a ligand such as H and CH$_{3}$. As
a result, the nuclei with non-zero spin are far from the optically
active valence electron, and hyperfine interactions are on the order
of only $\lesssim 1$~MHz in both the ground and excited electronic states.
Hyperfine interactions therefore play a negligible role in laser cooling, though
they might be important in single-quantum-state preparation and readout
as needed for many applications (see Sec.~\ref{sec:CoherentControl}). 

Some applications, for
example studying nuclear-spin-dependent parity violation (\cite{Norrgard2019}) or nuclear
magnetic quadrupole moments (\cite{hutzler2020}), require the optically active electron to be localized near a nucleus with spin $I>0$. This complicates the structure by splitting each $J$ level into
$2I+1$ states of distinct $F$ levels. Generally, this should have
no substantial effect on laser cooling as long as each $F$ level is optically
addressed by either spectral broadening or the addition of frequency
sidebands to optical cycling lasers. 

Another consequence of nuclear spins in polyatomic molecules stems from the connection between nuclear spin symmetry and molecular rotation that arises due to symmetry requirements on the total wavefunction, as described in \cite{BunkerJensen}.
In the simplest case of a molecule composed of two identical atoms
with vanishing nuclear spin, some rotational levels do not exist because
the wave function must be even under exchange of the identical nuclei\textendash for
example, the ground $^{3}\Sigma_{g}^{-}$ manifold of $^{16}$O$_{2}$
only has odd $J$ levels. Generically, rovibronic levels occur with
a ``statistical weight'' that corresponds to the number of nuclear
spin states that give the required totally symmetric state. For example,
in CaOCH$_{3}$ the ``para'' nuclear spin configuration (in which two hydrogen
nuclear spins are aligned) occurs in $K=1,2,4,5,\ldots$ rotational
states, but the ``ortho'' spin configuration (in which all nuclear
spins are aligned) occurs in $K=0,3,\ldots$ rotational states. Because of
the weak hyperfine coupling in CaOCH$_{3}$, conversion between
para and ortho configurations is highly suppressed; because of the lack of interconversion, the two nuclear spin configurations behave as though they were essentially independent species. Experimentally, laser cooling schemes
for both nuclear spin isomers were demonstrated by \cite{mitra2020direct} where the isomer cooled was selected spectroscopically, by driving transitions out of either the $K=0$ or $K=1$ rotational state.

\subsection{Transitions} \label{sec:Transitions}
In the following sections, we describe the properties and intensities of transitions between the various energy levels present in polyatomic molecules.

\subsubsection{Electronic transitions} \label{sec:ElectronicTransitions}
Transition intensities can be calculated from the square of the transition moment integral (\cite{BernathTextbook})
\begin{equation}
    \mathbf{M} = \int \psi'(\mathbf{r},\mathbf{R})^\ast \, \boldsymbol{\mu} \, \psi(\mathbf{r},\mathbf{R}) \, d\tau,
\end{equation}
where integration with respect to $\tau$ implies integrating over all electronic and nuclear coordinates, and single primes denote excited states. Invoking the BO approximation, we can write $\psi(\mathbf{r},\mathbf{R}) = \psi_n(\mathbf{R}) \psi_e(\mathbf{r};\mathbf{R})$. Then the transition moment integral becomes
\begin{equation}
    \mathbf{M} = \int \psi_{n'}(\mathbf{R})^\ast \left( \int \psi_{e'}^\ast(\mathbf{r};\mathbf{R}) \, \boldsymbol{\mu} \, \psi_{e}(\mathbf{r};\mathbf{R}) \, d\tau_e \right) \psi_{n}(\mathbf{R}) \, d\tau_n.
    \label{eq:FullMev}
\end{equation}
We have separated integration over all nuclear coordinates ($d\tau_n$) and electronic coordinates ($d\tau_e$). Let us now define an electronic transition dipole moment as $\Ree(\mathbf{R}) = \braket{\psi_{e'}|\boldsymbol{\mu}}{\psi_{e}}$, and we explicitly denote its dependence on the nuclear coordinates. We can imagine expanding $\Ree(\mathbf{R})$ in a Taylor series about some value of $\mathbf{R}$ and retaining only the first (constant) term, which we will denote by $\Ree$.\footnote{More details about the choice of point on which to center the expansion can be found in the textbook by \cite{BernathTextbook}. Often, the value of \Ree is inferred from measurements or $\Ree(\mathbf{R})$ can be found via \textit{ab initio} calculations and used in conjunction with numerical vibrational wavefunctions to compute the necessary integral.} This simplifies Eq.~\ref{eq:FullMev} and allows us to write
\begin{eqnarray}
    \mathbf{M} &=& \Ree \int \psi_{n'}(\mathbf{R})^\ast \psi_{n}(\mathbf{R}) \, d\tau_n.
\end{eqnarray}
The transition moment has factored into an electronic and nuclear portion, as would be expected from the BO approximation treatment. If we explicitly include both vibrational and rotational nuclear motions, we can write the intensities of rovibronic transitions as (\cite{BernathTextbook})
\begin{equation}
    I_{(e'v'J')\rightarrow(evJ)} = \abs{\Ree}^2 q_{v'\rightarrow v} \SJJ,
    \label{eq:FullLineStrength}
\end{equation}
where \Ree is the electronic transition dipole moment between electronic states $e'$ and $e$, \qvv is a Franck-Condon factor (FCF) between vibrational states $v'$ and $v$, and \SJJ is a \Honl-London factor between states $J'$ and $J$. We can interpret this factorization as stating that the intrinsic strength of some transition is set by $\abs{\Ree}^2$ while the distribution of that intensity among vibrational and rotational lines are set by \qvv and \SJJ. 

\subsubsection{Vibrational transitions} \label{sec:VibrationalTransitions}

  \begin{figure}[tb]
    \centering 
    \includegraphics[width=1\columnwidth]{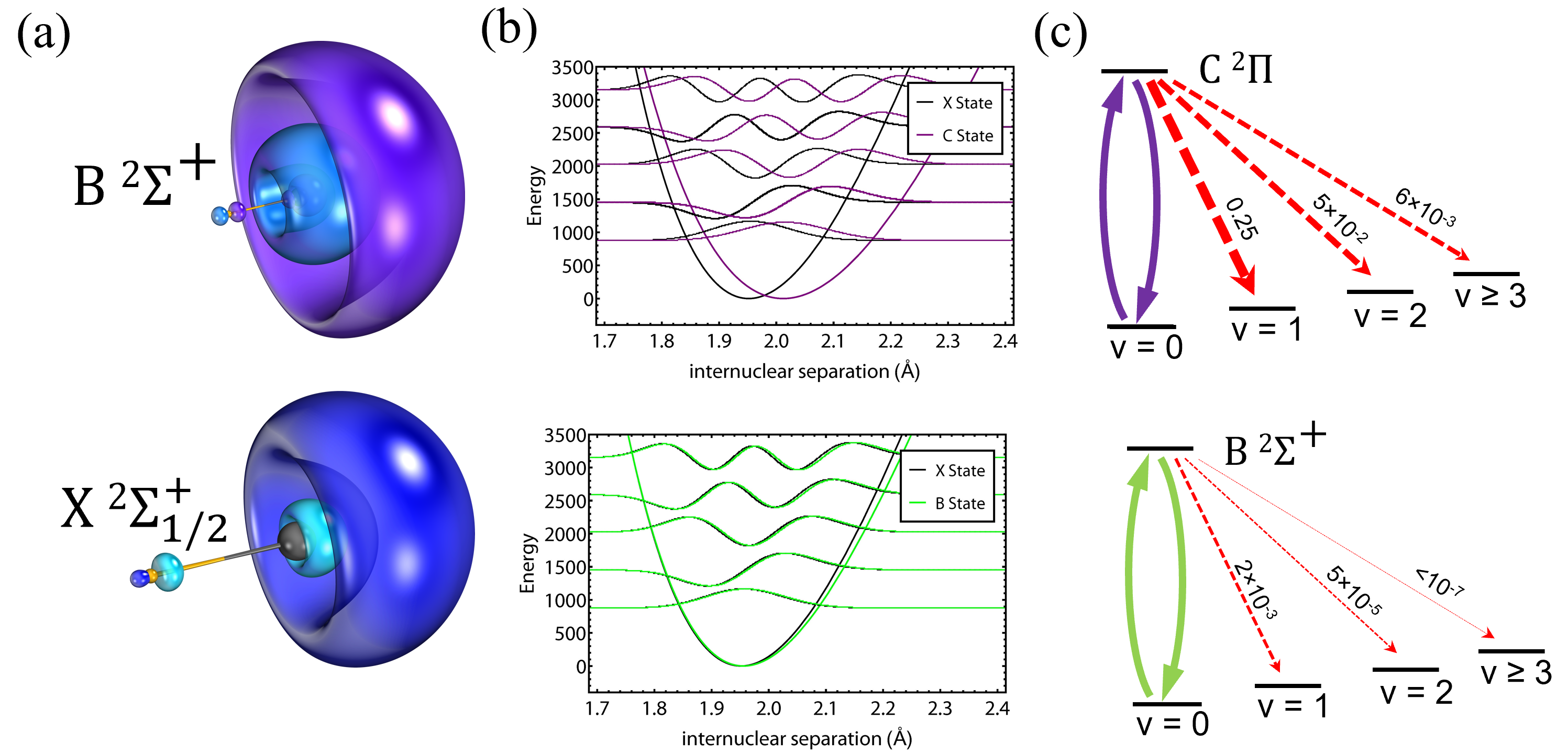}
    \caption{The Franck-Condon principle, as illustrated for diatomic molecules using CaF. (a) The $X\,^2\Sigma^+$ and $B\,^2\Sigma^+$ electronic orbitals of the prototypical laser cooling molecule CaF. Localization of the electron density near the Ca metal in both the ground and excited states leads to favorable Franck-Condon overlap.  (b) Vibrational wavefunctions of CaF. (c) The corresponding vibrational decay strengths for the $X\,^2\Sigma^+-B\,^2\Sigma^+$ and $X\,^2\Sigma^+-C\,^2\Pi$ transition of CaF.}
    \label{fig:Diadecays}
\end{figure}

Unlike rotational transitions, where angular momentum conservation
rigorously constrains the allowed decay channels, in vibrational transitions
no selection rules are absolute. Instead, vibrational branching is
governed in the Born-Oppenheimer approximation by the overlap of the
nuclear wave functions of the initial and final vibronic states, as
characterized by the Franck-Condon Factor,
\begin{equation}
    \qvv = \left\lvert \int \psi_{e,v'}^\ast \psi_{g,v} d\tau_n \right\rvert^2.
    \label{eq:FCFintegral}
\end{equation}  
where $v'$ denotes a vibrational state in the excited electronic
manifold $e$, and $v$ denotes a vibrational state in the ground
electronic manifold $g$. If the Franck-Condon factors of an electronic
transition are ``diagonal'' then $\qvv \approx \delta_{v',v}$;
in other words, a vibrational level $e(v')$ decays (almost) only to
$g(v)$. Geometrically, this will occur when all bond lengths, bond
angles, and vibrational constants are approximately identical between
the ground and excited electronic states. In practice, off-diagonal
FCFs are more sensitive to small fractional changes in bond lengths
compared to similarly small fractional changes in harmonic constants.
This can be understood from a simple model of wave function overlap
between displaced 1D harmonic oscillators with distinct harmonic constants. An example of this, for the diatomic molecule CaF, is shown in Fig.~\ref{fig:Diadecays}.

The vibrational branching ratios from a given vibronic excited state
are proportional to the FCFs, but contain an additional factor of
$\omega_{v',v}^{3}$: 
\begin{equation}
    b_{v'\rightarrow v} = \frac{q_{v'\rightarrow v} \omega^3_{v',v}}{\sum_{v} q_{v'\rightarrow v} \omega^3_{v',v}}.
\end{equation}
Thus, transitions to lower-lying vibrational states are slightly favored,
relative to what one might expect from considering only the FCF. In
practice, the VBRs and FCFs are quantiatively similar, but only the
VBR is important for determining and achieving a nearly-closed optical
cycle.

In the Born-Oppenheimer (BO) approximation, the symmetry of
a vibrational state is conserved in a vibronic transition. For example,
in linear molecules like CaOH and SrOH, $|\ell|$ would not be expected
to change. In practice, non-BO perturbations induce weak $\abs{\ell}$-changing
transitions at the level of $\sim$$10^{-3}$ vibrational branching
probability; see Sec.~\ref{sec:Perturbations}
for details.

\subsubsection{Measuring vibrational branching ratios} \label{sec:VBRmeasurements}

\begin{figure}
\centering{}\includegraphics[width=8cm]{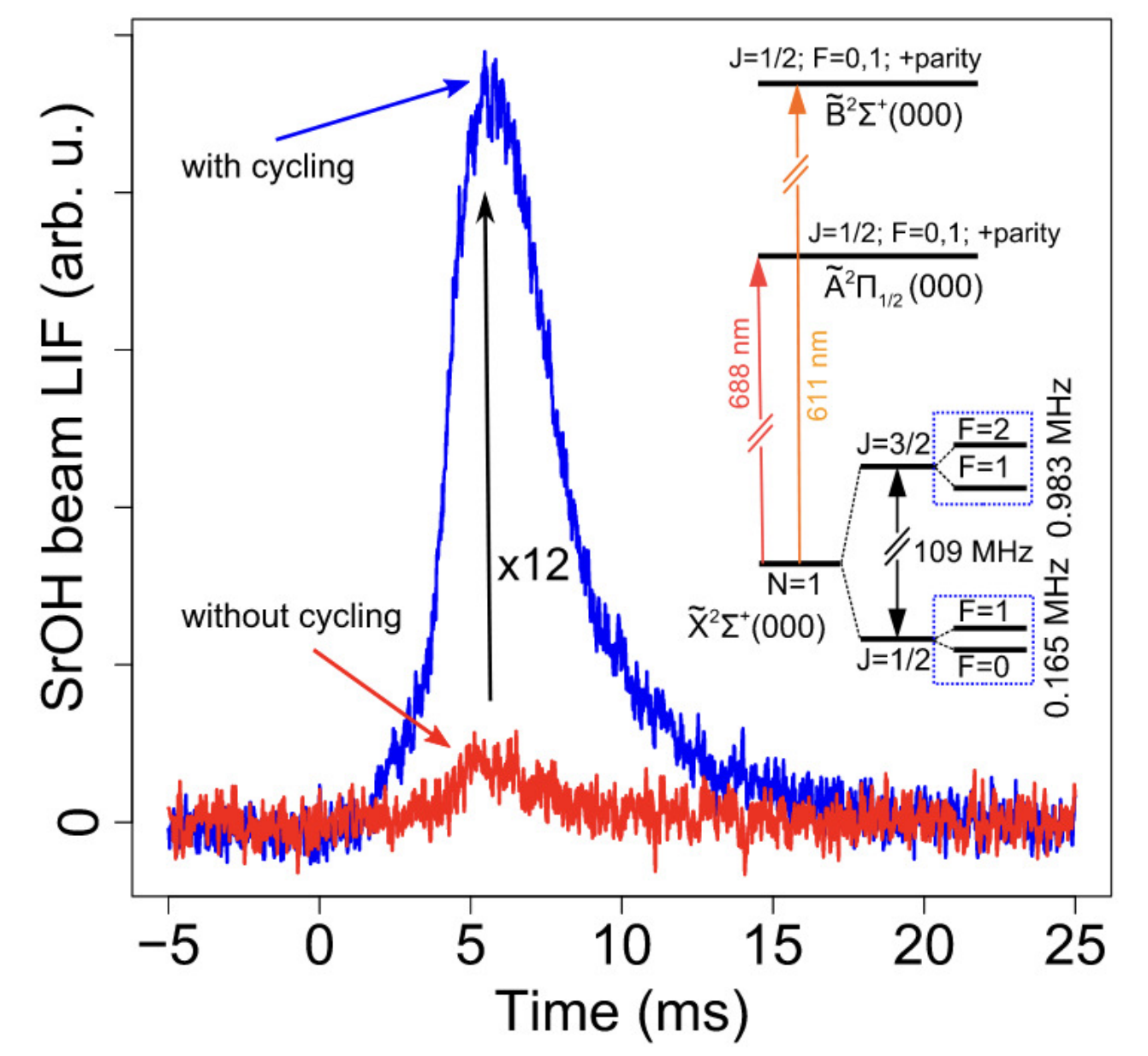}\caption{Demonstration of optical cycling with the polyatomic molecule SrOH. Molecular beam fluorescence with (blue) and without (red) addressing both spin-rotation components of the optical cycling rovibronic transition. When both spin-rotation components are addressed, population cycles between the ground $N=1$ and excited $J'=1/2$ manifolds, and repeated photon scattering increases the detected fluorescence by over an order of magnitude. Reproduced from~\cite{kozyryev2016radiation}.\label{fig:SR-closure}}
\end{figure}

Vibrational branching ratios can be measured in any of several ways.
The simplest method conceptually is to add repumping lasers (lasers that address vibrational loss channels and return them to the set of bright states) one at
a time and observe the total fluorescence collected. To determine
the diagonal vibrational branching ratio, one can measure the fluorescence
induced when addressing all spin-rotation and hyperfine components
of a rotationally-closed vibronic transition, and then compare this to the fluorescence
induced when only a single quantum state is addressed (for which
typically $\sim$1 photon is scattered before populating an unaddressed
quantum state, depending only on theoretically well-known rotational
branching ratios). See Fig.~\ref{fig:SR-closure} for an example in SrOH. In a similar manner, observing the fluorescence
collected when repumping $v$ in an optical cycle, as opposed to exhausting
a lossy optical cycle without repumping $v$, reveals the probability
that $v$ is populated among all possible loss channels. By sequentially
adding repumpers, in principle all excited vibrational branching ratios
can be determined in this way. In practice, this method may be suitable
to measure vibrational branching ratios in a long-lived trap (e.g.,
a MOT), but it is impractical in a molecular beam for VBRs smaller
than $\sim$1\% because molecules exit the fluorescence region before
exhausting a highly closed optical cycle.

An improved method to measure vibrational branching ratios below $\sim$1\%
is to optically cycle $\sim$100 photons in an ``interaction region,''
e.g. by addressing the diagonal vibronic transition and dominant one
or two repumping pathways, and to measure the fraction of molecular
population that is recovered to low-lying states when repumping a
state $v$ in a ``cleanup region'' (between the interaction region
and fluorescence detection region). If recovery of $\sim$1\% of the
molecular population can be resolved after $\sim$$10^{2}$ photons
are scattered in the interaction region, then vibrational branching
ratios of $\sim$$10^{-4}$ can be measured. This method was used
by \cite{Baum2021Establishing} to determine an optical cycling
scheme for CaOH with approximately $5\times10^{3}$ photons scattered
per molecule.

A major limitation of the methods described above is that a vibrational
branching ratio can only be measured for a state with known high-resolution
repumping transitions. Most polyatomic molecules that might be of interest
for future laser-cooling experiments have little preexisting
spectroscopic data, especially in excited vibrational states. Therefore,
an alternative method should be used to screen a potential laser-coolable
molecule for favorable VBRs without requiring high-resolution spectroscopy
of up to a dozen vibrational states.

\begin{figure*}
\begin{centering}
\includegraphics[width=0.75\columnwidth]{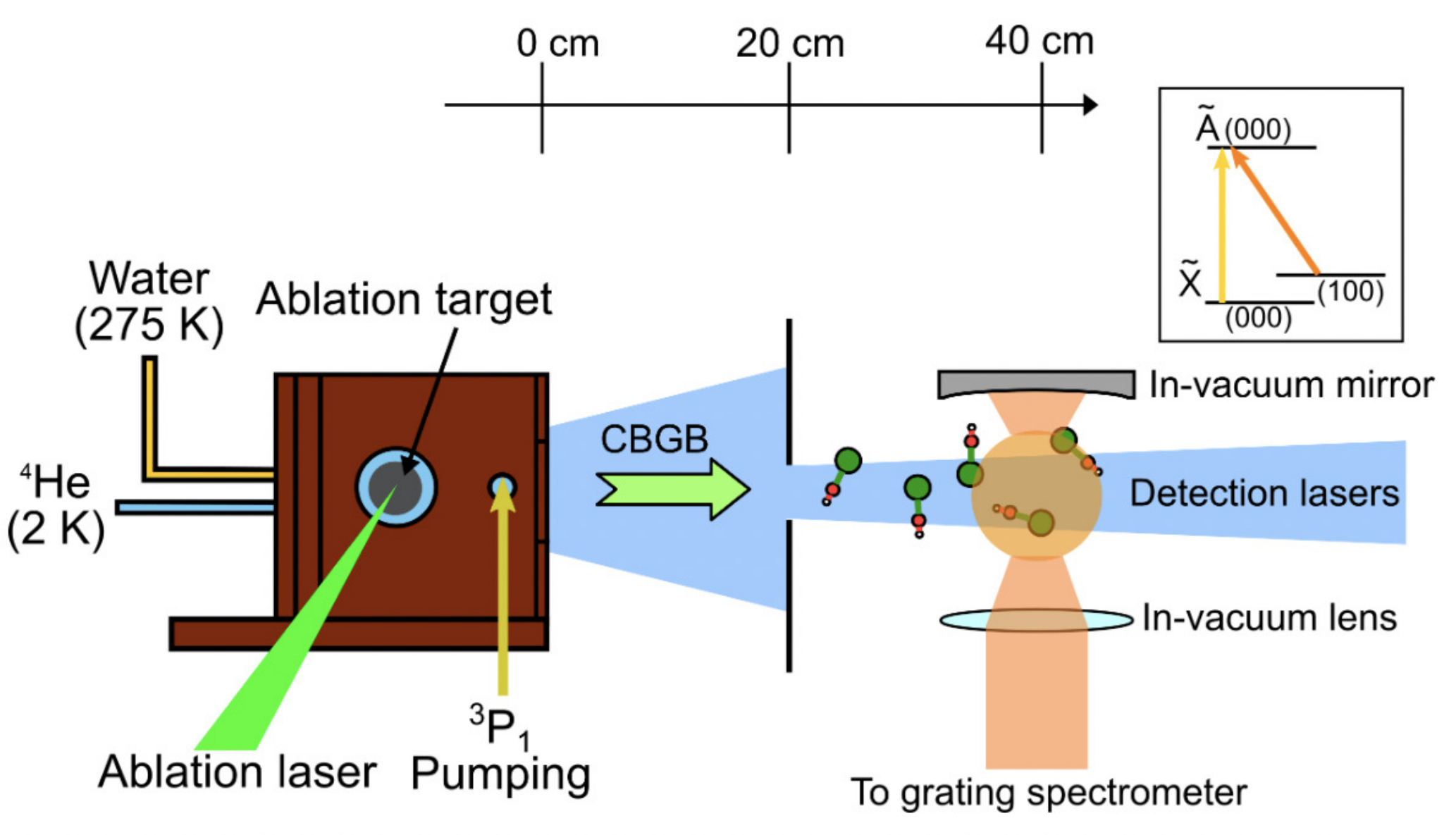}\caption{Dispersed laser-induced fluorescence apparatus, reproduced from~\cite{Zhang2021}. CaOH and YbOH molecules are produced in a CBGB
via ablation of a metal precursor in the presence of water vapor.
40 cm downstream, molecules are excited via an optical cycle with
50$-$100 photons scattered per molecule on average, increasing the
fluorescence yield. Fluorescence is collimated by an in-vacuum lens
and directed toward a Czerny-Turner monochromator, which disperses
light emitted in different vibronic transitions (generally separated
by many nm in wavelength) onto different regions of an EMCCD camera.\label{fig:DLIF-apparatus}}
\par\end{centering}
\end{figure*}

\begin{figure}
\centering{}\includegraphics[width=8cm]{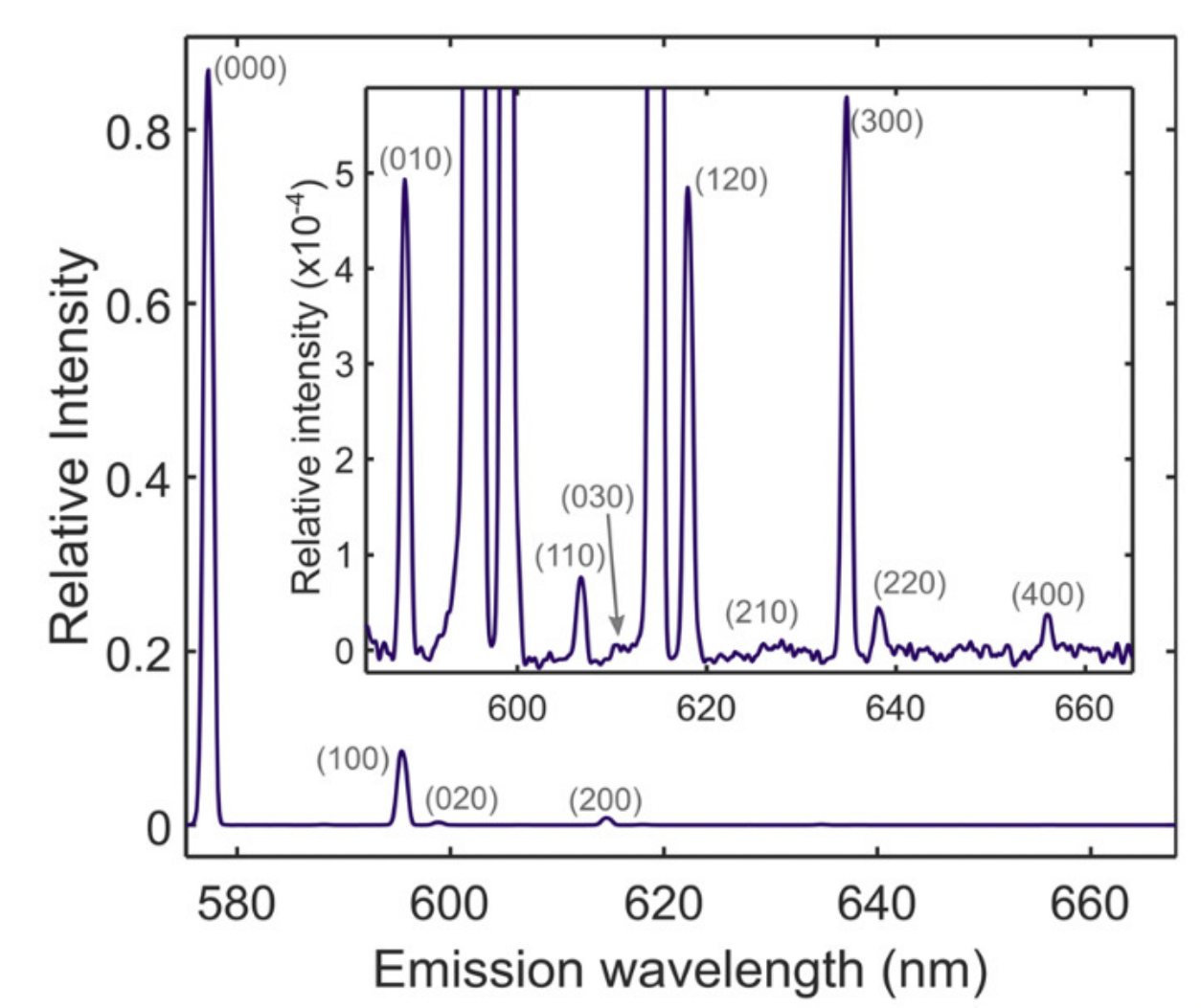}\caption{Dispersed fluorescence measurements of YbOH $\tilde{A}(000)$, showing 89\%
probability of decay on the diagonal vibronic transition, with progressively
weaker decays to higher-lying vibrational states. Decays with strength
as small as $2(1)\times10^{-5}$ relative probability can be measured
due to the increased fluorescence yield arising from optical cycling
excitation. Reproduced from~\cite{Zhang2021}.\label{fig:DLIF-data}}
\end{figure}

The standard approach to measure VBRs without directly repumping
vibrationally excited states is dispersed laser-induced fluorescence
(DLIF). See Fig. \ref{fig:DLIF-apparatus}
for a typical experimental configuration. Molecules in a molecular beam are driven to an excited state of interest,
$e$, and subsequently fluoresce to the vibronic ground states
$\{v_{i}\}$ with probabilities $\{P_{i}\}$. The fluorescence is
collected and focused into a spectrometer, in which a diffraction grating is used to spatially disperse the fluorescence wavelengths. The light is then imaged onto an EMCCD, producing an image that shows the relative intensity of each decay feature. Example data are shown for
emission from the $\tilde{A}(000)$ state of YbOH in Fig.~\ref{fig:DLIF-data}.

With $\sim$$10^{4}$ ablation pulses, vibronic decays with relative
probabilities of $10^{-2}-10^{-3}$ may be detected. By optically
cycling molecules in the detection region, up to 100 photons may be
scattered per molecule, directly increasing the number of fluorescence photons collected.
In cases where the detection sensitivity is limited by camera read-noise or clock-induced-charge, the sensitivity of the measurement is increased proportionally
to the number of photons scattered per molecule. In this
way, branching ratios on the order of $10^{-5}$ have been measured
for CaOH, YbOH, and SrOH by \cite{Zhang2021,Lasner2022}.
The unambiguous assignment of weak decays is greatly aided by high-quality
theoretical predictions, which must account for perturbations like those described in Sec.~\ref{sec:Perturbations}.

DLIF measurements of VBRs are a powerful method to directly observe
all decay channels (inside an observed wavelength range) without requiring
high-resolution spectroscopy for optical pumping and repumping of
many vibrational states. It is especially important for measuring
vibrational branching ratios of polyatomic molecules, where numerous
vibrational states may have VBRs around $10^{-5}-10^{-4}$, a level that is weak in an absolute sense but sufficiently strong to require repumping during
radiative slowing and magneto-optical trapping. For a polyatomic molecule with many vibrational modes, it is not necessarily clear in the absence of measurements what
the dominant loss channel will be for an optical cycle scattering many more than $\sim$100
photons per molecule. This makes a brute-force search for vibrational
repumpers inefficient in all but the exceptional cases where the molecule
is extensively well-understood from the outset.

\subsubsection{Repumping transition spectroscopy}

\begin{figure*}
\begin{centering}
\includegraphics[width=0.9\columnwidth]{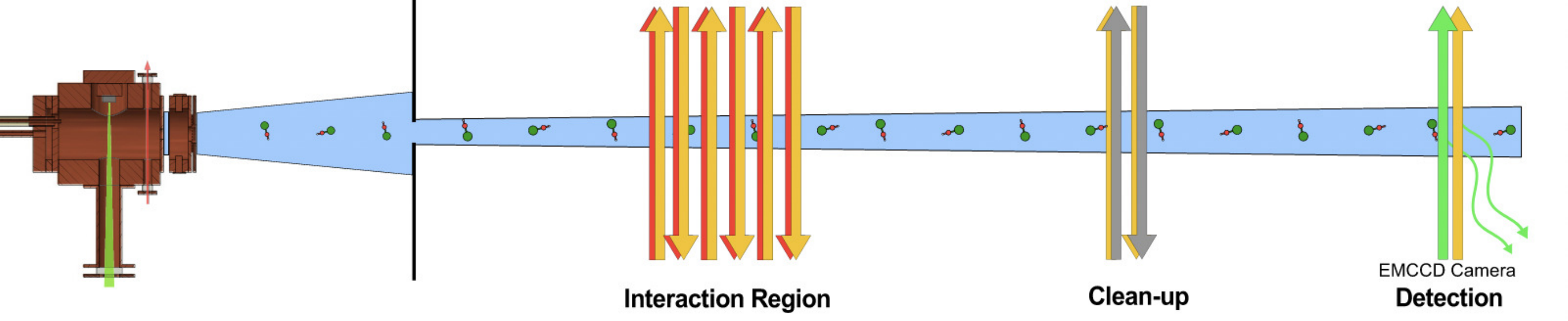}\caption{Beamline configuration for pump-repump spectroscopy of vibrationally
excited states. An upstream
interaction populates vibrationally excited states either through
optical cycling or direct off-diagonal excitation. A clean-up region
recovers population into a detected state (or set of states), which
is excited by a known laser transition to produce fluorescence that
is imaged onto an EMCCD camera. When a laser in the interaction region
is scanned over an excited state with the clean-up lasers off or out
of resonance, a dip in detected fluorescence is observed. This enables
the discovery of excited-state repumping pathways. When a laser in
the clean-up region is scanned over a repumping transition, detected
fluorescence is partially recovered. \label{fig:pump-repump}}
\par\end{centering}
\end{figure*}

Even in the best-studied laser-coolable molecules such as CaOH and
SrOH, the vibrational states that become populated after a molecule has scattered around
$10^{4}$ photons have not all been fully analyzed in the literature.
It is thus necessary to conduct spectroscopic searches for repumping
pathways. These measurements are often conducted in one of two ways:
fluorescence in a molecular beam, or absorption in a cryogenic buffer-gas cell.

When repumper spectroscopy is conducted using laser-induced fluorescence from a molecular beam (see Fig. \ref{fig:pump-repump}),
a vibrationally excited state can be prepared either by optically
cycling until vibrational dark states are populated, or by direct
off-diagonal excitation that decays preferentially to the vibrational
state of interest in the ground electronic manifold. Due to the rotationally-closed nature of the transition used for optical cycling, often only the $N=1$ level of the target vibrational state is populated. (Additional rotational levels may be populated in nonlinear polyatomic molecules where the rotational selection rules are slightly relaxed; see Tab.~\ref{tab:selectionrules}.) The frequency of a laser
in a downstream ``clean-up'' region can then be scanned to repump population
into (often lower-lying) detectable states, which are excited in a region
even farther downstream. Fluorescence in the detection region is observed
on a PMT or EMCCD camera. When the frequency of the repumping laser
passes through a resonance, an increase in detected population will be
observed. Since CBGBs sometimes do not efficiently thermalize the molecular
vibrational distribution, in some cases a molecular beam may have
sufficient natural population of a vibrationally excited state to
detect, provided the correct repumping frequency is addressed in the
clean-up region. The same geometry shown in Fig. \ref{fig:pump-repump}
can also be used to search for excited vibrational states in the
electronically excited manifold by scanning the frequency of a laser
in the interaction region and observing a dip in the detected population
of low-lying vibrational states.

This pump-repump method is best-suited to discovering the frequency
of \emph{specific} rovibronic transitions, since in practice at most
a few states can be simultaneously detected in the downstream region.
This means that, for a given experimental configuration, only a few
transitions (including the ``laser cooling transition'') in the
clean-up region can produce a spectroscopic signal, dramatically simplifying
the data analysis. A corresponding disadvantage, however, is that
the signal is spectroscopically sparse, and a good initial estimate
of the repumping frequency is required. This estimate can be made
either from the observed decay wavelengths obtained via DLIF, or by
high-quality theoretical predictions (e.g., made by extrapolating
from the known positions of lower-lying vibrational states). Repumper
frequency uncertainties obtained by high-resolution DLIF measurements
are typically on the order of 5 cm$^{-1}$, but can be improved by
using a diffraction grating with higher line density or by narrowing
the width of the entrance slit to the spectrometer.

A related technique can be used to locate repumping transitions after a MOT has
been achieved. Namely, one can apply laser light near a suspected repumping transition in the MOT region,
with all other trapping laser beams on. While scanning the frequency of the (unknown) repumping laser, one monitors the MOT lifetime. When
a resonance is reached, the lifetime of the MOT should increase. This
is analogous to the pump-repump method but with an interaction
time of tens or even hundreds of ms, which would be impractical in a molecular beam.

An alternative approach to identifying vibrational repumpers is high-sensitivity
absorption measurements inside a buffer gas cell (\cite{Pilgram2022}). Frequency-modulated (FM) absorption spectroscopy has been
used to observe the weak $\Delta v=-2$ repumping transition for states
as high as $(300)$ in YbOH. Unlike the pump-repump spectroscopy described
above, all thermally populated rotational states will be probed in
this way. This may be desirable in order to make a full spectroscopic
assignment of the molecular transitions and constants in a vibronic  transition, or to more quickly locate a spectroscopically active
frequency region (since a denser ``forest'' of lines appears, compared
with the signal in pump-repump measurements). On the other hand, the
spectrum must be fully assigned in order to identify which transitions
are connected to the laser cooling state. Aside from these broader
considerations, the experimental signal-to-noise ratio may favor either
the pump-repump or high-sensitivity absorption measurements depending
on factors like the vibrational quenching efficiency in the buffer
gas cell, the degree of scattered light suppression in the downstream
fluorescence region, the strengths of the probed transitions, and other technical factors.

\subsubsection{Designing an optical cycle}\label{sec:design-optical-cycle}

\begin{figure*}

\begin{centering}
\includegraphics[width=0.9\columnwidth]{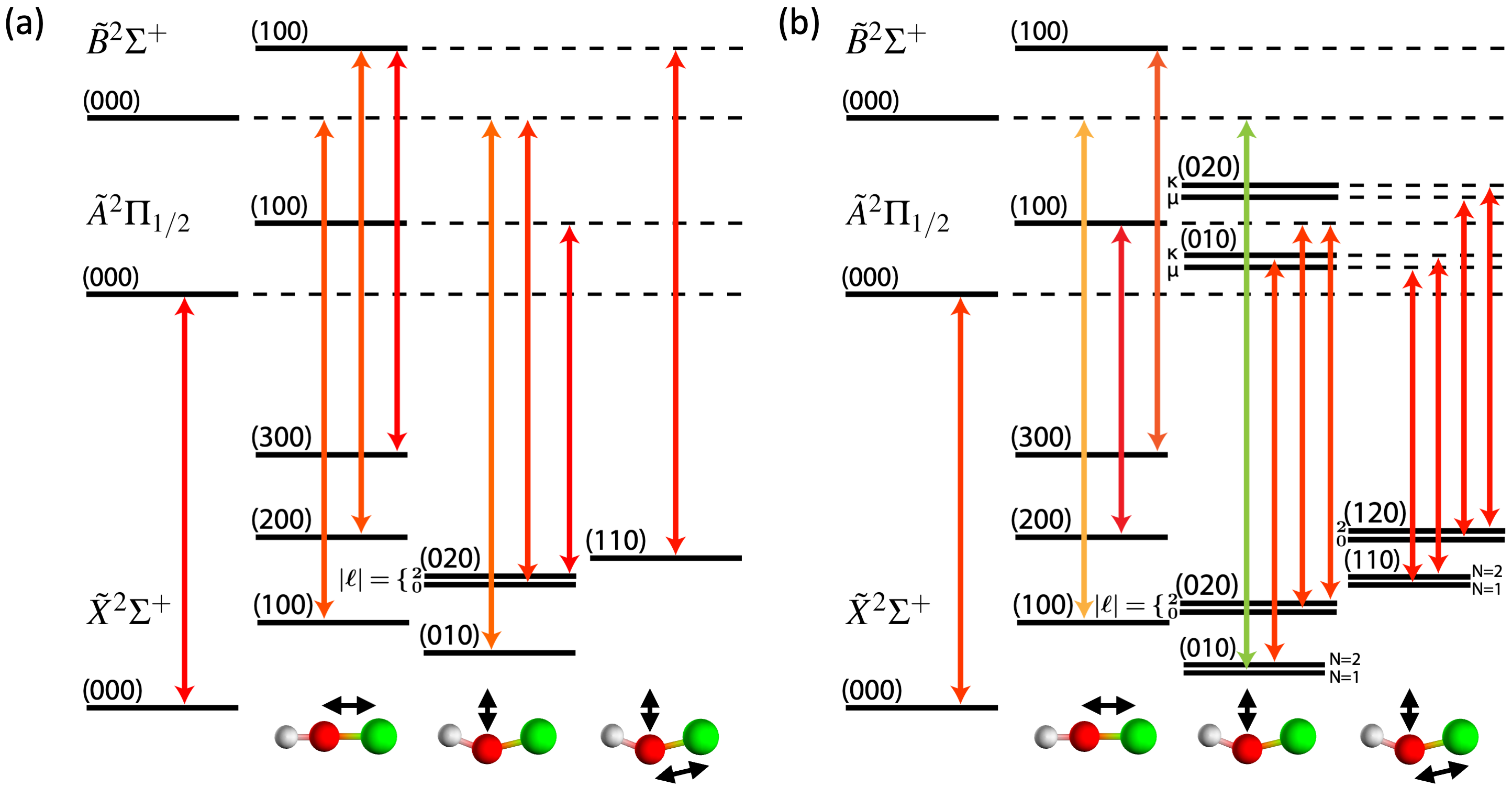}\caption{(a) Optical cycling scheme proposed for laser cooling and trapping SrOH,
reproduced from~\cite{Lasner2022}. (b) Optical cycling scheme used to produce a MOT of CaOH.  \label{fig:optical-cycles}}
\par\end{centering}
\end{figure*}

With known vibrational branching ratios (e.g., from high-resolution
DLIF measurements), one must carefully design the optical cycle. Several
factors are important:
\begin{itemize}
\item The number of lasers should be minimized, to reduce cost and experimental
complexity
\item The excited state that is coupled to the ground vibrational state\textendash e.g.,
(000) in an MOH molecule\textendash should not be coupled to any other
vibrational states if possible, so as to maximize the photon scattering
rate and optical forces (see Sec.~\ref{sec:EffectiveScatteringRate})
\item Lasers should be at convenient optical wavelengths and available at
high power
\item Optical transitions should be strong (e.g., $\Delta v=-1$ repumpers)
to increase the optical pumping or repumping rates for fixed laser
powers
\item Decays to states without existing spectroscopy should be eliminated
where possible
\end{itemize}
Not all of these factors are mutually compatible, and some
balancing is required (e.g., the number of lasers is generally minimized
when only a single excited state is used). Typically, only one or
two ``diagonal enough'' excited states are available to choose from
for the dominant excitation, and one of these may result in significantly
fewer states populated after the $\sim10^{4}$ photon scattering events
required for magneto-optical trapping. For example, in SrOH, both
$\tilde{A}(000)$ and $\tilde{B}(000)$ appear to have reasonably diagonal VBRs, but
high-resolution DLIF measurements show that $\tilde{B}(000)$ populates $(03^{1}0),$
$(12^{0}0)$, $(13^{1}0)$, and $(05^{1}0)$ at the $\sim10^{-4}$
level; none of these are significantly populated by $\tilde{A}(000)$, as demonstrated in \cite{Lasner2022}. Therefore, many additional repumpers
would need to be added to scatter $10^{4}$ photons primarily through
$\tilde{B}(000)$, as compared to scattering the same number of photons through
$\tilde{A}(000)$. For this reason, an optical cycle coupling $\tilde{A}(000)\leftarrow \tilde{X}(000)$
is favored. Because vibrationally excited states are populated only
after approximately 20 photons are scattered through $\tilde{A}(000)$, the $\tilde{B}(000)$
state may be used as a repumping pathway without limiting the optical
cycle to less than $1.5\times10^{4}$ photons scattered. Wherever possible,
the optical cycle proposed in~\cite{Lasner2022} favors repumping through the $\tilde{B}$ manifold due to the much more inexpensive
and convenient high-power sum-frequency generation (SFG) sources around 630$-$650 nm compared
with 690$-$710 nm. The only exception is that the $(02^{2}0)$ state
cannot be repumped through $\tilde{B}$ due to a near-vanishing transition
strength, and is instead coupled to $\tilde{A}(100)$. Wherever possible,
the strongest transition that decreases vibrational quantum
numbers from the ground to excited state is chosen, under the constraint that
$\tilde{A}(000)$ may not coupled strongly to any state except $\tilde{X}(000)$. The resulting
optical cycle is shown in Fig. \ref{fig:optical-cycles}, along with
the optical cycle used to magneto-optically trap CaOH.

To model the number of scattered photons before loss to unaddressed
vibrational states occurs, it is useful to construct an absorbing Markov
chain model similar to that described in~\cite{Baum2021Establishing}:
the states of the Markov chain represent vibrational levels in the
electronic ground manifold. For a given optical cycling scheme, the
transition probabilities from a state $v$ are given by the VBRs of
the excited state to which $v$ is coupled. Each step of the Markov
chain represents a single photon scatter. Unaddressed vibrational
states ``transition'' only to themselves, and are absorbing states
in the Markov chain. In this model, it is straightforward to calculate
many properties of interest to laser cooling, including the average
number of scattered photons before an absorbing (i.e., dark vibrational)
state is reached, how many times each transient state is visited on
average, and the distribution of population among dark states.

\subsubsection{Rotational transitions} \label{sec:RotationalTransitions}

Rotational transitions follow a generally well-behaved set of selection rules governed by angular momentum algebra. Unless otherwise stated, in this section we consider only electric dipole (E1) transitions. Higher-order transitions (e.g. M1, E2, etc.) are significantly weaker and are typically irrelevant for molecular laser cooling.

The general formula for generating a closed cycling transition in molecules was first proposed by \cite{Stuhl2008}, and relies on rotational and parity selection rules ($\Delta J = 0,\pm1$; $\Delta p = \pm 1$; $J=0 \rightarrow J'=0$ forbidden). The key observation was that, while a traditional ``type-I'' $J \rightarrow J+1$ transition cannot be rotationally closed (since a rotational level with $J+2$ necessarily exists in the ground state), driving the ``type-II'' $J=1 \rightarrow J'=0$ transition (or its closest analogue within the molecule of choice) forms a closed transition. One consequence of this choice of transition is that there are more ground states than excited states, $m_J > m_J'$, meaning that there will be dark states regardless of the laser polarization chosen. These can be remixed with magnetic fields or polarization modulation (\cite{Berkeland2002}).

\begin{table}[]
    \centering
    \begin{tabular}{ccc|l}
    \hline
    \hline
    Geometry & Transition axis & Basis$^*$ & Selection rules$^*$  \\
    \hline
    Linear & $\parallel$ & b $\rightarrow$ b & $\Delta \Lambda = 0$; $\Delta \ell = 0$; $\Delta N = 0, \pm1 $ \\
    Linear & $\perp$ & b $\rightarrow$ a & $\Delta \Lambda =\pm1$; $\Delta \ell = 0$ \\
    \hline
    STM & $\parallel$ & b $\rightarrow$ b & $\Delta K = 0$; $\Delta N = 0, \pm1 $ \\
    STM & $\perp$ & b $\rightarrow$ a & $\Delta K_R = 0$; $\Delta K = \pm 1$\\
    \hline
    ATM & a type & b $\rightarrow$ b & $\Delta K_a = 0$; $\Delta K_c = \pm1$; $\Delta N = 0,\pm 1^{**}$ \\
    ATM & b type & b $\rightarrow$ b & $\Delta K_a = \pm 1$; $\Delta K_c = \pm1$; $\Delta N = 0,\pm 1^{**}$  \\
    ATM & c type & b $\rightarrow$ b & $\Delta K_a = \pm 1$; $\Delta K_c = 0$; $\Delta N = 0,\pm 1^{**}$ \\
    \hline
    \hline
    \end{tabular}
    \caption{Summary of the main selection rules discussed in this review. All transitions follow the selection rules $\Delta J = 0,\pm1$ and $\Delta p = \pm 1$ in addition to those listed. Also note that at least one angular momentum quantum number must change, so transitions with $\Delta K = \Delta N = \Delta J = 0$ (or analogous) are forbidden.\\
    $^*$Note that the selection rules described here apply only when the states are exactly described by the basis listed; in many real scenarios the selection rules are only approximate.
    \\
    $^{**}\Delta N = 2$ transitions may be allowed in ATMs where the electronic angular momentum is not fully quenched.}
    \label{tab:selectionrules}
\end{table}

All polyatomic molecules that have been directly laser cooled (and most polyatomics proposed for laser cooling) have a single valence electron and therefore have spin-doublet ground states ($S=1/2$). In this case, the rotational angular momentum $N$ couples to the electron spin to form a total angular momentum quantum number $J$, which takes half-integer values.\footnote{Here we ignore hyperfine degrees of freedom, which are often unresolved in polyatomic molecules, though the selection rules discussed below can be easily generalized to include hyperfine structure.} In such a system, the electric dipole selection rules $\Delta J = 0,\pm1$ and $J=0 \nrightarrow J'=0$ always apply, as well as the requirement that the parity of the state changes, $\Delta p = \pm 1$. In this case, the excited state used to achieve rotational closure is no longer $J'=0$ but $J'=1/2$.

In addition to the universal selection rules on $J$ and parity, other selection rules exist for specific molecular geometries and angular momentum coupling cases. These are summarized in Table \ref{tab:selectionrules}. The selection rules are sorted by molecule geometry, as well as the axis along which the transition dipole moment is induced. For molecules with cylindrical symmetry, transitions can be either parallel ($\parallel$) or perpendicular ($\perp$), meaning the transition dipole moment is induced either along the principal molecular axis or perpendicular to it, respectively. For asymmetric top molecules, a transition can be induced along any of the three molecular axes, as described further below. Finally, different selection rules apply depending on the coupling of angular momenta within each electronic state involved. Typically, in nondegenerate states with zero electronic angular momentum Hund's case (b) applies, where both $N$ and $J$ are good quantum numbers. However, degenerate states with electronic angular momentum $\Lambda \neq 0$ are typically described by Hund's case (a), where the electron spin is strongly coupled to the molecular axis and $N$ is no longer a good quantum number.

Rotational transition strengths are described by the H\"onl-London factors $S_{J}^{J'}$ in eqn. \ref{eq:FullLineStrength}. They can be calculated using angular momentum algebra by taking matrix elements of the dipole operator (see, e.g., \cite{Hirota1985}), or looked up in tables.

\begin{figure}
\centering
\includegraphics[width=0.8\textwidth]{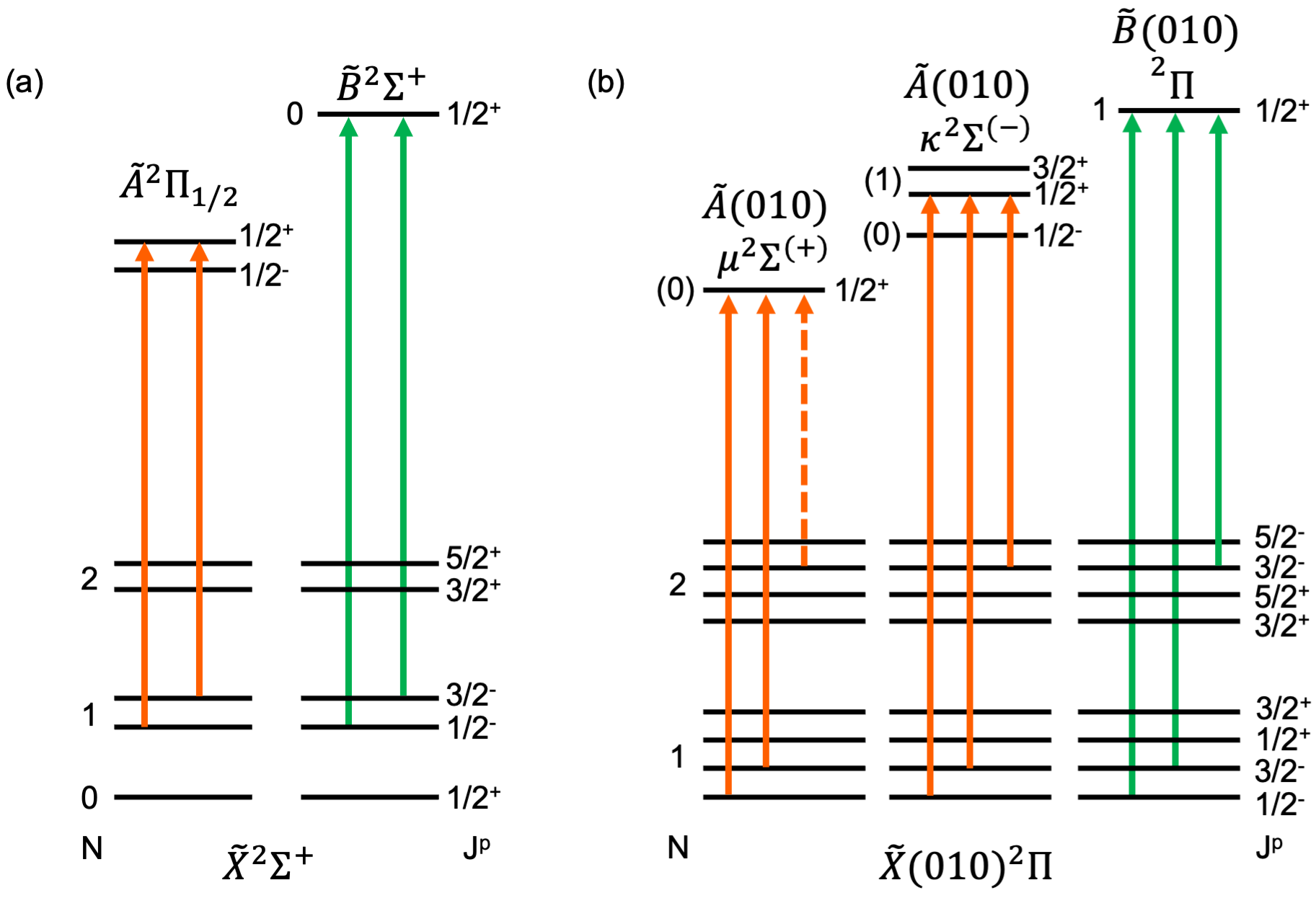}
\caption{Closed cycling transitions in alkaline-earth monohydroxides (e.g. CaOH) for laser cooling on the $\widetilde{X}^2\Sigma^+\rightarrow\widetilde{A}^2\Pi_{1/2}$ and $\widetilde{X}^2\Sigma^+\rightarrow\widetilde{B}^2\Sigma^+$ electronic transitions. (a) For nondegenerate ($\ell = 0$) vibrational levels of the ground state, two laser frequencies are required to address the $J=1/2^-$ and $J=3/2^-$ levels of the $N=1$ rotational state. (b) For repumping the $\widetilde{X}(010)$ bending mode, which has vibrational angular momentum $|\ell| = 1$, an additional color addressing the $N=2, J=3/2^-$ state is required, regardless of which excited state is chosen. The dashed line refers to a transition which is allowed only when the excited state does not adhere to the case (b) limit, i.e. when $N$ is not a good quantum number.}
\label{fig:linearRotational}
\end{figure}

\textbf{Linear molecules.} In nondegenerate vibrational states ($\ell = 0$) of $^2\Sigma^+$ linear polyatomic molecules like CaOH, SrOH, and YbOH, a closed optical cycling transition is formed by driving the $^PQ_{21}(J=1/2)$ and $P_1(J=3/2)$ transitions, which address both spin-rotation components of the $\widetilde{X}\,^2\Sigma^+(N=1^-)$ ground state and excite them to either the $\widetilde{A}\,^2\Pi_{1/2}(J'=1/2^+)$ or $\widetilde{B}\,^2\Sigma^+(N'=0,J'=1/2^+)$ state (Fig.~\ref{fig:linearRotational}a). The key ingredient for achieving rotational closure in this scheme is not the specific electronic state but the fact that it have a well-resolved $J'=1/2^+$ level. In this case the transition is guaranteed to be closed by the E1 selection rules $\Delta J = 0,\pm 1$ and $\Delta p = \pm 1$.

The inclusion of hyperfine structure threatens to void these selection rules because the $(J'=1/2,F=1^+)$ level of the excited state may, in principle, decay to the $(N=3,J=5/2,F=2^-)$ ground state level. However, this requires that the hyperfine interaction mix the $N=1$ and $N=3$ states. The mixing fraction is on the order of $\sim c^2/(10B)^2 \sim 10^{-10}$ for CaOH, where $c$ is the dipolar hyperfine constant and $B$ is the rotational constant. While this decay mechanism is negligible for CaOH and alkaline-earth monohydroxides with similar structure, it may be of some significance for species with large hyperfine structure, e.g., with a nuclear spin on the optical cycling center (see, e.g., \cite{Pilgram2021Fine}).

In degenerate vibrational states ($\ell \neq 0$) rotational closure is complicated by the appearance of parity doublets in the ground state (Fig. \ref{fig:linearRotational}b). In this case, the excited $J'=1/2^+$ state can not only decay to $N=1^-$ as before, but also to the $N=2,J=3/2^-$ sublevel. For the linear polyatomic molecules which have been laser cooled to date, the most important cases are $\ell = 1$ and $\ell = 2$. For example, in CaOH, YbOH, and SrOH, the $\widetilde{X}(01^10)$ and $\widetilde{X}(02^20)$ states typically require repumping at or above the $10^{-4}$ level (\cite{Zhang2021, Lasner2022}). As shown in Fig. \ref{fig:linearRotational}, $\ell = 1$ states require both an $N=1^-$ and an $N=2^-$ repumping laser. Unlike the spin-rotation splitting between the $J=1/2$ and $J=3/2$ states in $N=1$, which are typically spaced by $\sim 10-100$ MHz and can be addressed with rf modulation (e.g. AOMs or EOMs), the $N=1$ and $N=2$ repumpers in $\ell = 1$ bending modes are split by 10s of GHz in the alkaline-earth monohydroxides, meaning that either two separate lasers or high frequency EOMs need to be used to bridge the gap. The $\ell = 2$ states, meanwhile, require just a single repumping laser, which addresses the $N=2, J=3/2^-$ state.

For $\ell \neq 0$ bending modes, the excited state used for repumping also merits careful consideration. The primary concern is that many candidate states are best described by a Hund's case (b) basis, meaning that both $N'$ and $J'$ are good quantum numbers. Accordingly, while the $N=1^-$ ground state levels can always be repumped through the excited $J'=1/2^+$ state, in some cases repumping of $N=2^-$ is forbidden by an approximate $\Delta N = 0,\pm1$ selection rule. For example, in the laser cooling scheme used for CaOH by \cite{vilas2022magneto}, $\widetilde{X}(01^10)(N=1^-)$ is repumped through the $\widetilde{B}\,^2\Sigma^+(000)(N'=0,J'=1/2^+)$ state, but $N=2^-$ is not because of $\Delta N$ selection rules. Instead, the $N=2, J=3/2^+$ state is repumped through the $\widetilde{A}(010)$ electronic manifold, which has two components $\mu^2\Sigma^{(+)}$ and $\kappa^2\Sigma^{(-)}$ (Fig. \ref{fig:linearRotational}b; this state is described in detail in \cite{li1995bending}). These states are intermediate between case (a) and case (b), so $\Delta N$ selection rules are somewhat weakly enforced; however, the $\kappa^2\Sigma^{(-)}$ is the best candidate for $N=2^-$ repumping because its $J'=1/2^+$ state has dominantly $N'=1$ character, while the $\mu^2\Sigma^{(+)}(J'=1/2^+)$ state has predominantly $N'=0$ character. In CaOH, the $\mu^2\Sigma^{(+)}(J'=1/2^+)$ state has only a {$\sim7\%$ transition strength to $N=2^-$, while the $\kappa^2\Sigma^{(-)}(J'=1/2^+)$ state connects in approximately equal proportion to $N=1^-$ and $N=2^-$ in the $\ell=1$ bending modes.

In practice, it is often necessary to repump degenerate bending modes using transitions that do not satisfy the $\Delta \ell = 0$ selection rule (see section \ref{sec:design-optical-cycle}). In this case, rotational transition strengths are challenging to calculate because they may be altered by vibronic perturbations. These must be considered on a molecule-by-molecule basis. For repumping the $\ell = 1$ bending modes in CaOH, it is empirically known that repumping on the $\widetilde{X}^2\Sigma^+(010)(N=1^-) \rightarrow \widetilde{B}^2\Sigma^+(000)(N'=0,J'=1/2^+)$ transition is sufficiently strong (\cite{Baum2020,Baum2021Establishing,vilas2022magneto}). However, in CaOH the $N=2^-$ states are repumped using $\Delta \ell = 0$ transitions (\cite{vilas2022magneto}).

\begin{figure}
\centering
\includegraphics[width=0.8\textwidth]{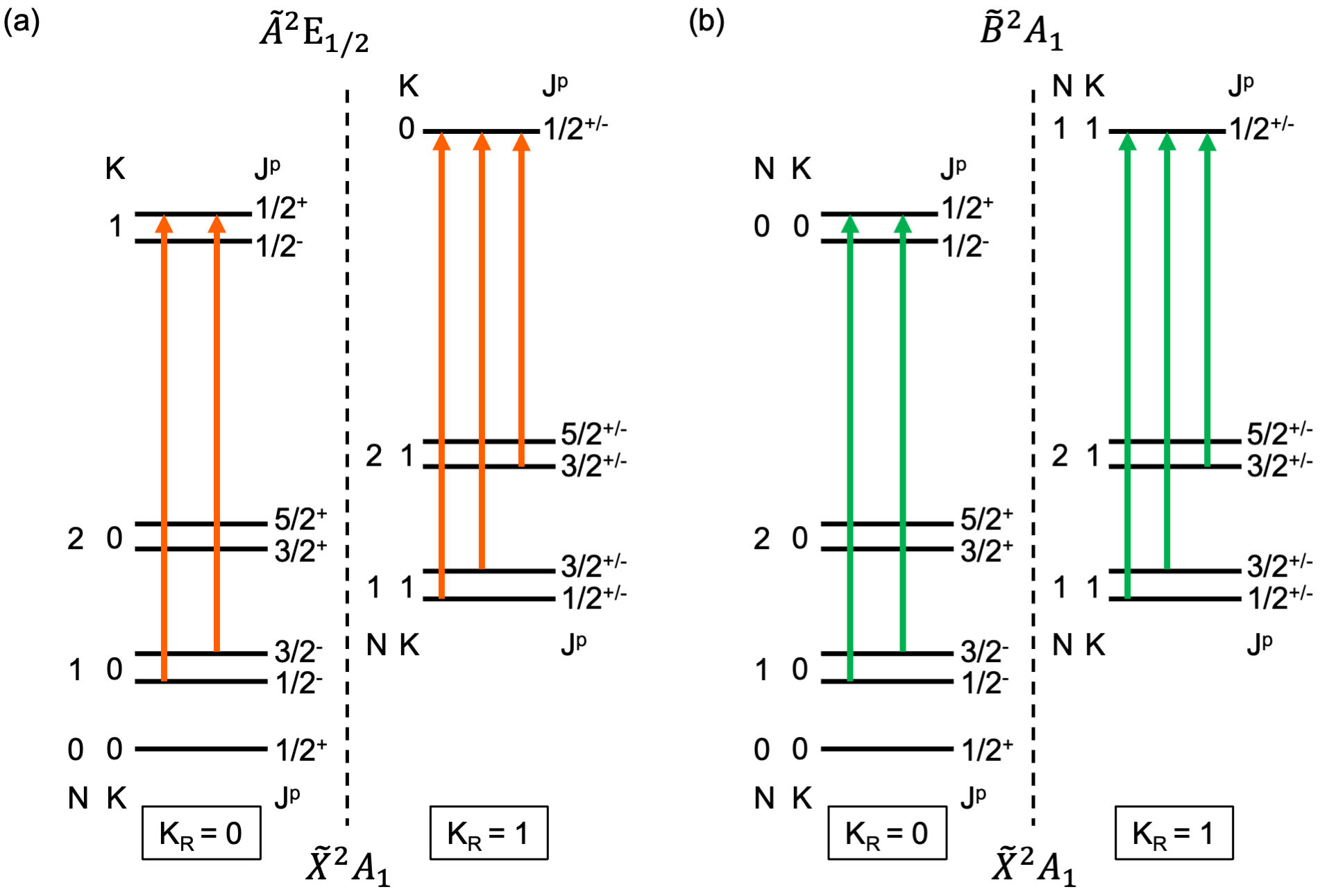}
\caption{Rotationally closed cycling transitions for symmetric top molecules with structure similar to CaOCH$_3$ on the (a) $\widetilde{X}^2A_1 \rightarrow \widetilde{A}^2E$ and (b) $\widetilde{X}^2A_1 \rightarrow \widetilde{B}^2A_1$ electronic transitions. Cycling transitions exist within both the $K_R = 0$ stack and the $K_R = 1$ stack, though in the latter case an $N=2$ rotational repumping laser is required.}
\label{fig:STMRotational}
\end{figure}

\textbf{Symmetric top molecules.} The rotational cycling transitions for ${}^2\!A_1$ symmetric top molecules (STMs) (e.g. CaOCH$_3$ or YbOCH$_3$) are shown in Fig.~\ref{fig:STMRotational}. In addition to the $J$ and parity selection rules discussed above, there are also $K$ selection rules for STMs.

For parallel transitions between two nondegenerate electronic states (e.g. $\widetilde{X}\,^2A_1 \rightarrow \widetilde{B}\,^2A_1$), the selection rule $\Delta K = 0$ holds. It is useful to divide the molecule into ``$K$ stacks'', each of which is well isolated during optical cycling. A canonical $N=1^- \rightarrow N'=0^+$ cycling transition can therefore be formed in the $K=0$ stack, while cycling in the $K=1$ stack requires an additional $N=2$ repumping laser, as shown in Fig.~\ref{fig:STMRotational}b.

For perpendicular transitions between a nondegenerate and a degenerate electronic state (e.g. $\widetilde{X}\,^2A_1 \rightarrow \widetilde{A}\,^2E$), the photon adds one unit of electronic angular momentum ($\Lambda$) about the molecular axis, while the rigid body rotation of the molecule (i.e. $K_R$, the projection of $R$ onto the molecular axis) is unchanged. Therefore $\Delta K = \Delta K_R + \Delta \Lambda = \pm 1$, but $\Delta K_R = 0$.\footnote{Note that the electronic angular momentum is quenched in symmetric top molecules, so $\Lambda$ is not necessarily an integer. Instead, the orbital angular momentum about the symmetry axis is labeled by $\zeta_e$, and both it and $K_R$ are only approximate quantum numbers. Nonetheless, the selection rules discussed above based on $K_R$ stacks are expected to hold in general, though the $K_R$ notation is simply a convenience for molecules where $\zeta_e \approx 1$ (see \cite{Brazier1989}).} In this case we can instead divide the structure into isolated ``$K_R$ stacks'', noting that $K=K_R$ in the nondegenerate ground state. A closed $N=1^- \rightarrow J'=1/2^+$ cycling transition can be found in the $K_R=0$ stack, while cycling in the $K_R = 1$ stack requires an $N=2$ repumping laser because each rotational level contains parity doublets (Fig. \ref{fig:STMRotational}a). Note that the degenerate excited state is typically well described by a Hund's case (a) basis, so $N'$ is not a good quantum number. See \cite{HerzbergVol3, brown1971, Cerny1993, Brazier1989} for additional details on rotational selection rules and $K_R$ stacks in STMs.

The selection rules described above have been tested in experiments with CaOCH$_3$ (\cite{mitra2020direct}), where $\sim$120 photons were scattered in the $K_R = 0$ stack and $\sim$30 photons were scattered in the $K_R=1$ stack. It is empirically unknown whether the $K_R$ selection rules hold beyond this number of photons.

\begin{figure}
\centering
\includegraphics[width=0.8\textwidth]{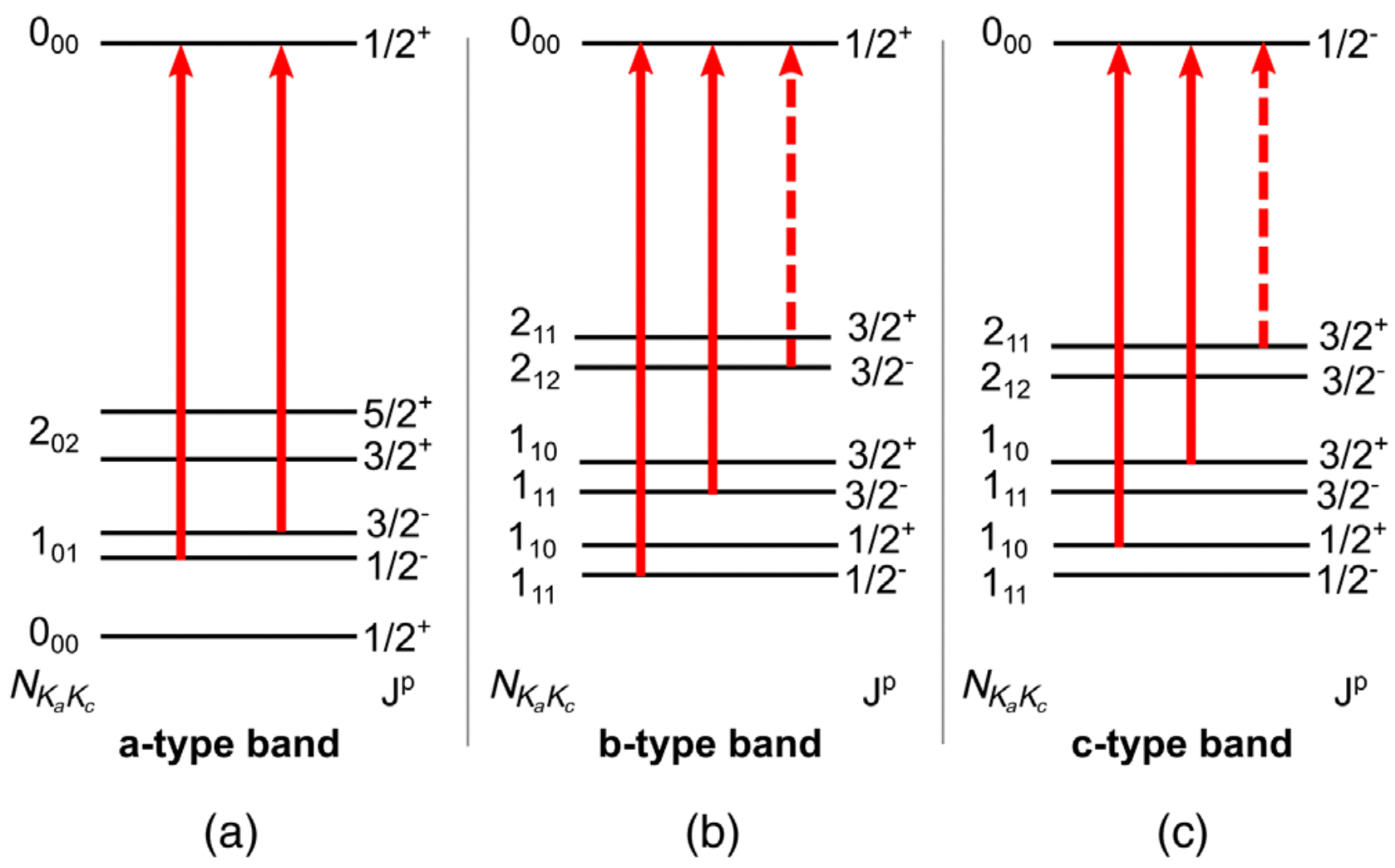}
\caption{Rotationally closed cycling transitions for $a$, $b$, and $c$-type bands in asymmetric top molecules, as described in the text. Dashed lines correspond to transitions that are allowed in ATMs with unquenched electronic angular momentum, which is expected to be true for molecular geometries near the symmetric-top limit. Reproduced from \cite{Augenbraun2020ATM}.}
\label{fig:ATMRotational}
\end{figure}

\textbf{Asymmetric top molecules.} In asymmetric top molecules (ATMs), there are three types of electronic transitions to consider, corresponding to induced dipole moments along either the $a$, $b$, or $c$ principal axis of the molecule. These are analogous to parallel and perpendicular transitions in linear molecules or STMs, where parallel transitions become $a$ ($c$)-axis transitions for prolate (oblate) molecules, and the other two axes play the role of perpendicular transitions. Each transition axis has its own selection rules, shown in Tab. \ref{tab:selectionrules}, and closed cycling transitions for each transition type are shown in Fig~\ref{fig:ATMRotational}. The transition dipole moment $\mu$ will in general have a projection onto each of the principal axes and inherit some of each of the selection rules (\cite{Augenbraun2020ATM}). It is therefore advisable to choose molecules whose transition dipole moments are well aligned with the molecular axis to limit the number of rotational states that require repumping.

\subsection{Perturbations} \label{sec:Perturbations}

\begin{figure}
    \centering
    \includegraphics[width=0.5\textwidth]{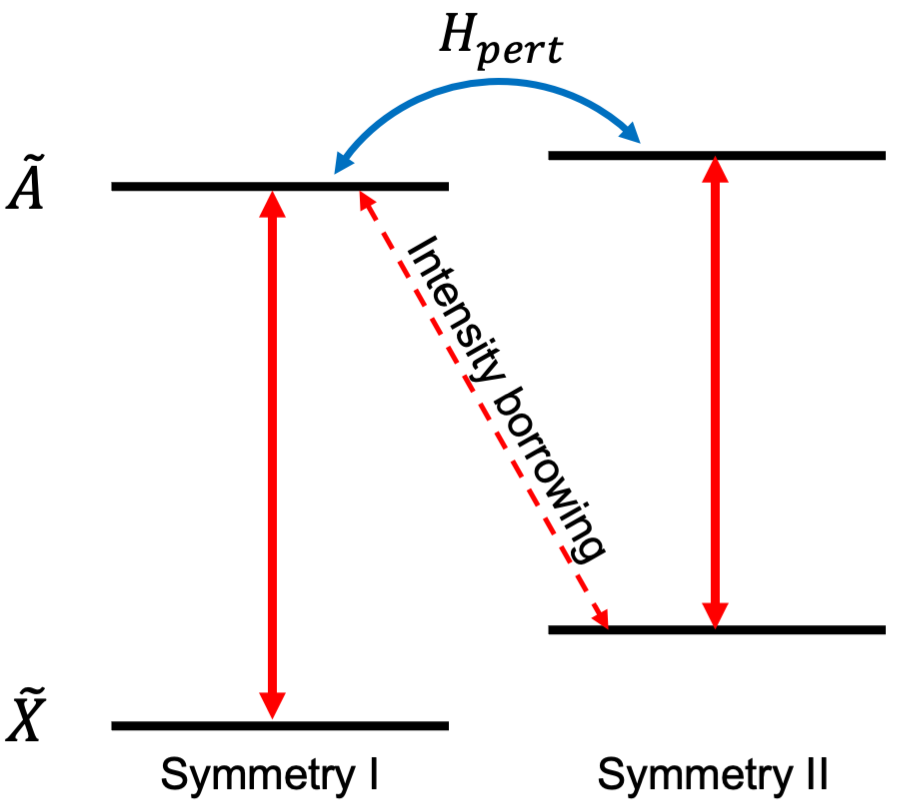}
    \caption{Schematic illustration of how perturbations among electronically excited states can lead to an ``intensity borrowing'' effect that induces nominally forbidden transitions (or increases the intensities of transitions that were expected to be very weak).}
    \label{fig:perturbations}
\end{figure}

We have thus far described the energy eigenstates of polyatomic molecules using a basis of well-defined quantum numbers. These, in turn, led to strict selection rules governing vibrational and rotational branching. In real molecules, however, there are mechanisms that perturbatively couple these basis states, e.g. via mixing of electronic and vibrational angular momentum. The result is an effect known as ``intensity borrowing'': nominally forbidden transitions become allowed because the energy eigenstates of the molecule contain a small admixture of basis states with different quantum numbers and/or symmetry. The effect is illustrated schematically in Fig.~\ref{fig:perturbations}. While these effects are typically small ($\sim10^{-3}$ level or below), they can become important when forming cycling transitions capable of scattering many thousands of photons, or when certain vibronic levels of electronically excited states have energy gaps that are ``accidentally'' small.

While perturbations can take many forms and must in general be considered on a molecule-to-molecule basis, below we will discuss a few known effects for molecules presently being laser cooled or proposed for laser cooling.

The Renner-Teller (RT) effect describes vibronic mixing between the electronic angular momentum $\Lambda$ and the vibrational angular momentum $\ell$ in linear polyatomic molecules with degenerate vibrational modes. In particular, it allows $\Lambda$ and $\ell$ to change while conserving the total spinless angular momentum projection $K = \Lambda + \ell$. The physical origin of this effect is that bending vibrations reduce the cylindrical symmetry of the molecule and can therefore break the degeneracy of the in-plane and out-of-plane electronic orbitals in states with $\Lambda > 0$. See \cite{Hirota1985} for a detailed description of this interaction.

The Renner-Teller effect is responsible for vibrational branching that violates the $\Delta \ell = 0$ selection rule in linear polyatomic molecules, as previously studied in detail for CaOH, SrOH, BaOH, and YbOH (\cite{Baum2021Establishing, Zhang2021, Lasner2022, KinseyNielsen1986}). In these molecules, there are two important types of RT-induced branching. The first is $\Delta \ell = \pm 1$ branching enabled by first-order RT coupling, which mixes states according to the selection rule $\Delta \Lambda = -\Delta \ell = \pm 1$. This enables direct vibronic coupling between the $\widetilde{A}^2\Pi(000)$ and $\widetilde{B}^2\Sigma^+(01^10)$ states in alkaline earth monohydroxides, thereby allowing $\widetilde{A}(000)$ to decay directly to $\ell = 1$ ground states (e.g. $\widetilde{X}(01^10)$) via intensity borrowing from the $\widetilde{B}(01^10)$ state. Likewise, $\widetilde{B}(000)$ can decay to $\widetilde{X}(01^10)$ via RT coupling with $\widetilde{A}(01^10)$. This coupling typically contributes at the $\sim10^{-3}$ to $10^{-4}$ range in the alkaline earth monohydroxides (\cite{Baum2021Establishing, Zhang2021, Lasner2022}). A smaller effect mixes $\widetilde{A}(000)$ and $\widetilde{A}(01^10)$ directly via contributions from both first-order RT and spin-orbit coupling, but it is not discussed further here. See \cite{Baum2021Establishing, Zhang2021} for more details.

The second effect of RT mixing is $\Delta \ell = \pm 2$ branching induced by second-order RT coupling, which mixes states according to the selection rule $\Delta \Lambda=-\Delta \ell = \pm 2$. This term directly couples $\widetilde{A}^2\Pi(000)$ to $\widetilde{A}^2\Pi(02^20)$, as it can mix the $|\Lambda = 1, \ell = 0\rangle$ and $|\Lambda = -1, \ell = 2\rangle$ components of the $\widetilde{A}$ state. Observed decays to $\widetilde{X}(02^20)$ and $\widetilde{X}(12^20)$ in alkaline earth monohydroxides are attributed to this mechanism (\cite{Baum2021Establishing, Zhang2021, Lasner2022, vilas2022magneto}).

An analogous interaction, called the (pseudo-)Jahn-Teller (JT) effect, is possible in nonlinear symmetric top molecules. A detailed review of Jahn-Teller physics is provided by \cite{Barckholtz1998}. For the purposes of laser-coolable polyatomic molecules such as CaOCH$_3$ or YbOCH$_3$, we can regard the (pseudo-)JT effect in the $^2E$ state of a nonlinear molecule as analogous to the RT effect in a $^2\Pi$ linear molecule. This effect can lead to mixing between a $^2E$ and $^2A_1$ electronic state which alters the vibrational structure of the $^2E$ state and changes the vibronic emission intensities associated with spontaneous emission from this state. In a molecule such as CaOCH$_3$ or YbOCH$_3$, the first electronically excited state ($\tilde{A}\,^2\!E$) is affected by mixing with the $\tilde{B}\,^2\!A_1$ level. Especially due to second-order spin-orbit--vibronic coupling, this interaction can dramatically increase the intensity of decays to vibrational bending modes that would have been symmetry-forbidden within the BO approximation. Experimentally, these decays have been observed in CaOCH$_3$ (\cite{AugenbraunThesis}) and in YbOCH$_3$ (\cite{Augenbraun2021Observation}). In both cases, it was possible to model the intensity of decay to vibrational bending modes on the basis of quantum chemical predictions of the JT parameters provided by \cite{Paul2019}. In both cases, the nominally symmetry-forbidden decays were significantly stronger than decays to symmetry-allowed levels of the same vibrational mode, directly indicating the role that vibronic coupling plays in this process. 

These observations point to two important considerations in the selection of laser-coolable nonlinear molecules. First, one must consider beyond-BO-approximation effects when deciding whether simple estimates of FCFs are justified in selecting a molecule for future experimentation. Second, experimental measurements of vibrational branching ratios are crucial to identify weak decays that violate expectations based on molecular symmetry. See Sec.~\ref{sec:VBRmeasurements} for details on such measurements.

\section{Experimental techniques} \label{sec:ExperimentalTechniques}

\subsection{Cryogenic buffer-gas beams} \label{sec:CBGBs}

Almost all applications of laser-cooled molecules benefit from long
interaction times, achieved either through a trap (enabling, in principle,
arbitrarily long hold times) or a molecular beam propagating over
a large distance (tens of centimeters to a few meters, enabling probe
times of $\sim$1$-$100 ms). In both cases, a large flux of initially
slow molecules is essential. In order to trap molecules, any motion
at the time of production must be removed in a manner whose difficulty
typically scales with the initial momentum or kinetic energy, whereas
for a fixed beam line the interaction time is inversely proportional
to the beam velocity.

An additional requirement for a practical molecular source is a low
rotational temperature. Rotationally excited states of small molecules
like CaF and SrOH obtain significant thermal population at temperatures
of $\sim$1 K, and the population in any single quantum state at room
temperature is suppressed by orders of magnitude compared to the low-temperature
limit. This problem grows exponentially for large asymmetric top molecules,
with multiple rotational modes that have small rotational constants
(owing to the large moments of inertia).

\begin{figure}

\begin{centering}
\includegraphics[width=7cm]{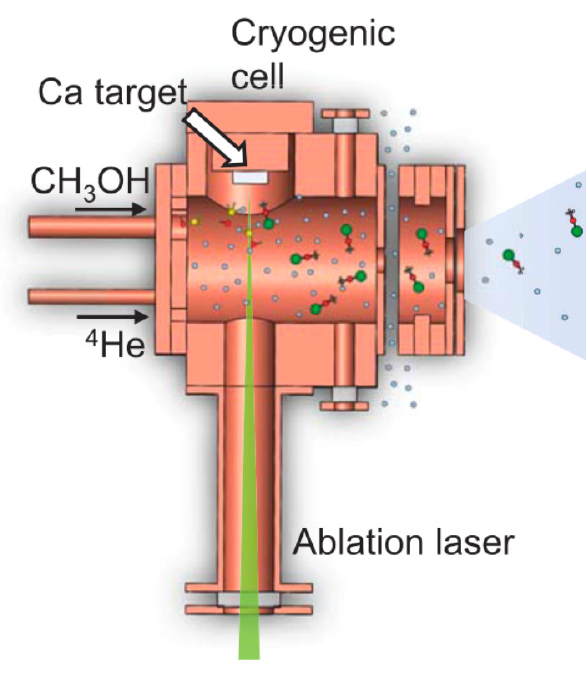}\caption{A representative cryogenic buffer gas cell, modified from \cite{mitra2020direct}. Hot reagent CH$_{3}$OH (methanol) gas is introduced via
a capillary, and cold $^{4}$He buffer gas is introduced via a second
capillary. A pulsed Nd:YAG laser ablates a ``target'' of Ca metal,
which reacts with methanol to form CaOCH$_{3}$ in the first-stage
cell. Additional windows at the downstream end of the first-stage
cell may be used to monitor molecular production via optical absorption
measurements. The helium thermalizes the molecules and entrains them
through a hole in the front of the cell. A second-stage cell, stood
off from the first stage to lower the buffer gas density, reduces
the forward velocity of the molecular beam. Buffer gas and molecules
are extracted through the front of the second-stage cell. \label{fig:buffer-gas-cell}}
\par\end{centering}
\end{figure}

Buffer-gas cooling in closed cells was developed by the atomic hydrogen community in the late 1970s/early 1980s. Early examples with molecules include the De Lucia group, which studied CO molecules (and, later, polyatomic molecules) in the presence of a $\mysim4$~K He buffer gas, as in \cite{Messer1984Measurement,Willey1988Very, Mengel2000Helium}. The Doyle group pioneered the use of buffer-gas cooling to load atoms and molecules into (superconducting) magnetic traps in \cite{Doyle1995Buffer,Weinstein1998, Campbell2007Magnetic, Doret2009Buffer}. The first molecular beam source generated from buffer-gas cells was developed in the Doyle group in collaboration with Prof. David DeMille, as described in \cite{Maxwell2005High}. Hydrodynamic cryogenic buffer-gas beams (CBGBs) were created by \cite{Patterson2007Bright} and developed by both the Doyle and DeMille groups and the Hinds/Tarbutt groups. Reviews by \cite{hutzler2012buffer,Barry2011} discuss the key features of such sources, which include low forward velocities and hydrodynamic enhancement effects that led to beams several orders of magnitude brighter than previous realizations. 

A primary benefit of a CBGB (beyond the low velocity and high flux) is that it allows the experiment to separate molecule \textit{production} (occurring in a cryogenic region with poor optical access) from molecule \textit{manipulation} (usually occurring in a room-temperature, ultra-high vacuum region). The CBGB was a key technological development that enabled direct laser cooling of molecules, as supersonic beams produce velocities so high as to make the deceleration of molecules to trappable velocities extremely difficult. 

A CBGB source typically
consists of four nested regions. The innermost region is a ``buffer
gas cell,'' usually made of high-purity copper and cooled to $\sim1-4$~K. A representative example is shown in Fig. \ref{fig:buffer-gas-cell}.
There, stable reagent molecules (methanol) are introduced into the
cell via a hot capillary, while Ca metal is ablated via a pulsed Nd:YAG
laser (typically with $\sim$10$-$40 mJ/pulse). The ablated atoms
and methanol gas react to form CaOCH$_{3}$. Simultaneously, He buffer
gas is flowed through a second capillary that pre-cools the He gas before it enters the cell. Inside the cell, the He thermalizes with the
cold walls of the cell and collisions between molecules and He atoms
cool the molecules. The flow of helium entrains molecules through
a hole in the front of the cell (typically $\sim$$3$-$7$ mm diameter).
The helium and molecules then flow into a lower-density ``second-stage
cell'' that is stood off from the first stage of the buffer gas cell
with a small gap for buffer gas to escape. This lowers the buffer
gas density in the second-stage cell and reduces the forward velocity
of molecules, at the expense of some reduction in the molecular
extraction (typically a factor of order unity). Molecules then emerge from the second cell with forward
velocities in the range of $\sim$40$-$200 m/s, depending on the
molecular mass, cell temperature, and buffer gas flow rate.

A number of variations on the buffer gas cell are used, depending
on experimental requirements and the molecular species of interest.
To produce cold beams of stable molecules such as ammonia or formaldehyde, no ablation
is necessary (\cite{VanBuuren2009}). On the other hand, even complex
radical molecules can be produced directly from ablation of pressed-powder
pellets (or ``targets'') of stable constituents, without reagent
gases. For example, a mixture of SrH$_{2}$ and naphthol powders produces
SrO-naphthyl radicals (\cite{Mitra2022pathway}). Because of the versatility
in molecular production methods, a wide variety of polyatomic molecules including both radical (\cite{AugenbraunYbOHSisyphus,Baum2020,mitra2020direct,Zhu2021,Mitra2022pathway} and non-radical~\cite{Maxwell2005,VanBuuren2009,Herschbach2009,Sawyer2011,Eibenberger2017,Spaun2016,Satterthwaite2019,patterson2013sensitive,Piskorski2014}) species can be created
in a CBGB. Molecules as large as nile red (C$_{20}$H$_{18}$N$_{2}$O$_{2}$)
have been produced and spectroscopically studied in a cryogenic buffer
gas cell without forming helium-molecule clusters (\cite{piskorski2014cooling}).

Both helium and neon buffer gas are commonly used in CBGBs, with
neon requiring cell temperatures above approximately 16 K to maintain
a suitable vapor pressure. Depending on the requirements for molecular
flux and the initial beam velocity, the second-stage cell may be omitted.
The lowest forward velocities can be achieved by cooling the second
stage with a He-3 pot at temperatures around or below 1 K (\cite{Augenbraun2021zeeman}). In that work, the heat loads arising from
ablation and gas flows, in the range of tens to hundreds of mW, made it impractical to
cool the first-stage cell using a He-3 refrigerator. Nevertheless, by cooling the second-stage alone, beams of Yb (Ca) with peak forward velocities as low as about 20~m/s (40~m/s) have been observed by \cite{AugenbraunThesis, Augenbraun2021zeeman, Sawaoka2022}.

Surrounding the buffer gas cell is a cold box, usually constructed
of high-purity copper and held at $\sim$4 K by a pulse tube cryocooler
to serve as a cryopump. When helium buffer gas is used, charcoal sorbs
are thermally anchored to the 4 K box in order to achieve adequate
helium (cryo)pumping speeds; other buffer gasses, such as neon, are
efficiently cryopumped directly using copper surfaces held at 4 K. 

The cryopumping box is contained within another box, typically made
of aluminum or copper, held at $\sim$50 K by the first stage of a
pulse tube cryocooler. This box shields the cryopumping box and buffer
gas cell from room-temperature black body radiation. The radiation
shields are housed within a vacuum chamber at $\sim$$10^{-7}$ Torr
or better.

We summarize several important molecular beam parameters under typical
conditions in Tab.~\ref{tab:CBGB-parameters}. Experiments seeking
to trap molecules usually operate toward the lower range of buffer
gas flow rates, ablation energy, and ablation repetition rate, to
enable lower temperatures and forward molecular beam velocities at
the expense of beam brightness and experimental duty cycle.

\begin{table}
\centering
\begin{tabular}{|c|c|}
\hline 
Parameter & Typical range\tabularnewline
\hline 
\hline 
Forward velocity & 40$-$200 m/s\tabularnewline
\hline 
Solid angle FWHM & 0.2$-$1 sr\tabularnewline
\hline 
Rotational temperature & 1$-$4 K\tabularnewline
\hline 
Brightness & $10^{8}$-$10^{11}$ sr$^{-1}$ pulse$^{-1}$\tabularnewline
\hline 
Ablation repetition rate & 1$-$50 Hz\tabularnewline
\hline 
Ablation energy & 10$-$40 mJ\tabularnewline
\hline 
Buffer gas flow rates & 2$-$40 sccm\tabularnewline
\hline 
\end{tabular}\caption{Representative operating and performance parameters of molecular CBGBs.
\label{tab:CBGB-parameters}}

\end{table}

\subsection{Optical cycling} \label{sec:OpticalCycling}

\begin{figure}

\centering{}\includegraphics[width=7cm]{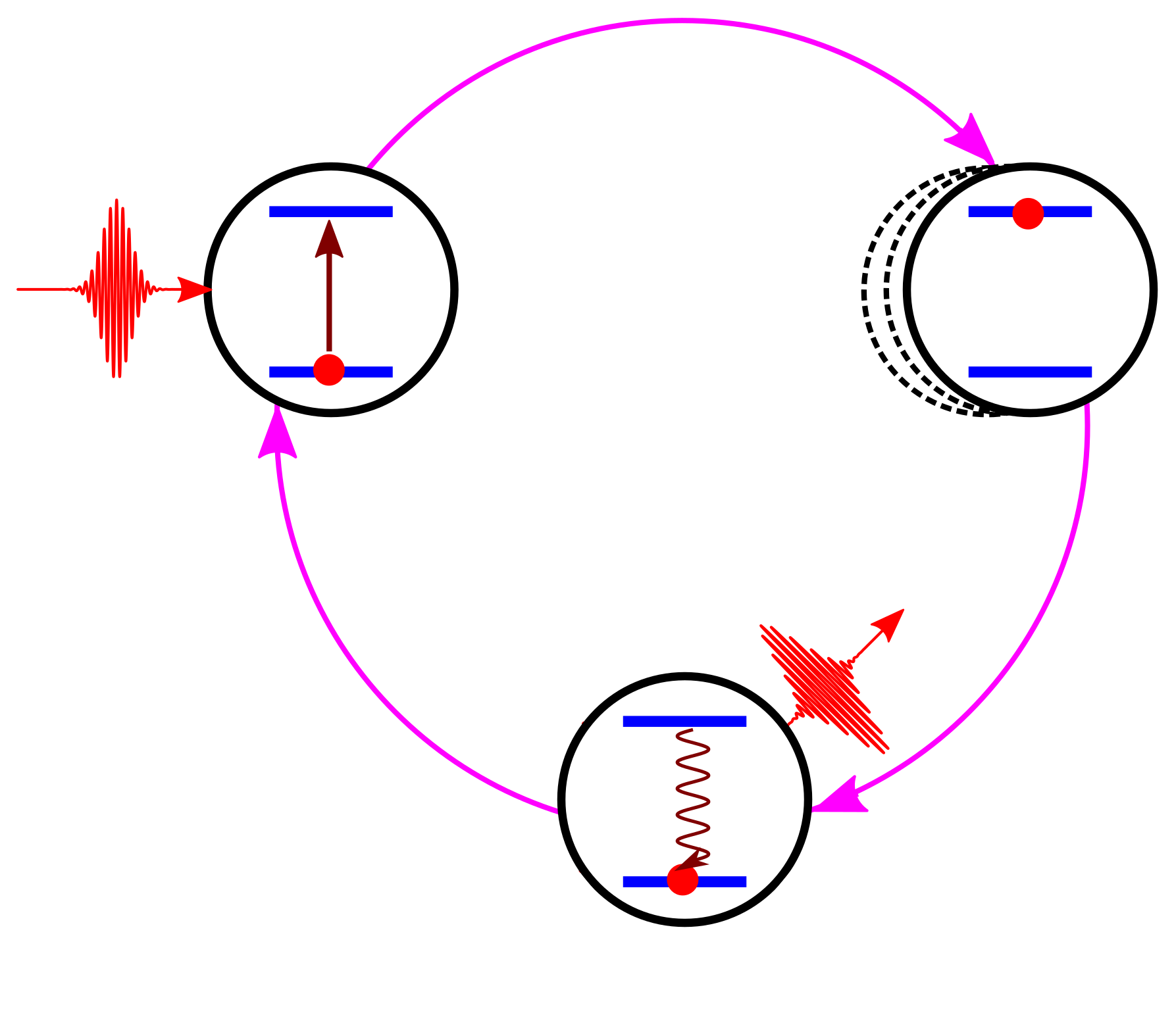}\caption{Optical cycle for an ideal two-level system, reproduced from~\cite{Augenbraun2020ATM}. Three phenomena repeat up to tens of
thousands of times: (1) A photon is absorbed, driving a molecule
from the ground to excited state and imparting a momentum recoil to
the molecule, (2) subsequently, a photon is spontaneously emitted
in a random direction, and (3) the molecule returns to its ground state.
\label{fig:generic-optical-cycle}}
\end{figure}

The key ingredient to laser cooling is the near-closure of an optical
cycle, so as to approximate an ideal two-level system, as depicted in Fig. \ref{fig:generic-optical-cycle}.
A molecule initially in its ground state absorbs a photon with
energy $\hbar\omega$, which excites the molecule to a higher quantum
level and imparts a momentum recoil $p_{{\rm recoil}}=\hbar k$. The
molecular state subsequently spontaneously decays, ideally back to the ground
state, emitting a photon in a random direction. By using optical cycling, the molecule's external motion can be controlled through various cooling schemes such as Doppler or polarization gradient cooling, $\Lambda$-enhanced grey molasses, etc. In real molecules,
an optical cycle consists of many ``ground'' states (which are generally in the ground electronic state, but excited vibrationally or rotationally) and many excited states (which are electronically, vibrationally, and/or rotationally excited).
Thus, any chosen pair of ground and excited states will fail
to form a closed optical cycle. 

However, by careful selection of a
group of ground states and excited states (manifolds), a nearly-closed optical cycle
can be formed. For example, in molecules like CaOH and SrOH, parity and angular momentum selection
rules ensure that the excited $J^{P}=1/2^{+}$ states in the lowest-lying
excited electronic state $\tilde{A}\,^{2}\Pi_{1/2}$ can \emph{only} decay to
the $N^{P}=1^{-}$ rotational manifolds in the ground electronic state
$\tilde{X}\,^{2}\Sigma^{+}$. Each $J^{P}=1/2^{+}$ state contains four optically
unresolved hyperfine levels, while each $N^{P}=1^{-}$ manifold contains
a $J=1/2$ manifold and a $J=3/2$ manifold, which are split from
each other by spin-rotation splittings of $\sim$10--100 MHz that are
easily spanned with optical frequency modulation techniques. These
$J=1/2$ and $J=3/2$ manifolds contain 4 and 8 hyperfine states,
respectively. This nearly-closed molecular cycling
transition contains 12 ground and 4 excited states. A scheme to achieve a rotationally and vibrationally closed optical cycle (and eventually laser cooling) of a polyatomic molecule was first discussed by \cite{kozyryev2015collisional, kozyryev2016radiation} using the example of SrOH.

For any practical
application of optical cycling or laser cooling, it is necessary to
add repumping lasers to also address excited vibrational
states of the ground electronic manifold, as when the molecule decays from the electronically excited state there are no selection rules on the vibrational quantum number. How many vibrational repumpers are required depends on the vibrational branching ratios
(VBRs) and on the optical cycling scheme chosen, see Sec.~\ref{sec:design-optical-cycle}. The exact optical cycling scheme required
for a particular application depends on the details of the molecular structure. For example,
the optical cycle used to laser cool the symmetric top molecule CaOCH$_{3}$,
(with $C_{3v}$ symmetry) is depicted in \cite{mitra2020direct}. Similarly, an overview
of the optical cycles required to achieve optical cycling with respect to the rotational structure in asymmetric
tops are shown in \cite{Augenbraun2020ATM}.

\subsection{Optical forces}

\begin{figure*}

\centering{}\includegraphics[width=0.95\columnwidth]{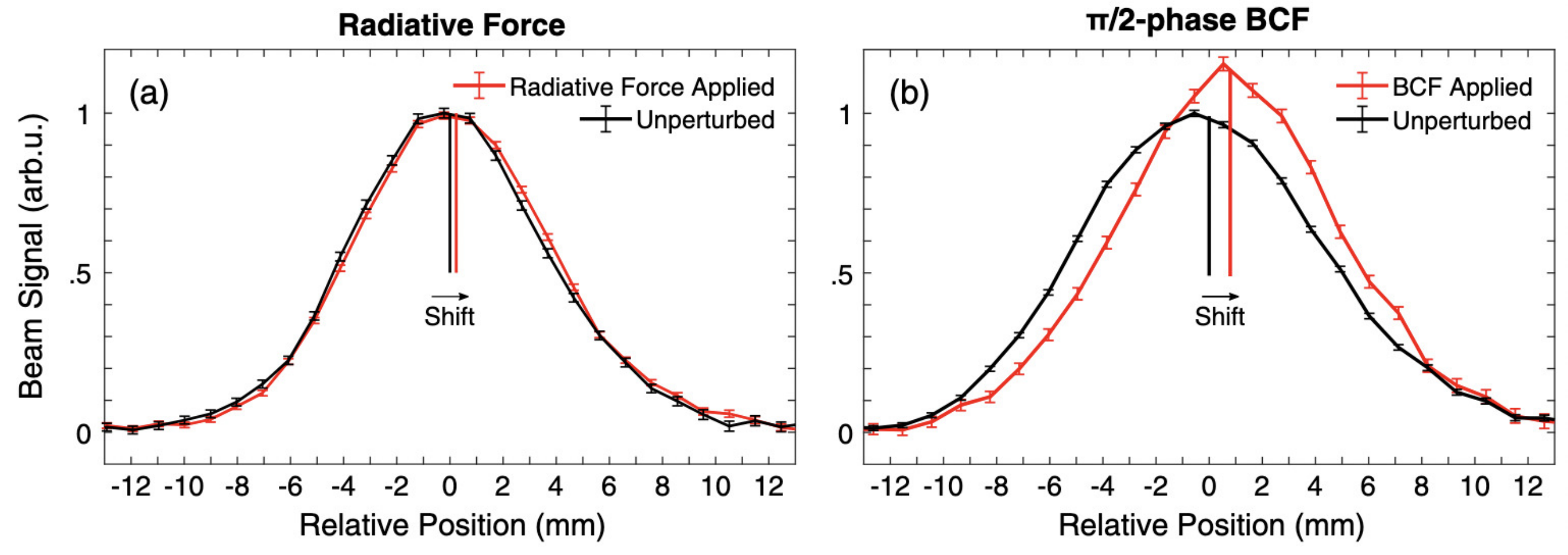}\caption{Comparison of molecular beam deflection via radiation pressure force
(left) and bichromatic force (right) in SrOH, reproduced from~\cite{kozyryev2017BCF}. Under these conditions, the bichromatic
force is greater by a factor of 3.7, as seen by the shift in the position
of the molecular beam.\label{fig:optical-forces}}
\end{figure*}

The radiative force on a molecule is determined by the momentum recoil,
$p_{{\rm recoil}}$, and the scattering rate $\gamma$, as $F_{{\rm recoil}}=p_{{\rm recoil}}\gamma$.
Thus to achieve large optical forces, it is important to maximize
the scattering rate by judicious selection of optical cycling transitions
and saturation of all laser powers. The first demonstration of the radiation
pressure force on a polyatomic molecule was performed
with SrOH, with a single vibrational repumper, scattering $\sim$100
photons per molecule in a single direction orthogonal to the molecular
beam propagation axis. In this experiment, \cite{kozyryev2016radiation} observed a resultant 0.2$^{\circ}$ deflection of
the molecular beam. This work served
as an initial proof of principle that polyatomic molecules could experience
significant optical forces in a practical experimental configuration, and thus that direct laser cooling of polyatomic molecules should be possible.
Extending optical cycling to $\sim$$10^{4}$ photon scattering events would allow for radiation pressure slowing of molecular beams and capture into a magneto-optical trap.

The magnitude of the radiation pressure force is determined by the spontaneous decay rate,
$\Gamma$, of the excited state. For a saturated two-level system,
the force is $F_{{\rm recoil}}=\hbar k\Gamma/2$. Larger optical forces
can be applied using a coherent process, sometimes in combination with magnetic or electric field interactions, enabling many units $p_{{\rm recoil}}$
of momentum transfer per photon scattered. This approach was demonstrated
for SrOH using the bichromatic force, in which two phase-locked laser beams
(at different frequencies) pass through the molecular beam transversely
and are retroreflected. Due to the presence of two laser frequencies,
beat notes are formed and by tuning the distance between the molecular
beam and retroreflecting mirror, the relative phase of the counterpropagating
beats can be controlled. The resulting light field induces alternating cycles of stimulated
absorption of a photon travelling in one direction, and stimulated
emission of a photon travelling in the opposite direction. For a fixed
detuning $\delta$ between frequency components, the maximum force is set by $F_{{\rm BCF}}=\hbar k\delta/\pi$, which can vastly exceed $F_{{\rm recoil}}$
at large $\delta$. Bichromatic force deflection was demonstrated
using SrOH by \cite{kozyryev2017BCF}. Calculations show
that by employing four laser frequencies in a four-level system, which
consists of two coupled two-level subsystems in bichromatic force
configurations, large optical molasses forces should also be possible, as simulated by \cite{Wenz2020large}.

\subsection{Transverse cooling}

\begin{figure}

\begin{centering}
\includegraphics[width=7cm]{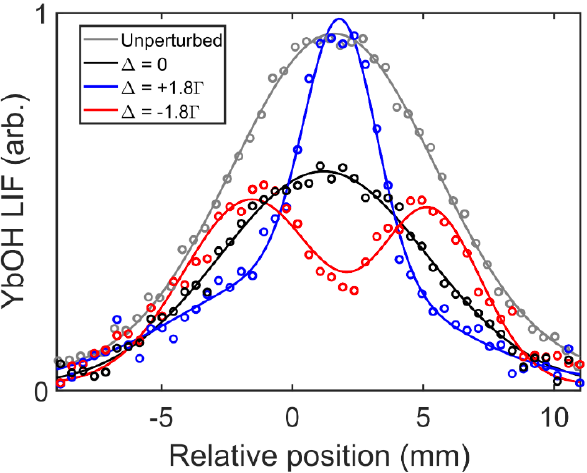}\caption{Spatial distribution of a YbOH molecular beam along the transverse
direction, in the presence of Sisyphus heating ($\Delta=-1.8\Gamma$)
and cooling $(\Delta=+1.8\Gamma$). In the cooling configuration,
the on-axis flux is increased compared to an unperturbed beam. The
double-lobed structure in the heating configuration arises from a
balancing effect between Sisyphus heating and conventional Doppler
cooling. Reproduced from~\cite{AugenbraunYbOHSisyphus}.\label{fig:sisyphus}}
\par\end{centering}
\end{figure}

A more efficient method of imparting forces to polyatomic molecules,
for a fixed number of scattered photons, is Sisyphus cooling. Specifically,
magnetically-assisted transverse 1D Sisyphus cooling has been demonstrated
for SrOH, YbOH, and CaOCH$_{3}$ (\cite{kozyryev2016Sisyphus,AugenbraunYbOHSisyphus,mitra2020direct}). In all cases,
the experimental configuration was very similar to that shown in Fig.
\ref{fig:pump-repump}, except that the lasers in the interaction
region were partially overlapped with their retroreflections to form
a standing wave. The bright molecular states (i.e., those addressed
by the laser) are AC Stark shifted at the antinodes of the standing
wave, but not at the nodes where the laser intensity vanishes. As
a molecule in a bright state traverses the standing wave along the
direction orthogonal to the primarily molecular beam direction, it
gains or loses kinetic energy, depending on whether the laser detuning
is red- or blue-detuned, respectively. To transversely cool a molecular beam, therefore,
blue-detuned light is used. A molecule is preferentially likely to
be optically pumped at higher laser intensities, and has a probability
of order unity to populate a dark state, whose energy is unaffected
by the laser light. As the molecule then traverses the region of
the standing wave node, the bright and dark states come to near degeneracy
and are mixed by magnetic fields, converting the dark state to a bright
state with non-zero probability. The molecule is then free to ``ride
up'' the potential hill again (in the blue-detuned configuration),
losing more energy. The energy loss per photon scatter is limited
only by the depth of the AC Stark shift, which is in turn limited
by available laser power. At high laser intensities, therefore, Sisyphus
cooling can be far more efficient per photon scatter at removing energy
than conventional Doppler cooling. 

The effect of Sisyphus cooling (or heating) can be observed by the
reduced (or increased) thermal expansion of the molecular beam along
the standing wave axis, as it propagates far downstream. An example
of the effect of Sisyphus cooling on YbOH molecules, reported by \cite{AugenbraunYbOHSisyphus}, is shown in Fig.
\ref{fig:sisyphus}. 
The authors also demonstrated Doppler cooling of the YbOH molecular beam, although the Sisyphus laser cooling produced a colder sample and was more efficient at cooling molecules on a per-photon basis. 
In future experiments, Doppler or sub-Doppler cooling could be used to increase the flux in molecular beam experiments or the loading efficiency into magneto-optical traps, as advocated by \cite{Alauze2021}.

\subsection{Molecular deceleration} \label{sec:Deceleration}
Common trapping techniques used in laser cooling experiments generally have capture velocities that are considerably lower than the forward velocities of molecular beams, even those produced in CBGBs. Molecular MOTs, in particular, can typically only trap molecules with speeds below $\sim$$15~\text{m/s}$ (\cite{Tarbutt2015b, Williams2017, langin2022improved}), while the forward velocities of molecular beams produced by CBGBs fall in the range of $\sim$$50~\text{m/s}$ to several $100~\text{m/s}$ (\cite{hutzler2012buffer}). Under certain optimized conditions, CBGBs with output velocities as low as 30~m/s--50~m/s have been observed for species such as CaF or CaOH; see \cite{Augenbraun2021zeeman,Lu2014}. For heavier species, like YbOH, peak beam velocities as low as 20~m/s can be achieved, as shown in \cite{AugenbraunThesis}. The molecules must be slowed as they travel from the beam source to the trap to a velocity at or below the capture velocity of the chosen trapping method. Significant work has been devoted to the slowing of atomic beams since the initial demonstration of slowing a beam of sodium atoms in 1981 by \cite{phillips1982}. These efforts have helped guide recent efforts in slowing of molecular beams. 

Slowing of atoms is often based on the radiation pressure forces due to a laser beam counterpropagating to the flow of atoms in an atomic beam. This laser addresses a closed electronic transition to continuously scatter photons. Importantly, as the atomic beam is slowed, the transition frequency of the cycling transition is shifted due to a gradually changing Doppler shift. This Doppler shift must be compensated during slowing, and three principal methods to accomplish this have been demonstrated with atoms. Two of these techniques, white-light slowing (\cite{zhu1991}) and chirped slowing (\cite{ertmer1985}) have been successfully extended to diatomic molecules (\cite{Barry2012, Zhelyazkova2014,Hemmerling2016, Yeo2015, Truppe2017b}). In white-light slowing, the slowing light is frequency-broadened to create ``white light'' that covers the full range of the Doppler shift, whereas in chirped slowing, the frequency of the slowing light is narrow-band but rapidly shifted, or ``chirped'' to match the changing Doppler shift as the particles decelerate. Zeeman slowing (\cite{phillips1982}) is one of the most efficient and popular techniques used for atoms. This method uses the Zeeman shifts induced by a spatially varying magnetic field to compensate for the changing Doppler shift, keeping the laser light resonant with atoms as they are slowed. The type of transitions that are typically used for laser cooling of molecules, namely transitions in which $J' \geq J$, makes it challenging to adapt Zeeman slowing to molecules, although this approach is potentially viable and being pursued by \cite{Petzold2018b,Petzold2018a}. The review paper by \cite{Fitch2021LaserCooled} provides a comprehensive overview of laser deceleration of diatomic molecules.

\subsubsection{Radiative slowing} \label{sec:CaOHWhiteLight}
Due to the presence of many rotational and vibrational degrees of freedom, beams of polyatomic molecules are generally more difficult to decelerate than are atoms or diatomic molecules. Radiative slowing requires scattering of thousands of photons per atom/molecule, which is difficult to achieve for polyatomic molecules due to vibrational branching, as described previously. While sufficient optical cycling for radiative slowing has been achieved for diatomic molecules by repumping just 1 or 2 vibrational stretching modes, a much larger number of vibrational modes must be addressed for polyatomic molecules. By choosing to work with polyatomic molecules that have favorable VBRs, it is possible to use a reasonable number of laser wavelengths to scatter sufficiently many photons to achieve radiative slowing of a molecular beam, e.g., to the capture velocity of a magneto-optical trap. White-light slowing was demonstrated for CaOH molecules, as reported by \cite{vilas2022magneto} and reviewed in detail below.

In order to achieve radiative slowing of any species, a photon cycling scheme capable of scattering the required number of photons must first be established. Assuming a molecular mass of $m$, an initial beam velocity of $v_\text{beam}$, and wavenumber $k$ for the scattered photons, the number of photons needed for slowing is of order $n_\text{slowing} \sim v_\text{beam} / v_\text{recoil} = m v_\text{beam} / \hbar k$. For smaller polyatomic molecules, typical values of $n_\text{slowing}$ are of order $\sim$$10^4$. Radiative slowing of a polyatomic molecule therefore will typically require repumping enough rovibrational decays to limit branching to dark states to the $\sim$$10^{-4}$ level. Considering CaOH and using the values $m_\text{CaOH} = 57~\text{amu}$, $v_\text{beam} = 140~\text{m/s}$, and $k = 2\pi / (626~\text{nm})$, the number of photons required for slowing is $n_\text{slowing} \sim 12{,}500$. The electronic transition chosen for cycling in CaOH is the $\widetilde{A}^2\Pi_{1/2}(000)(J'=1/2,p'=+) \leftarrow \widetilde{X}^2\Sigma^+(000)(N=1,p=-)$ transition, which is both rotationally closed and has favorable vibrational branching ratios: by repumping spontaneous decay to 11 rovibrational states, an average of $\sim$$12{,}000$ photons are scattered per molecule before a $1/e$ fraction of the molecules has decayed to unaddressed dark states. In other words, about a $1/e$ fraction of the molecules in the molecular beam will, in principle, scatter enough photons to be slowed to zero velocity. A diagram for the corresponding optical cycling scheme is shown in Fig.~\ref{fig:optical-cycles}. In the slowing of CaOH molecules, all 11 transitions were addressed by separate repumping laser beams that were overlapped and coaligned with the main slowing laser beam counterpropagating to the molecular beam. This required combining a total of 12 lasers of different wavelengths varying from $566~\text{nm}$ to $651~\text{nm}$ into a single beam, which was accomplished using a series of dichroic beamsplitters. The overlapped beams were passed through an electro-optic modulator (EOM) that produced a frequency-broadened spectrum able to address all velocity classes between about $0$~m/s and $\sim$$140~\text{m/s}$ (the initial velocity of the molecular beam).

The primary slowing force came from driving the $\widetilde{A}^2\Pi_{1/2}(J'=1/2,p'=+) \leftarrow \widetilde{X}^2\Sigma^+(N=1,p=-)$ transition. As will be discussed in Sec.~\ref{sec:MOT}, this is a type-II transition, characterized by the existence of dark states in the ground state manifold. For any chosen polarization of the slowing light, there exist dark states into which a molecule may be optically pumped; after such an optical pumping event, that molecule will stop scattering photons from the slowing light. These dark states must be destabilized for continued photon scattering during slowing. In the experiment described by \cite{vilas2022magneto}, this was achieved by switching the polarization of the slowing light between two orthogonal polarizations. Crucially, the dark states of the two polarizations are distinct, so continued photon scattering can be achieved. The photon scattering rate achieved for CaOH molecules in the deceleration scheme was observed to be around $1{-}2~\text{MHz}$, similar to that achieved for diatomic molecules. 

\begin{figure}[tb]
    \centering 
    \includegraphics[width=1\columnwidth]{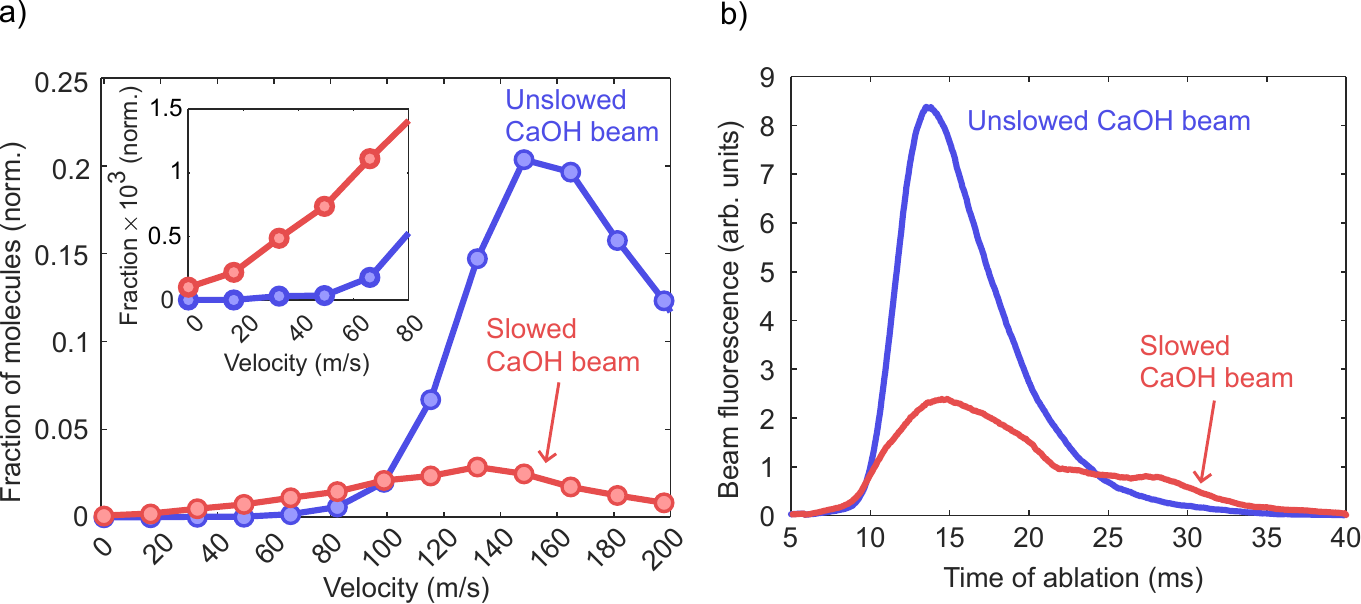}
    \caption{(a) Fraction of molecules detected as a function of velocity for both an unslowed (blue) and slowed (red) beam of CaOH molecules. Populations are detected by a Doppler-sensitive fluorescence signal. Inset: Accumulation of population at low velocities when slowing light is applied. (b) Time-resolved laser-induced fluorescence from unslowed (blue) and slowed (red) beams of CaOH molecules.}
    \label{fig:Fig1_Slowing}
\end{figure}

Figure~\ref{fig:Fig1_Slowing}(a) shows a characteristic velocity profile for an unslowed beam of CaOH molecules, as well as the velocity profile of the molecular beam following slowing. While the population of molecules with initial velocities below $50~\text{m/s}$ is negligible, the slowed distribution shows significant accumulation of molecules at velocities as low as $10{-}20~\text{m/s}$. Time-resolved measurements of the molecular beam with and without slowing light applied are shown in Fig.~\ref{fig:Fig1_Slowing}(b). Here, slowing of the molecular beam is evident from the late arrival of a significant portion of the slowed beam, as compared to the unslowed beam. Finally, we note that, while a large number of different laser frequencies was used to address the various vibrational repumping transitions, the total slowing power ($\sim$$2.5~\text{W}$) is similar to that used for diatomic molecules. This is because most of the repumping laser beams contained relatively low power ($10{-}100~\text{mW}$).

\subsubsection{Zeeman-Sisyphus deceleration} 
Despite the success of radiative slowing for CaOH molecules, the large number of photons that must be scattered off of each molecule to be slowed makes the method difficult to implement in general. Larger molecules, or molecules with more complex level structure, may have insufficiently diagonal FCFs to support such a photon budget. For this reason, it is desirable to develop other methods of molecular beam deceleration that do not depend on the radiation pressure force. The alkaline-earth-containing polyatomic molecules that we have focused on are polar radicals, implying that interactions with either electric or magnetic fields can lead to energy-level shifts that could be used to manipulate molecular motion. 

One example of such a method that has been experimentally demonstrated for laser-coolable polyatomic molecules is Zeeman-Sisyphus (ZS) deceleration. In a ZS decelerator, originally proposed by \cite{comparat2014} and expanded on by \cite{Fitch2016}, one leverages the large energy shifts induced by Tesla-scale magnetic fields. The principle of the ZS decelerator is depicted in Fig.~\ref{fig:ZSCartoon}. The deceleration scheme is highly reminiscent of the process used by \cite{Lu2014} to slow and load CaF molecules into a magnetic trap. In brief, molecules in a weak-field-seeking (WFS) state are incident on a region of increasing magnetic field magnitude and decelerate as they climb the potential. Near the magnetic field maximum, the molecules are optically pumped through an electronically excited state to a strong-field-seeking (SFS) state and continue to decelerate as they exit the high-field region. In this way, an energy $\DEstage \approx 2 \muB \Bmax$ can be removed from molecules passing through each deceleration stage, where \muB is the Bohr magneton and $\Bmax$ is the maximum magnetic field in the high-field region. This process can be repeated to remove additional energy. Furthermore, the deceleration applies to all molecules regardless of their arrival time, and thus is effective for continuous (or long-pulsed) molecular beams. Because a fixed \textit{energy} is removed in each stage, the decelerator will slow to rest $1\muB$ molecules of any mass produced at or below the same threshold temperature.

\begin{figure*}[tb]
    \centering 
    \includegraphics[width=0.55\textwidth]{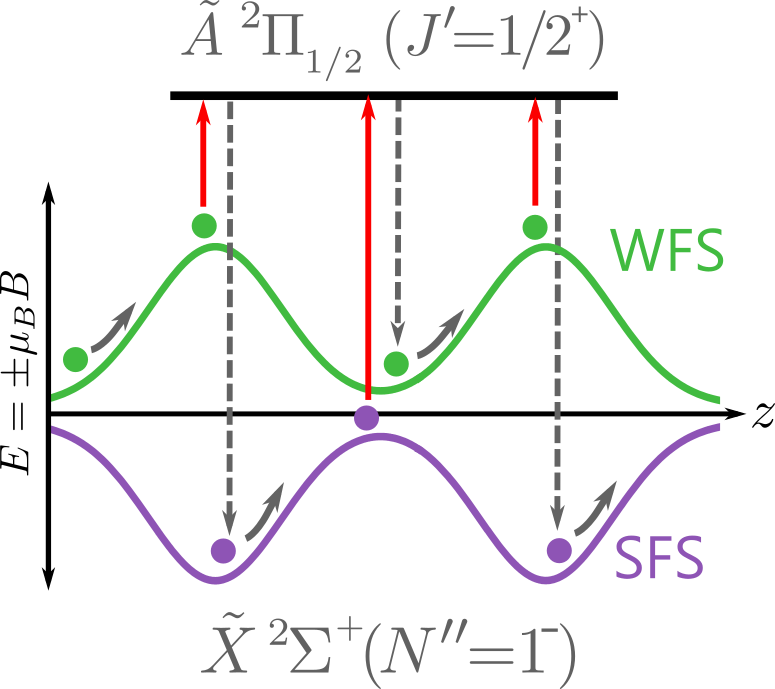}
    \caption[Schematic overview of Zeeman-Sisyphus deceleration.]{Overview of the Zeeman-Sisyphus deceleration scheme. Molecules enter the magnetic field region in a weak-field-seeking state and decelerate as they travel toward the peak magnetic field. At the peak magnetic field, molecules are optically pumped to a strong-field-seeking (SFS) state and continue to decelerate. Near the field minimum, molecules are pumped back to the weak-field-seeking (WFS) state and the process can be repeated for additional deceleration.}
    \label{fig:ZSCartoon}
\end{figure*}

ZS deceleration was first tested experimentally using CaOH molecules by \cite{Augenbraun2021zeeman}. The experimental setup comprised a set of superconducting coils, shown in Fig.~\ref{fig:ZSRender}. Our particular decelerator comprises two magnets with $\Bmax \approx 2.8$~T, leading to $\DEstage \approx 3.8$~K. The total energy removal for two stages ($\sim$7.6~K) is therefore well matched to CBGBs, which can have typical kinetic energies $E_\text{kin} \lesssim 8$~K. There is some overhead associated with superconducting coils (principally the use of a cryocooler). Nonetheless, the use of superconducting coils leads to a number of technical advantages as compared to permanent magnet designs. First and foremost, stronger magnetic fields can be achieved over a larger bore. The superconducting design is necessary to achieve peak fields $\Bmax \lesssim 4$~T over bores a few cm in diameter. That is, a superconducting coil can simultaneously enable greater deceleration per stage and larger spatial acceptances. Second, the cryogenic apparatus required to support the superconducting coils naturally leads to excellent vacuum due to high-speed cryopumping. Third, superconducting coils can easily be designed with transverse optical access in order to drive laser transitions only at particular positions along the solenoids. Performing optical pumping in a spatially selective way eliminates concerns about ``accidental" resonances along the slowing path (which could lead to population loss) and minimizes Zeeman broadening (which reduces the laser power requirements).

\begin{figure*}[tb]
    \centering 
    \includegraphics[width=0.95\textwidth]{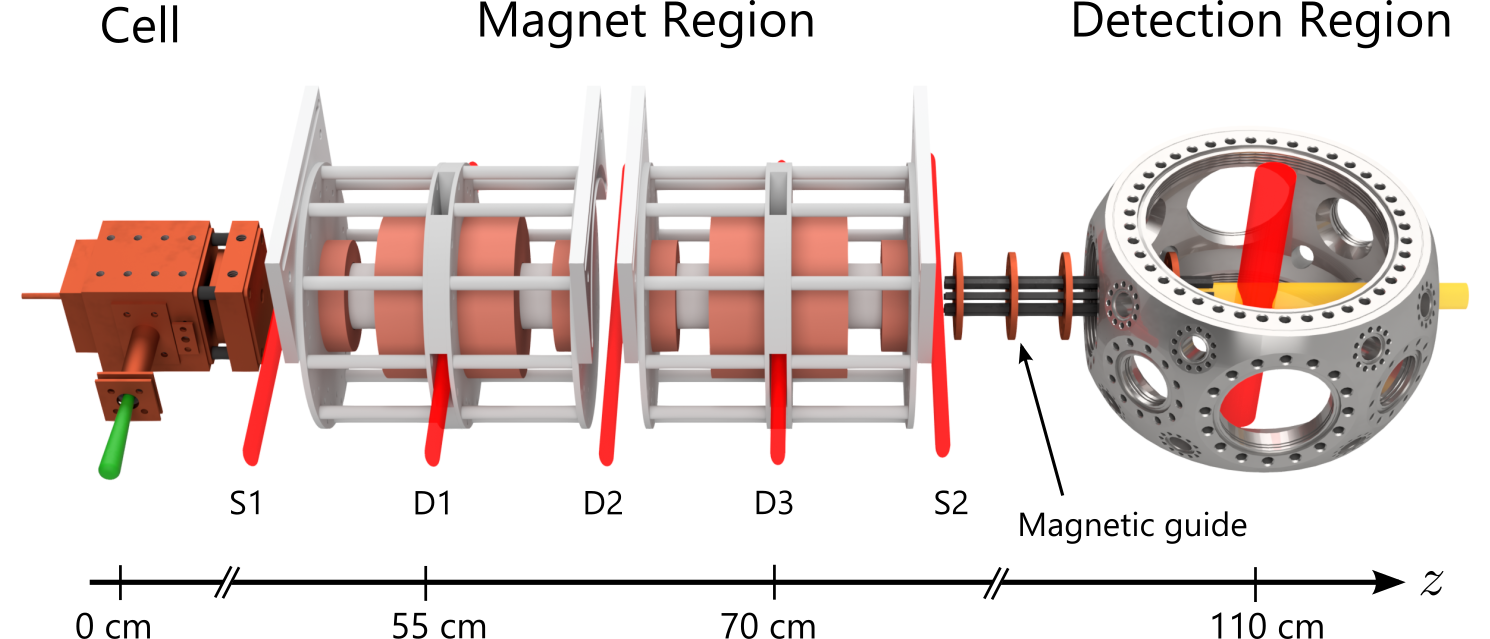}
    \caption[Rendering of decelerator beamline.]{Schematic of the Zeeman-Sisyphus decelerator (not to scale). Molecules are produced in a two-stage cryogenic buffer-gas beam. They travel through two superconducting magnets in Helmholtz configuration and are optically pumped in three deceleration pumping regions (D1, D2, D3) by transverse laser beams at 626~nm. State-preparation regions (S1 and S2) pump molecules into WFS states in order to populate magnetically guidable states. Molecules are detected via laser-induced fluorescence.}
    \label{fig:ZSRender}
\end{figure*}

The results of ZS deceleration of CaOH under optimal optical pumping performance in all pumping regions is shown in Fig.~\ref{fig:CaOHZSWithPR0} (\cite{Augenbraun2021zeeman}). Compared to the unperturbed molecular beam, when the optical pumping light is applied the fraction of slow molecules rises to $24(3)\%$ below 20~m/s and $3.5(5)\%$ below 10~m/s. The fraction of slow molecules is therefore enhanced by at least two orders of magnitude following deceleration. Based on the calibration of the fluorescence collection and the estimated number of molecules in the unperturbed beam, this means that approximately $3\times10^4$ molecules per pulse are found in velocity classes capturable by traps (e.g., MOT or magnetic). The solid lines in Fig.~\ref{fig:CaOHZSWithPR0} are the results of Monte Carlo simulations that take as input experimentally measured laser parameters and accurate, three-dimensional magnetic field profiles for both the superconducting coils and the magnetic guide. We find excellent agreement between the simulations and experimental results, indicating that the details of the slowing process are modeled accurately. 

\begin{figure}
    \centering
    \includegraphics[width=0.68\linewidth]{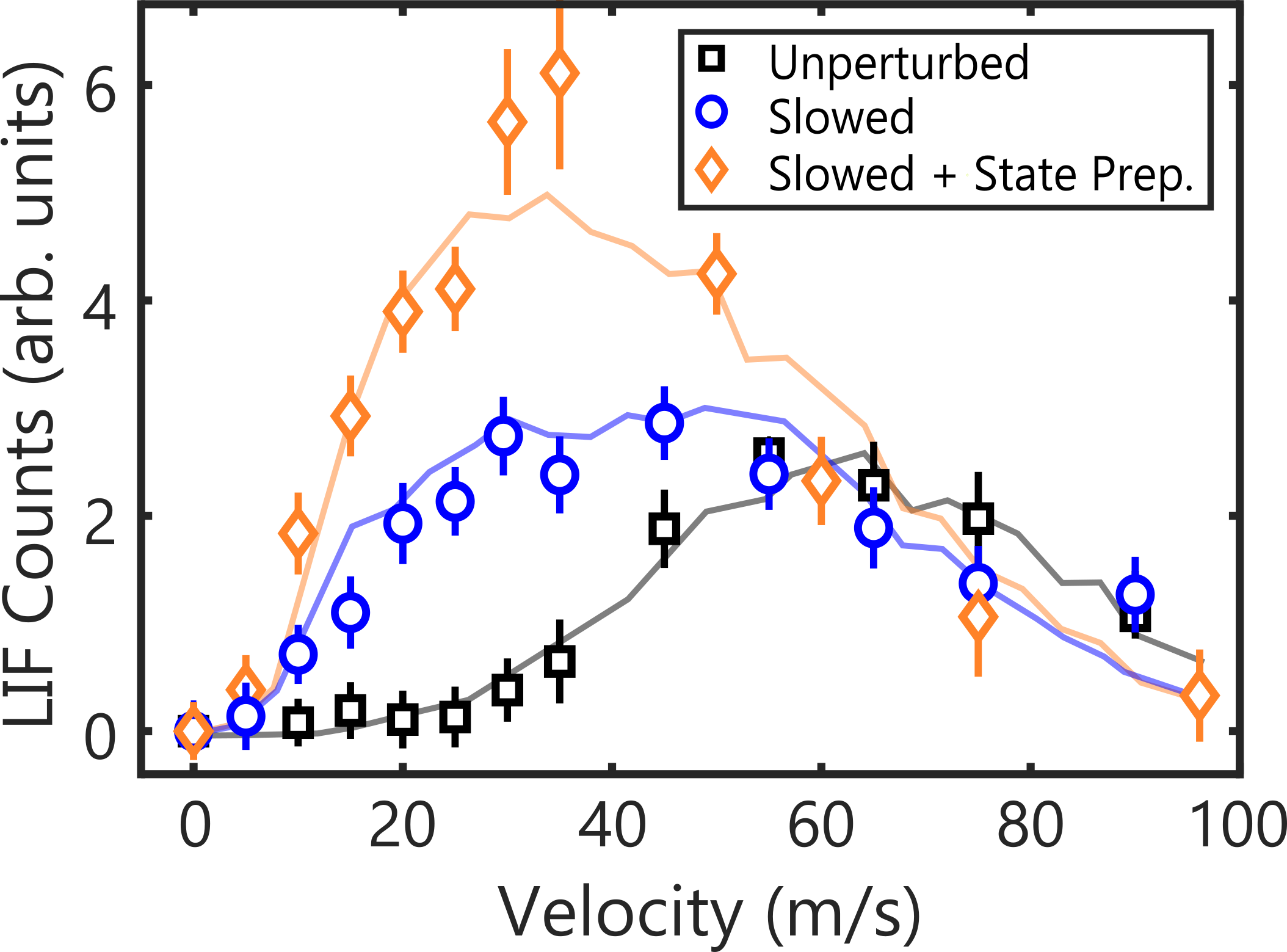}
    \caption[Zeeman-Sisyphus deceleration of CaOH including pumping in S1]{CaOH velocity distributions with (blue circles) and without (black squares) Zeeman-Sisyphus deceleration applied. Also shown (orange diamonds) is the velocity distribution when all molecules are pumped into WFS states in region S1 as they enter the decelerator. Solid lines are the results of Monte Carlo trajectory simulations that take into account the three-dimensional field profile inside the decelerator.}
    \label{fig:CaOHZSWithPR0}
\end{figure}

Based on measurements of the optical pumping into higher-lying vibrational states, it was found that fewer than 10 photons were scattered per molecule in the optical pumping steps to decelerate a CaOH molecule near the peak of the distribution by $\Delta v_f \approx 35$~m/s. By contrast, the radiative force due to 10 scattered photons would slow a CaOH molecule by just $\Delta v_f \approx 0.1$~m/s. This clearly indicates the promise of ZS deceleration to slow molecules for which radiation pressure force slowing would be impractically difficult due to to a limitation on the number of photon scattering events that can be realized. More recently, ZS deceleration has been extended to the complex polyatomic molecule YbOH by \cite{Sawaoka2022}. Again molecules with velocities below 20~m/s were produced.

\subsection{Magneto-optical trapping} \label{sec:MOT}
The magneto-optical trap (MOT) is an essential tool in atomic physics and, in particular, a key ingredient for producing ultracold samples of atoms with sufficiently high numbers (and densities) to be useful for subsequent science experiments. In brief, MOTs are based on radiation forces from six orthogonal laser beams, combined with a quadrupole magnetic field to tune the lasers into/out of resonance depending on position (\cite{phillips1998nobel,chu1998nobel}). By choosing the polarization appropriately, the trapped particles can be made to scatter photons preferentially from lasers that push them toward the center of the trap. This results in simultaneous trapping and rapid cooling to temperatures near the Doppler limit. An essential challenge in realizing MOTs for molecules is that they rely on the ability to cycle a large number of photons, typically at scattering rates of at least $\sim$$100~\text{kHz}$. Nonetheless, over the past decade magneto-optical trapping has been achieved for a number of diatomic molecules. In this section, we describe recent developments in extending magneto-optical trapping to polyatomic molecules, particularly the recent demonstration of a MOT for CaOH. The following discussion is based on the same experiment that was discussed in section~\ref{sec:CaOHWhiteLight} on radiative slowing of CaOH.

\begin{figure}[tb]
    \centering 
    \includegraphics[width=1\columnwidth]{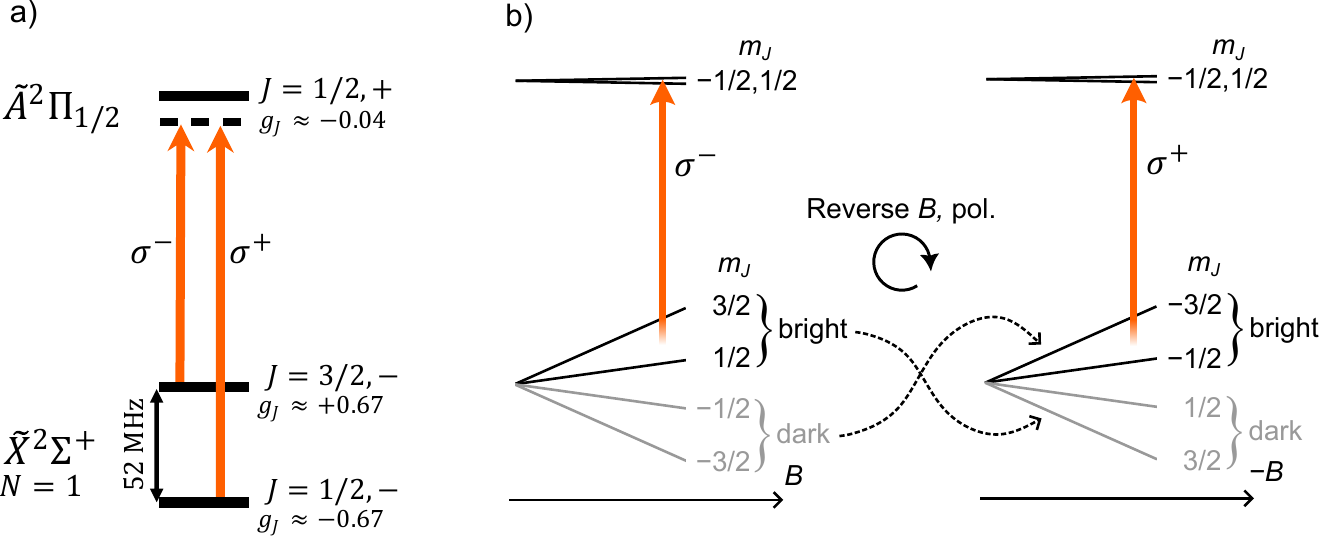}
    \caption{Overview of the CaOH laser cooling and RF MOT scheme. (a) The level structure of the main cycling transition used in the CaOH MOT, and the polarization configuration of the two frequency components of the MOT light. Since the $J=1/2$ and $J=3/2$ components have $g$-factors of opposite sign, each component is addressed with opposite polarization. Reproduced from \cite{vilas2022magneto}. (b) The magnetic sublevels of the $J = 3/2 \leftarrow J' = 1/2$ transition for the two configurations of the RF MOT. Magnetic dark states of each configuration are indicated.}
    \label{fig:Fig1_MOT}
\end{figure}

The CaOH MOT used the same rotationally-closed cycling transition as was used for laser slowing, i.e., $\widetilde{A}^2\Pi_{1/2}(000)(J'=1/2,p'=+) \leftarrow \widetilde{X}^2\Sigma^+(000)(N=1,p=-)$. The photon cycling scheme discussed in Sec.~\ref{sec:CaOHWhiteLight} allows CaOH molecules to scatter $\sim$$12{,}000$ photons in the MOT before $1/e$ of the population remains in bright states. In the experiment, the 11 repumping lasers that were used for slowing also covered the volume of the MOT; the same laser beams were used for both slowing and the MOT. Most of the repumping lasers used to achieve radiative slowing and a MOT for CaOH did not require much laser power, see Fig.~\ref{fig:Fig2_MOT}(b). While laser power requirements in laser cooling experiments with polyatomic molecules will generally depend on the exact photon cycling scheme used for the given molecule, the observed powers needed for CaOH provide a preliminary indication of the technical requirements for laser systems in laser cooling of polyatomic molecules. 

Coupling of the electron spin and rotational degrees of freedom splits the $\widetilde{X}^2\Sigma^+(000)(N=1)$ level into two spin-rotation (SR) components spaced by $\sim$$52~\text{MHz}$, with total angular momenta of $J=1/2$ and $J=3/2$. To address both SR components in the MOT, the MOT laser light includes two frequency components separated by the SR splitting, as shown in Fig.~\ref{fig:Fig1_MOT}(a). The small hyperfine splittings in $\widetilde{X}^2\Sigma^+(000)(N=1)$ of $\sim$$1~\text{MHz}$ were unresolved by the MOT light, so additional laser frequencies were not required to address hyperfine sublevels.

The transitions addressed in the CaOH MOT, $J'=1/2 \leftarrow J=1/2$ and $J'=1/2 \leftarrow J=3/2$, are both type-II transitions, characterized by $J' \leq J$. A defining characteristic of type-II MOTs is the presence of magnetic dark states, which arise because the ground state has more sublevels than the excited state. (A detailed description of the physics of type-II MOTs, and their comparison to type-I MOTs, was given by \cite{Tarbutt2015a,Tarbutt2015b}.) As described in Sec.~\ref{sec:CaOHWhiteLight}, these dark states must be destabilized in order to generate trapping forces in the MOT, which derive from repeated photon scattering. A common strategy to achieve this dark-state remixing is to rapidly switch the polarization of the MOT light while simultaneously alternating the direction of the magnetic field gradient. The switching occurs at a rate similar to the rate at which molecules pump into dark states, typically $1-2$~MHz, leading to the moniker radio-frequency (RF) MOT (\cite{norrgard2015sub}). Figure~\ref{fig:Fig1_MOT}(b) illustrates the situation for CaOH, showing how the RF switching destabilizes dark states to provide a trapping force and allows for continuous scattering from the laser beams that provide a restoring force toward the center of the MOT. A separate approach, known as a dual-frequency MOT, has been used for diatomic molecules and is likely to be possible for polyatomic molecules (\cite{Tarbutt2015b,Truppe2017,Ding2020Sub}). 

\begin{figure}[tb]
    \centering 
    \includegraphics[width=1\columnwidth]{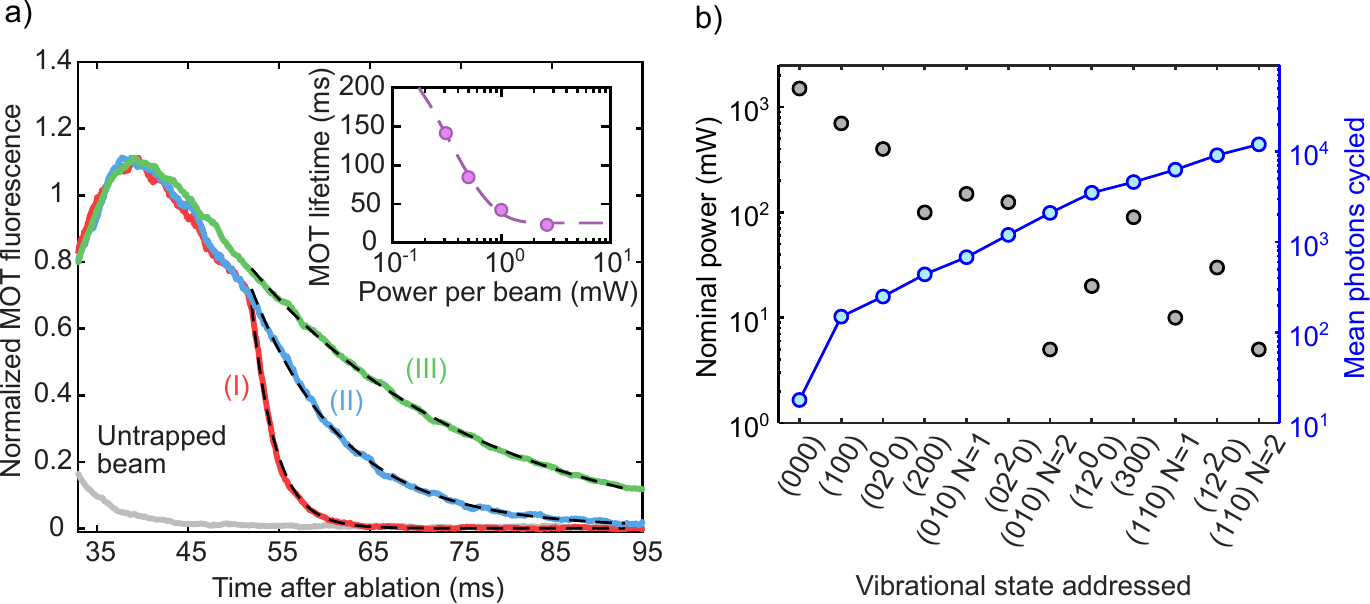}
    \caption{(a) Fluorescence signal from the CaOH MOT as a function of time after the molecules were produced. In the three curves (I), (II), and (III), different repumpers were turned off at $\sim$$50~\text{ms}$ to limit the average number of scattered photons per molecule to $1200$, $4600$, and $12{,}000$, respectively. Dotted lines are exponential fits to extract lifetimes, of $2.60(3)~\text{ms}$, $10.1(2)~\text{ms}$, and $25.7(6)~\text{ms}$ for (I)-(III). Inset: MOT lifetime as a function of MOT beam power. (b) Laser power required by each of the repumping lasers (denoted by the ground vibrational state addressed) used in the CaOH photon cycling scheme. Reproduced from \cite{vilas2022magneto}.
    }
    \label{fig:Fig2_MOT}
\end{figure}

\cite{vilas2022magneto} produced an RF MOT of CaOH molecules containing $2.0(5) \times 10^4$ molecules trapped at a peak density of $n = 3.0(8)\times10^8~\text{cm}^{-3}$. The temperature of the molecules ($T = 870(50)~\mu\text{K}$), the damping constant ($\beta = 455(85)~\text{s}^{-1}$), and oscillation frequency ($\omega = 2\pi \times 59(4)~\text{Hz}$), were all comparable to values characteristic of MOTs of diatomic molecules.\footnote{Here, the damping constant $\beta$ and oscillation frequency $\omega$ result from the common approximation of the MOT forces as a damped harmonic oscillator, $F/m = -\beta v - \omega^2 r$, where $v$ and $r$ are the velocity and position in the trap, respectively.} The lifetime of the CaOH MOT was limited by photon scattering causing population to accumulate in rovibrational states that are not addressed by the photon cycling scheme, as shown in Fig.~\ref{fig:Fig1_MOT}(a). The maximum achieved lifetime was $\sim$$150~\text{ms}$, similar to the timescale observed for diatomic molecules but much shorter than the lifetimes that can be realized in atomic MOTs. The characteristic damping time of the CaOH MOT---the time to compress the captured cloud of molecules---was an order of magnitude shorter than the lifetime, enabling full cooling and compression before the trapped molecules were lost to vibrational dark states. 

The temperature of the CaOH MOT was relatively high compared to the Doppler limit (around 150~$\mu$K), a situation that is common in type-II MOTs. This prevents direct loading from the MOT into conservative traps, such as optical traps, which have relatively low trap depths. The elevated temperature arises from polarization-gradient forces due to the presence of dark states in type-II transitions; we discuss these forces in more detail in Sec.~\ref{sec:SubDoppler}.

\subsection{Sub-Doppler cooling} \label{sec:SubDoppler}

As discussed in Sec.~\ref{sec:MOT}, dark states present in the optical cycling scheme of molecular MOTs can lead to relatively high temperatures for the trapped molecules. A variety of techniques have been developed to cool the molecules to lower temperatures, and many of these have been applied to polyatomic molecules. We discuss these methods here, using CaOH as a representative test case.

\subsubsection{Grey molasses}
  \begin{figure}[tb]
    \centering 
    \includegraphics[width=.9\columnwidth]{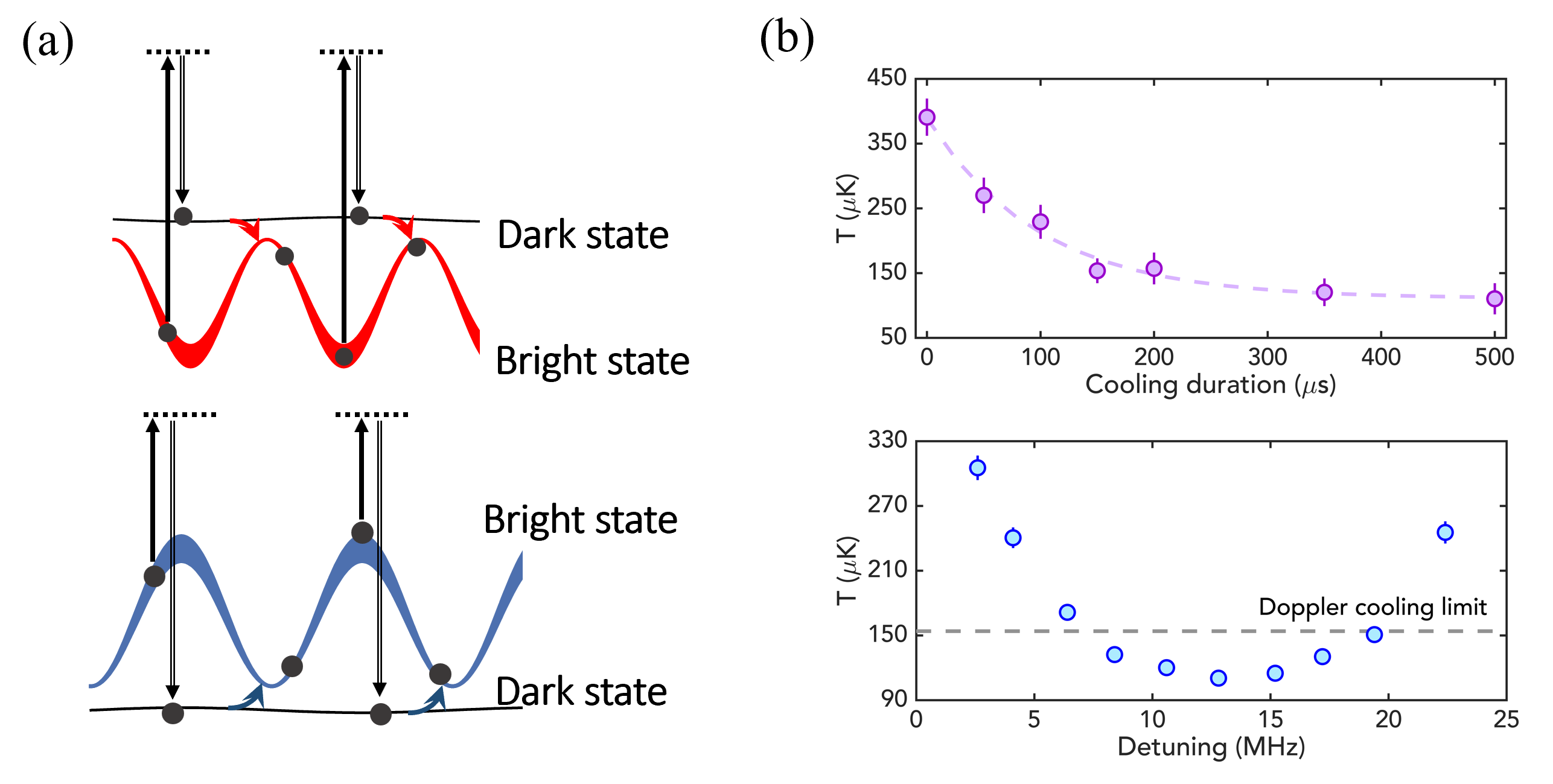}
    \caption{(a) Illustration of the grey molasses effect. Molecules are optically pumped into dark states near the antinodes of an intensity standing wave and then transit from dark to bright states near the standing wave's nodes. (b) Grey molasses cooling of CaOH. Top: Temperature as a function of cooling duration Bottom: Temperature as a function of sub-Doppler detuning $\Delta_\text{SD}$. Reproduced from \cite{vilas2022magneto}.}
    \label{fig:caohgm}
\end{figure}

As mentioned previously the temperature limitation in the MOT is due to a Sisyphus-like heating effect. This can be turned into a cooling effect by detuning the cooling light to the blue of resonance. Because photon cycling is achieved on an $F\rightarrow F$ and  $F\rightarrow F-1$ transition, dark states are present. This means the ground state manifold has both bright and dark states. The bright states, when blue detuned, are shifted above the dark state and are sinusoidaly modulated due to the AC stark shift, Fig.~\ref{fig:caohgm}(a). An atom or molecule in the ground state will move across this spatially varying potential, undergoing motional coupling from the
dark state to be bright state. The atom or molecule then moves up the potential hill and is driven to the excited state where it will
then decay back into the dark state. This cycle repeats, each time cooling the particle by an energy proportional to the amplitude of the AC stark modulation. This is commonly referred to as grey molasses cooling as it is a mix of a bright and dark molasses.

Grey molasses cooling was first successfully demonstrated in molecules by \cite{truppe2017CaF}, who focused on the molecule CaF. This cooling method was extended to polyatomic molecules by \cite{vilas2022magneto}, who used CaOH. It was found that, in the CaOH experiment, optimal cooling occurs at a detuning of 13~MHz (see Fig.~\ref{fig:caohgm}(b)). The temperature increases for higher detuning due the cooling light being red-detuned relative to the $J=1/2$ state. The observed (exponential) timescale for cooling is less than about 0.5~ms, indicating that the cooling is relatively rapid.

\subsubsection{$\Lambda$-enhanced grey molasses}
  \begin{figure}[tb]
    \centering 
    \includegraphics[width=.8\columnwidth]{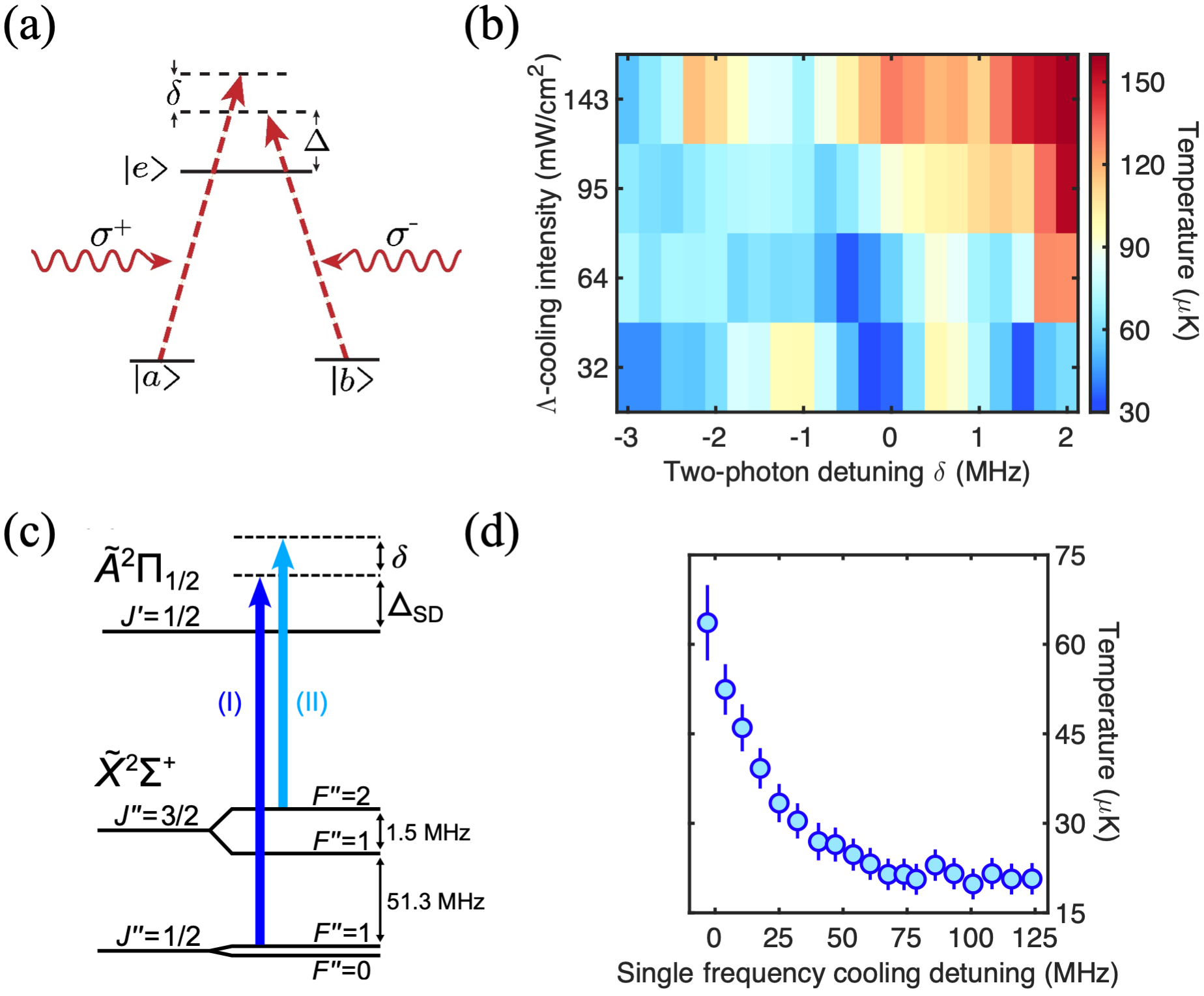}
    \caption{(a) Simple picture of the level structure used in $\Lambda$-enhanced grey molasses cooling. Ground states $|a\rangle$ and $|b\rangle$ are coupled to excited state $|e\rangle$ with single-photon detuning $\Delta$ and two-photon detuning $\delta$. (b) Temperature vs. two-photon detuning and cooling intensity for $\Lambda$-enhanced grey molasses cooling of CaOH. (c) Level scheme used in $\Lambda$-enhanced grey molasses and single-frequency cooling. $\Lambda$-enhanced cooling uses both frequencies, while only laser (I) is used for single-frequency cooling. (d) Temperature vs. detuning for single-frequency cooling of CaOH. Reproduced from \cite{Cheuk2018lambda} and \cite{Hallas2022}.}
    \label{fig:caohlambda}
\end{figure}

Standard grey molasses is often limited to temperatures $\sim$100~$\mu$K and more advanced schemes are required to reach lower temperatures. Two such techniques that have been successfully demonstrated in ultracold molecules are $\Lambda$-enhanced grey molasses (\cite{Cheuk2018lambda,langin2021}) and single frequency (SF) cooling (\cite{Caldwell2019}). Both techniques rely on creating velocity-dependent dark states.

$\Lambda$-enhanced grey molasses cooling combines grey molasses with velocity selective coherent population trapping (VSCPT) (\cite{aspect_1988}). The VSCPT mechanism relies on the creation of coherent dark states present in multi-level systems, as shown in Fig.~\ref{fig:caohlambda}(a). With counter-propagating circularly polarized laser beams, a superposition of two states (called $\ket{a}$ and $\ket{b}$ for generality) can be formed where the transition amplitudes destructively interfere to form a dark state. The dark state does not persist to large non-zero velocities because in that case the two beams are not at the same frequency (due to opposite Doppler shifts). 
While VSCPT cooling can reach sub-recoil temperatures in atoms, it is slow and inefficient, relying on random walks to cool toward zero velocity. However, by combining VSCPT cooling with grey molasses, the cooling no longer relies solely on a random walk, as the grey molasses provides a restoring force towards zero velocity and the VSCPT effect traps the atoms or molecules near zero velocity. However, the presence of grey molasses forces reduces the extent to which the zero-velocity state is ``dark,'' raising the minimum temperature achievable by this type of cooling. This $\Lambda$-enhanced grey molasses cooling method was first demonstrated on the D1 lines of alkali atoms (\cite{Grier2013}) and later in CaF molecules (\cite{Cheuk2018lambda}).

$\Lambda$-enhanced grey molasses cooling can be naturally extended to polyatomic molecules. In CaOH, this was achieved by coupling two hyperfine levels in the $\widetilde{X}^2\Sigma^+(N = 1)$ manifold to the $\widetilde{A}^2\Pi_{1/2}(J' = 1/2)$ excited state (\cite{Hallas2022}). Specifically, the $\widetilde{X}^2\Sigma^+(J=3/2, F=2)$ and $\widetilde{X}^2\Sigma^+(J=1/2)$ levels are addressed with $\sigma^-$ and $\sigma^+$ polarization, respectively. Both frequency components are blue-detuned by a common amount $\Delta_\text{SD}$ of about 12 MHz, and the two-photon detuning $\delta$ is varied. 

The dependence of $\Lambda$-enhanced grey molasses cooling on both $\delta$ and $I_\text{SD}$ is shown in Fig.~\ref{fig:caohlambda}(b). The lowest measured temperature, $T_\text{min}=34~\mu\text{K}$, occurs at $\delta \approx 0 \text{ MHz}$. A second, slightly higher local temperature minimum is observed at $\delta \approx 1.5 \text{ MHz}$, which corresponds to the two-photon resonance for the $\Lambda$-system consisting of $\widetilde{X}^2\Sigma^+(J=3/2, F=1)$ and $\widetilde{X}^2\Sigma^+(J=1/2)$. At higher intensities, the temperature is minimized at increasingly negative $\delta$ because of the AC Stark shifts of the hyperfine levels coupled in the $\Lambda$-enhanced grey molasses: for higher intensities, the levels move further apart, and smaller $\delta$ is required to satisfy the two-photon resonance condition.

\subsubsection{Single-frequency cooling}

A primary limitation of $\Lambda$-enhanced grey molasses cooling is the fact that the dark states are destabilized by off-resonant scattering from the two laser frequencies interacting with nearby states. Single-frequency cooling solves this by creating dark states with a single laser frequency at large detuning, reducing this effect. This cooling method was first demonstrated for molecules by \cite{Caldwell2019}, who used CaF molecules in their experiment.
 
Single-frequency cooling can also be implemented in polyatomic molecules, as was shown with CaOH by \cite{Hallas2022}. By applying light blue-detuned by an amount $\Delta_\text{SD}$ (about 70~MHz) from the $\widetilde{A}\,^2\Pi_{1/2}(J = 1/2) \leftarrow \widetilde{X}\,^2\Sigma^+(N = 1, J=1/2)$ transition, a minimum temperature $T_\text{min} = 20~\mu\text{K}$ was realized (Fig.~\ref{fig:caohlambda}(c-d)). The cooling was observed to be insensitive to detuning above a certain value ($\Delta_\text{SD} \approx 70 \text{ MHz}$), Fig.~\ref{fig:caohlambda}(d). This insensitivity is beneficial for cooling molecules into an ODT, where trap-induced light shifts can affect the cooling efficiency.

\subsection{Optical trapping} \label{sec:ODT}

  \begin{figure}[tb]
    \centering 
    \includegraphics[width=.95\columnwidth]{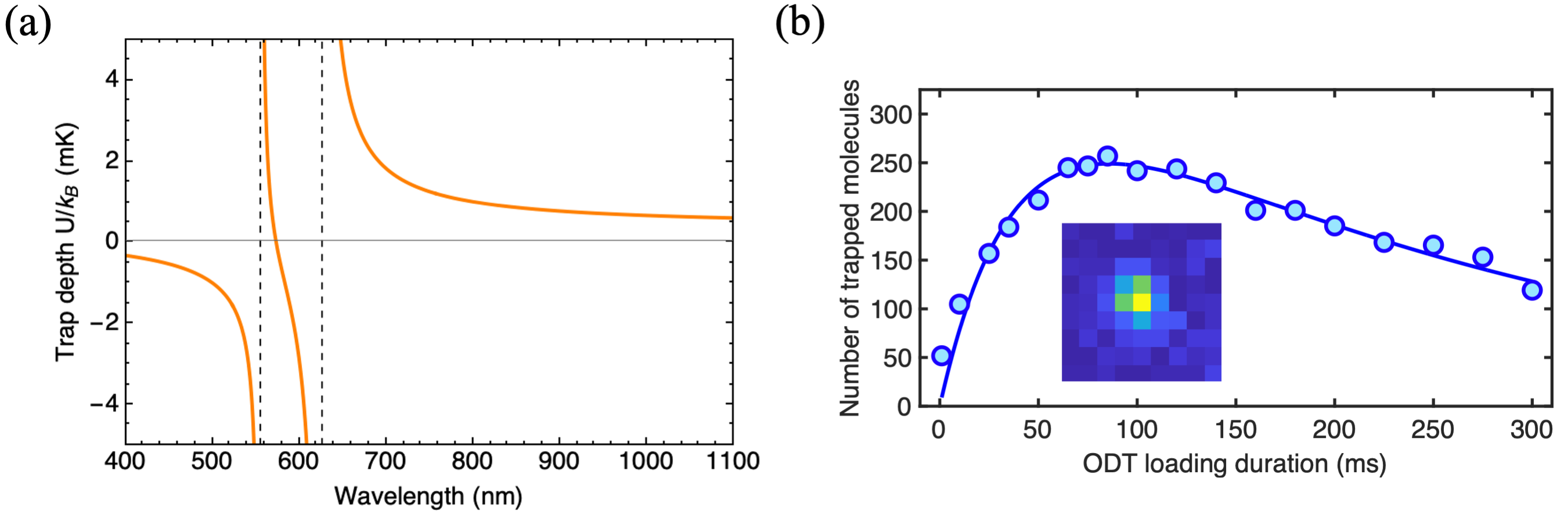}
    \caption{(a) Trap depth vs. wavelength for an optical dipole trap (ODT) of CaOH generated from a 13~W laser with a 25~$\mu$m waist. (b) Loading of CaOH molecules into an ODT. Reproduced from \cite{Hallas2022}}
    \label{fig:odt}
\end{figure}

Due to the availability of high-power fiber lasers, optical trapping has become a popular method to trap ultracold atoms and diatomic molecules. Optical trapping has several key advantages, including the ability to trap atoms and molecules irrespective of their internal state, and the ability to greatly increase phase space density due to the small trap volume. However, laser cooling of molecules inside an optical trap can be hindered by a variety of effects, including differential AC Stark shifts. Optical dipole traps consist of tightly focused Gaussian beams, which in combination with the induced Stark shift on the molecules, creates a harmonic confining potential for the molecules. This potential is wavelength-dependent, and the dependence can be quite complicated due to the many levels present in molecules. In the limit of large detuning from an electronic transition, the strength of the trapping potential is inversely proportional to the detuning. \cite{Caldwell2020sideband} provide details on calculating Stark shifts for molecules in trapping laser fields. The trap depth as a function of wavelength for the polyatomic molecule CaOH, in the large-detuning limit, is shown in Fig.~\ref{fig:odt}(a).

Optical trapping of directly laser-cooled molecules was first demonstrated with CaF (\cite{anderegg2018laser}). To efficiently load molecules into the optical dipole trap, the molecules must be cooled into the trap. This was accomplished with CaF by overlapping a far detuned 1064~nm laser with the cloud of molecules during grey molasses cooling. As the molecules traversed the optical trapping light, the grey molasses cooled the molecules, loading them into the trap and increasing their density. Using $\Lambda$-enhanced grey molasses, the transfer efficiency into the ODT was greatly improved (\cite{Cheuk2018lambda}). 

Optical trapping of CaOH was demonstrated in much the same fashion (\cite{Hallas2022}). Figure~\ref{fig:odt}(b) shows the loading of CaOH molecules into an ODT from the single-frequency grey molasses cloud. ODT loading is relatively inefficient, typically transferring only $1-10\%$ of the molecules from the molasses. This could be improved by employing new cooling and trapping techniques, such as blue-detuned optical traps (\cite{lu2022}) and blue-detuned MOTs (\cite{jarvis2018}), where molecules may be directly loaded from the MOT. Recent work by \cite{burau2022blue} has demonstrated the advantages of a blue-detuned MOT for the diatomic radical YO. The additional substructure present in polyatomic species has not been found to hinder the optical trapping process in any significant way. Generically, however, the increased density of states in larger polyatomic molecules may increase the likelihood of unwanted excitations, e.g. due to ``accidental'' resonances or via Raman or multi-photon processes.

  \begin{figure}[tb]
    \centering 
    \includegraphics[width=.95\columnwidth]{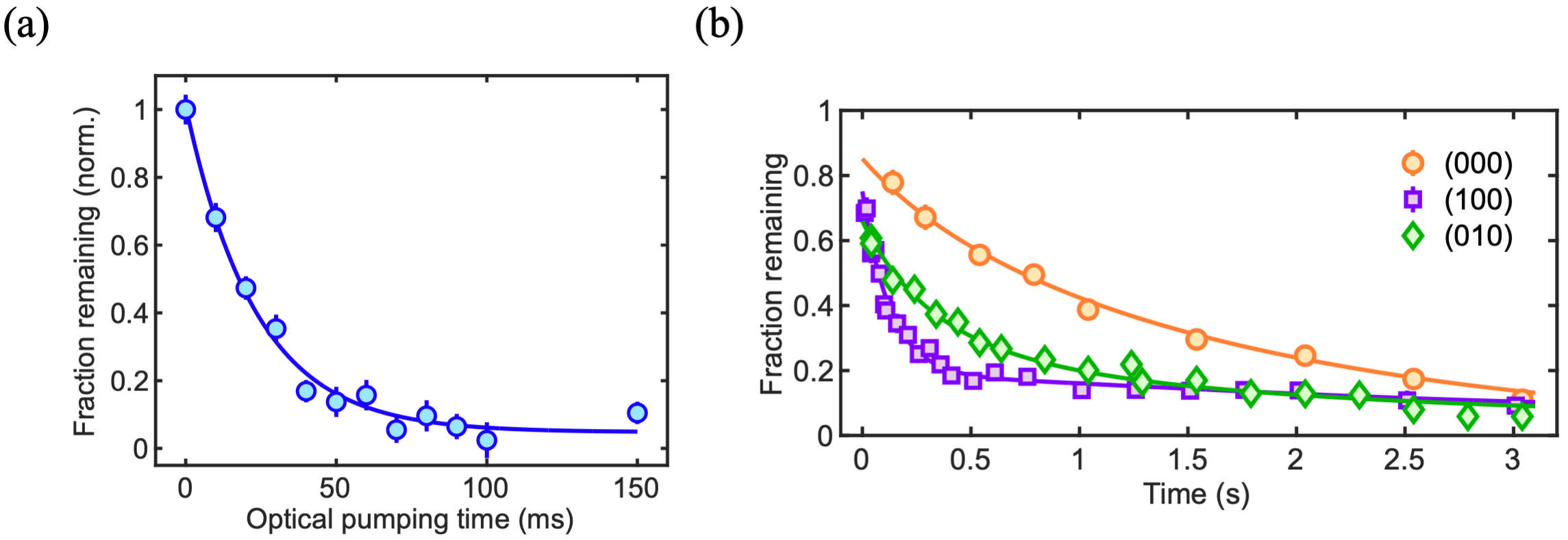}
    \caption{(a) Optical pumping of optically trapped CaOH molecules into the $\widetilde{X}(010)$ bending vibrational mode via single frequency cooling with the $\widetilde{X}(010)$ repumping laser removed. (b) Lifetime of optically trapped CaOH molecules in the $\widetilde{X}(000)$, $\widetilde{X}(010)$, and $\widetilde{X}(100)$ vibrational levels. The solid curves are fits to a rate equation model capturing blackbody excitation and radiative decay along with vacuum losses. Reproduced from \cite{Hallas2022}}
    \label{fig:odtlifetime}
\end{figure}

Along with trapping of CaOH molecules in the $\widetilde{X}(000)$ vibrational ground state, \cite{Hallas2022} demonstrated trapping of CaOH in the excited $\widetilde{X}(010)$ bending vibrational state and the $\widetilde{X}(100)$ stretching vibrational level. Trapped molecules were optically pumped into these states by applying a single-frequency molasses while turning off the corresponding vibrational repumping laser. For the $\widetilde{X}(010)$ bending mode the optical pumping requires 1200 photon scattering events before the average molecule vibrationally decays, corresponding to an optical pumping timescale of 23 ms (Fig. \ref{fig:odtlifetime}(a)). Optical pumping into the $\widetilde{X}(100)$ stretching mode was much faster due to the large branching ratio to this state ($\sim$5\%). Fig. \ref{fig:odtlifetime}(b) shows measurements of the lifetime of CaOH molecules trapped in each of the $\widetilde{X}(000)$, $\widetilde{X}(010)$, and $\widetilde{X}(100)$ states. It was found that the ground state lifetime was limited primarily by room-temperature blackbody excitation to excited vibrational levels and by imperfect vacuum, while the excited state lifetimes were shorter due to spontaneous, radiative decay back to the vibrational ground state (\cite{Hallas2022}). The lifetimes of all three states could be improved by cooling the surrounding environment to reduce blackbody radiation-induced losses.

\subsection{Preparation and coherent control of single quantum states} \label{sec:CoherentControl}
\begin{figure}[tb]
    \centering 
    \includegraphics[width=.65\columnwidth]{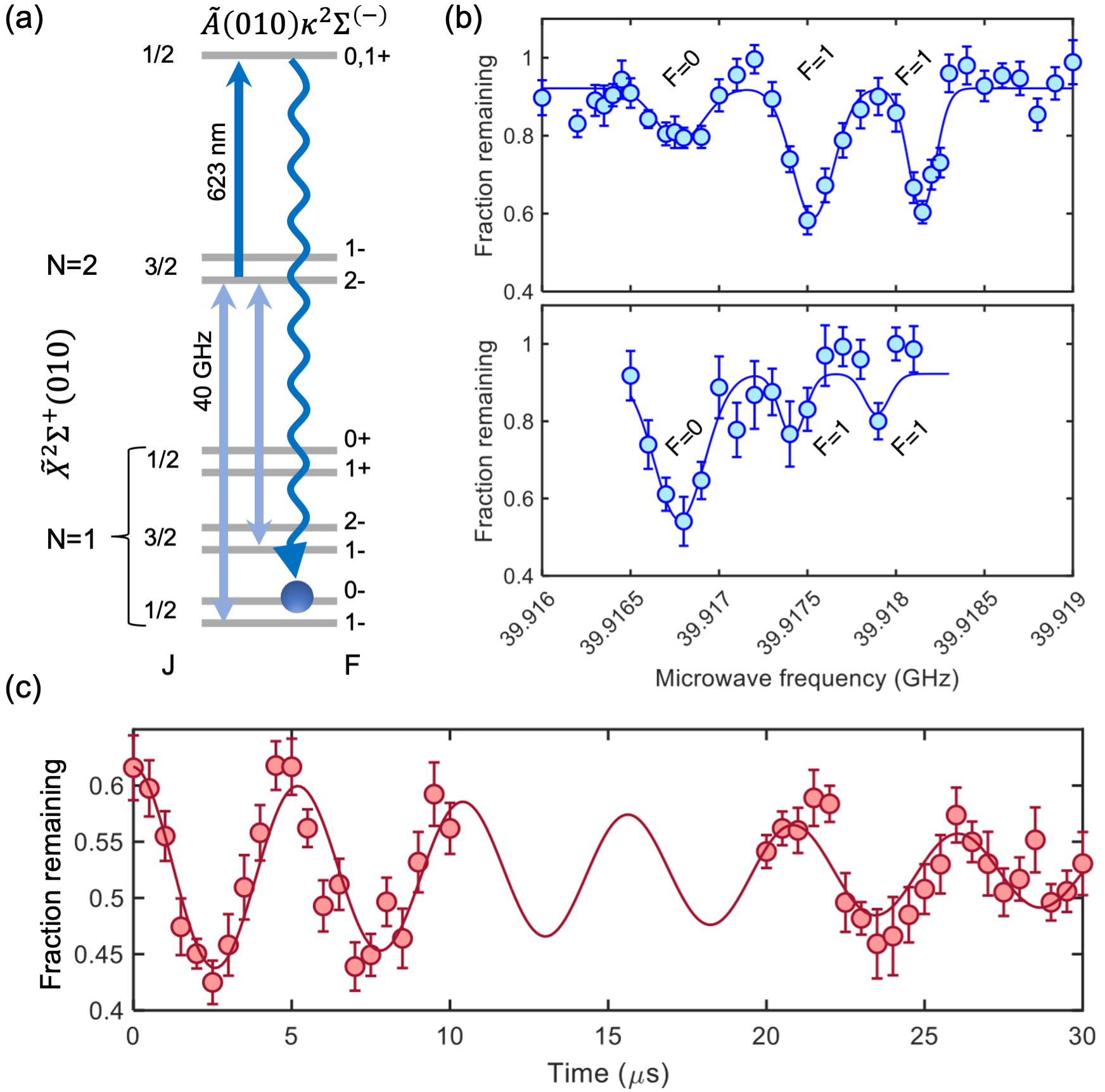}
    \caption{Demonstrations of coherent control of CaOH molecules. (a) Level diagram showing the microwave/optical pumping steps used to transfer population into a single quantum state. (b) Microwave spectroscopy showing $F=0$ and $F=1$ population before and after optical pumping into $F=0$. (c) Coherent Rabi oscillations driven with 40~GHz microwaves between the $N=1$ and $N=2$ levels of optically trapped CaOH molecules in the $\widetilde{X}(010)$ bending mode. Panel (a) is adapted from \cite{anderegg2023quantumcontrol}.
    }
    \label{fig:rabi}
\end{figure}

Following sub-Doppler cooling and ODT loading, the trapped molecular population is distributed over multiple hyperfine sublevels. The many internal states present in polyatomic molecules complicate the task of transferring this population into a single quantum state. An optical pumping sequence for CaOH, shown in Fig.~\ref{fig:rabi}(a), is used to populate a single quantum state in the $\widetilde{X}\,^2\Sigma^+(010)(N=1^-)$ vibrational bending mode. This bending mode is of interest because it is precisely the state proposed for use in various quantum computation or precision measurement experiments (\cite{kozyryev2017PolyEDM, Yu2019}).
Following optical pumping into the bending mode as described in the previous section, the molecular population is spread across twelve hyperfine states. To prepare the molecules in a single quantum state, a microwave-optical pumping sequence is employed; microwave transitions allow hyperfine splittings below the linewidth of optical transitions to be resolved, while optical excitation provides the dissipation necessary for single state preparation. In CaOH, molecules can be prepared in the $(N=1, J=1/2^-, F=0)$ state using the following sequence. Microwaves are first used to drive population from the $(N=1, J=3/2^-)$ state to the $(N=2, J=3/2^-)$ level. A small electric field is applied to mix states of opposite parity and thereby lend transition strength to this nominally forbidden transition. An optical transition then drives population from ($N=2, J=3/2^-$) to the excited $\widetilde{A} (010) \kappa^2 \Sigma^{(-)}(J=1/2^+)$ electronic state, which decays to the $F=0$ (the target state) and $F=1$ levels of the ($N=1, J=1/2^+$) manifold. This sequence is then repeated, but with the microwaves driving population in ($N=1, J=1/2^+, F= 1$) to ($N=2, J=3/2^-$), where molecules are again optically excited and pumped into the target $F=0$ state. Spectroscopy scans of the ($N=1, J=1/2^+)$, $F= 0$ and $F=1$ states before and after optical pumping are plotted in Fig.~\ref{fig:rabi}(b), showing that the population of the $F=0$ state is greatly enhanced. The population in the $F=0$ state can then be transferred to the desired target state and any remaining molecules that were not successfully transferred can be pushed out of the trap using resonant laser light. Rabi oscillations between states can also be driven, as shown in Fig.~\ref{fig:rabi}(c). In CaOH, hyperfine splittings are approximately 1 MHz, meaning that Rabi frequencies $<$1 MHz are necessary in order to avoid off-resonant excitation.

\section{Outlook and challenges} \label{sec:Outlook}
\subsection{Toward larger molecules} 
One of the underlying trends in the quest to laser cool polyatomic molecules has been a drive to control increasingly large and complex molecules. Nearly from the first proposals to laser cool polyatomic molecules, by both \cite{isaev2015polyatomic} and \cite{kozyryev2016MOR}, authors were identifying molecules with five or more atoms that appeared to have FCFs sufficiently diagonal to admit direct laser cooling. These proposals relied critically on the concept eventually dubbed an ``optical cycling center,'' such as the MO (M=Ca,Sr,Yb, etc.) moiety that has formed the core of the laser cooling experiments described throughout this review. 

More recently, theoretical work has identified an even wider range of aromatic molecules that can be adorned with optical cycling centers, including phenols (and derivatives; \cite{Dickerson2021,Ivanov2020Toward}), polycyclic arenes (\cite{Dickerson2021Optical}), and fully saturated hydrocarbons (\cite{Dickerson2022}). Remarkably, \cite{Dickerson2021} were even able to show that substitutions around a cyclic hydrocarbon could be used to \textit{tune} the FCFs of the metal-centered excitations using simple principles from organic chemistry---a capability that is only possible in large polyatomic species. Experimental verification of these theoretical predictions has been obtained by \cite{Zhu2021,Mitra2022pathway,Augenbraun2022,Lao2022}, who have synthesized both phenol and naphthol derivatives adorned with Ca- and Sr-based optical cycling centers and shown that these molecules indeed have the properties desired of laser-coolable species (namely diagonal FCFs and localized metal-centered electronic excitations).

\begin{figure}
    \centering
    \includegraphics[width=0.63\linewidth]{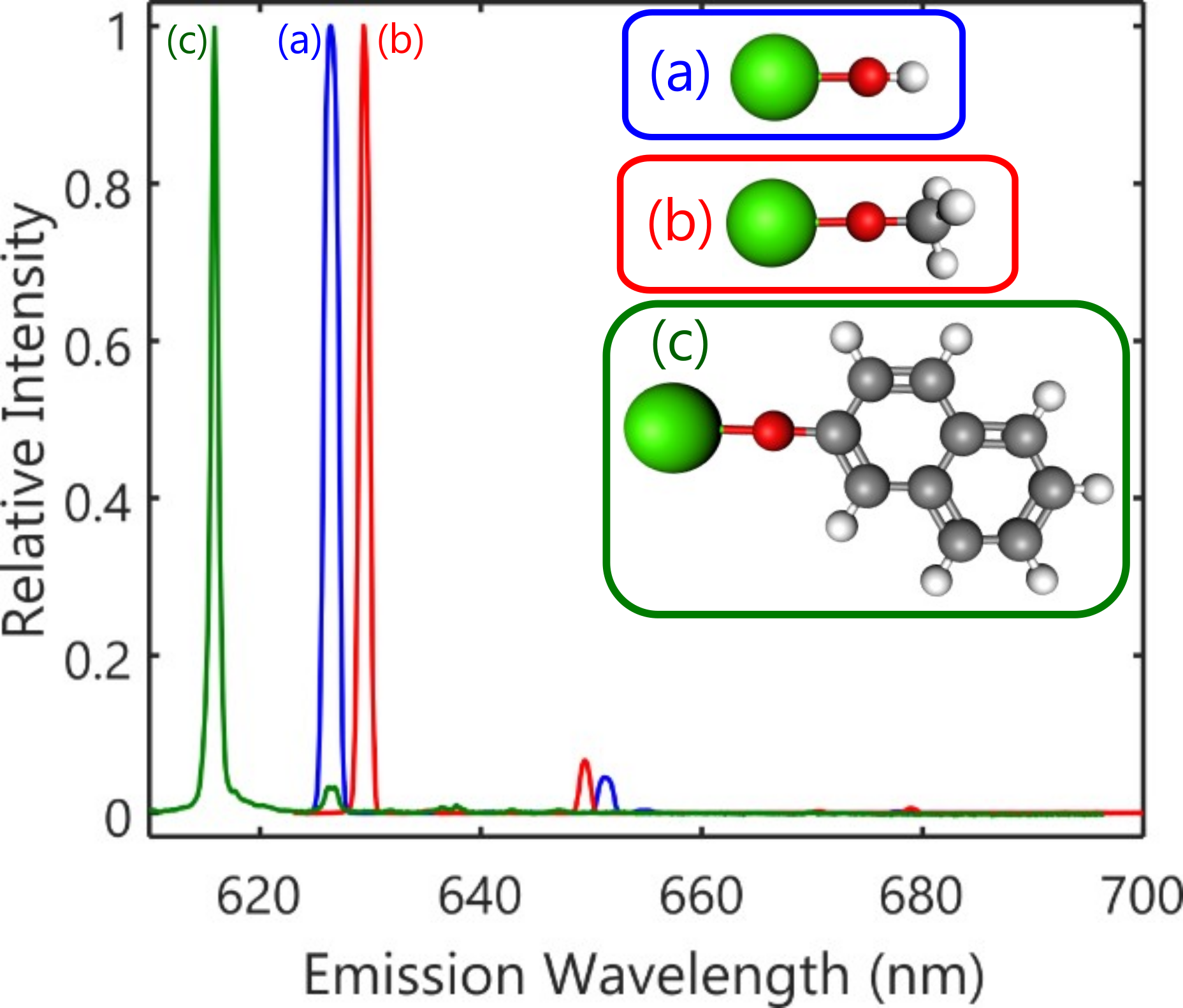}
    \caption[Comparison of DLIF spectra for CaOH, CaOCH$_3$, and \CaONp]{Comparison of DLIF spectra following excitation of (a) CaOH, (b) CaOCH$_3$, and (c) \CaONp to the $\tilde{A}$ excited state. Spectra (a) and (b) were recorded using fluorescence from a molecular beam, and spectrum (c) was recorded by collecting fluorescence from inside a buffer-gas cell. The diagonal fluorescence features are normalized to unity.}
    \label{fig:LargeMoleculeDLIFComparison}
\end{figure}

Figure~\ref{fig:LargeMoleculeDLIFComparison} compares the DLIF spectra for three Ca-containing molecules of increasing complexity: CaOH, CaOCH$_3$, and \CaONp (based on measurements reported by \cite{Zhang2021,AugenbraunThesis,Mitra2022pathway}). Despite the fact that \CaONp contains over an order of magnitude more vibrational modes than CaOH, the gross structure of their DLIF spectra is largely similar. In all cases, the Ca-O stretching mode is the dominant off-diagonal decay channel. \CaONp shows some activity in a handful of additional modes at the $\mysim0.1-1\%$ level, indicating that achieving optical cycling will be more challenging--- but not necessary prohibitive.

 The laser cooling and full quantum control of larger molecules (containing a dozen or more atoms) is at the very frontier of the the field, so much so that ideas of what to do with them are just beginning to be explored. Fundamentally, the larger number of atoms in the molecule, the larger the number of vibrational modes and hyperfine states. The concept of ``internal motion'' also starts to enter, e.g. a spinning ligand. Such modes can naturally be used to store quantum information, but exactly how this can be done and how useful this will be has not been fully explored. With a large enough molecule, one may be able to completely separate the laser cooling and readout section of the molecule (through an ``optical cycling center'') from the physics end, perhaps containing an exotic atom such as a heavy radioactive species. Thus, one might be able to realize a ``configurable'' molecular framework that allows targeted substitution of scientifically interesting components.

\subsection{Other molecular motifs}
To date, all laser-cooled polyatomic molecules are of the form MOR, in which an alkaline-earth-like metal hosting a localized valence electron is bonded to a linker oxygen atom and electronegative radical. However, other molecular structures may also be suitable for laser cooling. Closely related to MOR molecules are other ML molecules, where L is an electronegative ligand such as NC, SH, or CH$_{3}$ (\cite{Norrgard2019nuclear,Augenbraun2020ATM}). We describe two more dramatic departures from the MOR model, which appear to be favorable for future laser cooling experiments.

First, polyatomic molecules could be functionalized with multiple optical cycling centers, for example the linear molecules SrCCSr and YbCCCa, or asymmetric top molecules consisting of two metals linked by a benzene ring (\cite{ORourke2019,Ivanov2020}). These systems are expected to exhibit enhanced scattering rates due to the presence of two optical cycling centers, and offer separation of functions between distinct metals (for example, precision measurement localized on a Yb atom, and co-magnetometry localized on a Ca atom). Symmetric molecules like SrCCSr possess a structure analogous to Sr$_{2}$, which is a leading candidate for a molecular clock with applications to precision measurement (\cite{zelevinsky2008precision}).

Second, polyatomic molecules with multivalent optical cycling centers, for example AlSH, should also be possible (\cite{yu2022multivalent}), in a manner analogous to diatomic molecules like AlF, AlCl, TlF, BH, and CH (\cite{Hofsass2021optical,Daniel2021spectroscopy,Grasdijk2021centrex,Hendricks2014,Schnaubelt2021cold}). Although the generalization from MF to MOH molecules has been highly successful for alkaline-earth-like metals (e.g., M=Ca,Sr,Yb), calculations show that for p-block metals (e.g., M=Al,Si,P), MOH molecules are bent and undergo a large bond angle deflection upon electronic excitation. This phenomenon can be mitigated with the use of a different linker atom like S or Se despite their lower bond polarity, due to reduced bond repulsion. Thus by careful tuning of the competition between bond repulsion and bond polarity, optical cycling of polyatomic molecules also appears feasible for species with p-block metals. Generalization of optical cycling to other structures may also be possible, but remains so far unexplored.

\subsection{Challenges and possibilities for other polyatomic molecules}
As larger and more complex molecular species are explored, new difficulties and limitations of quantum state control are likely to arise. Larger polyatomic molecules offer richer internal structures, a promising prospect for encoding qudits (higher dimensional analogues of qubits), and unique rovibrational modes, a potential platform to use for searches for fundamental symmetry violations. However, controlling these structures will potentially be more difficult than is the case for small polyatomic species. Several open questions remain. For example, in the case of ATMs identified by \cite{Augenbraun2020ATM}, it is generally the case that the excited electronic states have large and anisotropic $g$-factors; it will be necessary to understand how that structure affects the magnetic-field-dependent forces that are necessary in a MOT. In addition, nonlinear molecules are subject to many symmetry-breaking effects such as the Jahn-Teller effect and other vibronic couplings, and it will be necessary to understand the extent to which these features affect the rovibrational selection rules that aided laser-based control of smaller and higher symmetry species.

Other challenges may arise in the quest to gain quantum-state control over larger species. From a practical standpoint, because buffer-gas cooling techniques are generally most applicable at temperatures above about 0.5 K, large molecules will be distributed over a substantially larger number of internal states. Moreover, the energy levels in large molecules are separated by smaller intervals, which generically complicates the task of achieving coherent control of these molecules. With smaller spacings, Rabi frequencies must be reduced in order to suppress off-resonant excitation and unwanted state transfer. In quantum simulation/quantum information processing applications, for example, reduced Rabi frequencies may become an impediment to high-speed gate operations. To complicate the problem, the larger number of states in these complex polyatomic molecules lead to more pathways present for blackbody excitations and spontaneous decay, limiting the coherence times of experiments.

There may also be fundamental challenges due to the complex level structure of large polyatomic molecules. For example, larger molecules are potentially more susceptible to non-radiative loss channels that may interrupt optical cycling (\cite{Bixon1968}). We present here a simple model that, while speculative, conveys our sense of the structural questions that must be understood in order to achieve laser cooling of increasingly large polyatomic molecules. The essential details follow the treatment provided by \cite{Uzer1991}. Readers should also consult the excellent, and highly pedagogical, overview of intramolecular vibrational energy redistribution presented by \cite{Nesbitt1996}. Our model begins with the observation that the density of vibrational states at some energy above the absolute ground state grows very rapidly with molecule size, especially for molecules that contain low-frequency ($\nu \lesssim 100$~\wavenumber) vibrational modes.  In many cases, laser cooling transitions involve excitation to an excited electronic state $\tilde{A}$ with energy below the dissociation threshold of the ground electronic state $\tilde{X}$, meaning that $\tilde{A}$ is embedded in a dense manifold of highly-excited vibrational levels of $\tilde{X}$. See Fig.~\ref{fig:IVR} for calculations of the density of vibrational states for molecules of the form CaOR obtained using the method presented in \cite{Haarhoff1964}. For ligands, R, such as C$_6$H$_5$ or C$_{10}$H$_7$, the density of vibrational states at the location of the $\tilde{A}(v=0)$ level can be as large as $10^{11}$ or $10^{15}$ states per \wavenumber, respectively. For CaOC$_6$H$_5$ (CaOC$_{10}$H$_7$) this means there are around $10^{8}$ ($10^{12}$) dark states within the frequency range spanned by the natural linewidth of a typical bright state ($\sim$30~MHz).

\begin{figure}
	\centering
	\includegraphics[width=0.97\linewidth]{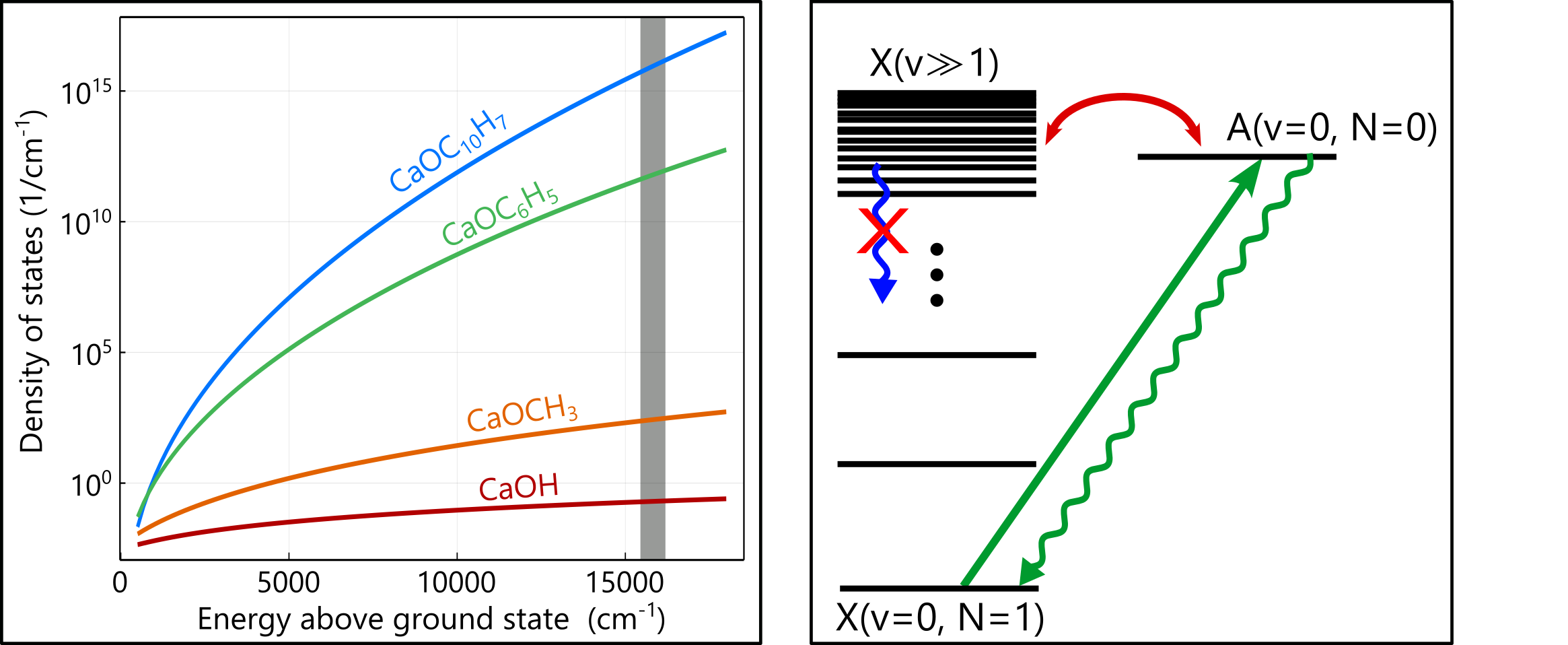}
	\caption{(Left) Density of states as a function of energy above the absolute ground state $\tilde{X}(v=0)$ for CaOR molecules with ligands of increasing size. The vertical gray band indicates the typical energy range of the $\tilde{A} \leftarrow \tilde{X}$ electronic transition in these species. (Right) Schematic diagram highlighting the levels important to our hypothesized loss mechanism.}
	\label{fig:IVR}
\end{figure}

Consider a near-resonant excitation that transfers molecular population from a ground rovibronic state $\ket{\phi_0}$ (in practice a single rotational level of the $\tilde{X}(v=0)$ manifold) to a ``bright'' rovibronic excited state $\ket{\phi_b}$ that has natural lifetime $\tau_b$ (typically a single rotational level of the $\tilde{A}(v=0)$ or $\tilde{B}(v=0)$ manifolds). If $\ket{\phi_b}$ were an energy eigenstate, the only time evolution that would occur following excitation is decay back to $\ket{\phi_0}$ at a rate $\Gamma_b = 1/\tau_b$. If, however, $\ket{\phi_b}$ is coupled via some vibronic interaction to the highly-excited levels of $\tilde{X}$ via a matrix element $V_{bd}$, then $\ket{\phi_b}$ is not an energy eigenstate; the energy eigenstates can be expressed as
\begin{equation}
	\ket{\psi_n} = a_{bn} \ket{\phi_b} + \sum_{j=1}^{N} a_{d_j n} \ket{\phi_{d_j}}.
\end{equation}
Thus, laser excitation that selectively targets $\ket{\phi_b}$ actually populates numerous energy eigenstates, meaning the excited state that is prepared will evolve in time and lead to population building up in the dark manifold $\lbrace \ket{\phi_d} \rbrace$ with a characteristic timescale $\tau_{bd}$. Because the states in $\lbrace \ket{\phi_d} \rbrace$ are very closely spaced, the timescale for population to return to $\ket{\phi_b}$ ($\tau_\text{return}$) may be very long. Under conditions where $\tau_{bd} \ll \tau_{b}$ (so that leakage to the dark manifold occurs more rapidly than photon emission) and $\tau_\text{return} \gg \tau_{b}$ (so that population remains in the dark manifold over many natural lifetimes), this process will appear to cause loss from the laser cooling experiment. The timescale during which population is primarily found in the dark manifold limits the achievable photon scattering rate, directly impeding optical cycling and laser cooling. Under a crude estimate based on Fermi's golden rule, if $\tau_{bd}$ is to compete with spontaneous emission (for a typical excited-state lifetime of about 30 ns), we must require
\begin{equation}
	\frac{1}{\tau_{bd}} \approx 2\pi \langle V^2 \rangle \rho,
\end{equation}
where $\sqrt{\langle V^2 \rangle}$ is the root-mean-squared coupling matrix element between the bright and dark manifolds and $\rho$ is the density of states. For a molecule with vibrational density of states $\rho \approx 10^{14}$ per \wavenumber, an average coupling matrix element as small as $\sqrt{\langle V^2 \rangle} \sim 10^{-9}$~\wavenumber would be sufficient to satisfy this condition. It is not currently clear whether coupling matrix elements of this magnitude are present in the large CaOR molecules being proposed for laser cooling applications. It is also possible that the average coupling matrix element could be tuned through judicious molecular design choices that reduce interaction between the metal-centered valence electron and the portion(s) of the molecule that contribute most to the vibrational density of states. It is critical that new theoretical and experimental studies be pursued to study these questions.

\section{Conclusion} \label{sec:Conclusion}
In this review, we have described how the field of ultracold polyatomic molecules is poised to impact many of the frontiers in modern atomic, molecular, and optical physics. Recent experimental results have shown that much of the toolbox developed for ultracold atoms can be applied to molecules with equal success. As we have described, the control of polyatomic molecules depends on a careful understanding of their internal structure. However, the class of molecules that has garnered the most experimental attention (alkaline-earth psuedohalides) share a set of common features such as diagonal FCFs and manageable rotational selection rules, that points toward a generic way to ``design'' laser-coolable molecules with diverse geometries, atomic constituents, and responses to external fields. The breadth of molecular species and structures that appear amenable to laser cooling is extremely large, and continues to grow as researchers explore increasingly complex systems.

\section{Acknowledgments}
We gratefully acknowledge valuable comments from Profs. David DeMille, Nicholas Hutzler, Michael Tarbutt, and Jun Ye, as well as Parul Aggarwal and Calder Miller. We also thank Prof. Robert W. Field and Bryan Changala for insightful input on IVR coupling mechanisms in larger molecules. The work described in this paper that was conducted within the Doyle group was supported by the AFOSR, the NSF, and the Heising-Simons Foundation.   

\bibliography{references}

\end{document}